\documentclass[a4paper,fleqn]{cas-sc}

\usepackage[authoryear,longnamesfirst]{natbib}
\usepackage{graphicx}
\usepackage{epstopdf, epsfig}
\graphicspath{{figures/}} 
\usepackage{xcolor}
\usepackage{amsmath}
\usepackage{amssymb}
\usepackage{subcaption}
\usepackage[percent]{overpic}

\def\tsc#1{\csdef{#1}{\textsc{\lowercase{#1}}\xspace}}
\tsc{WGM}
\tsc{QE}
\tsc{EP}
\tsc{PMS}
\tsc{BEC}
\tsc{DE}

\begin{document}
\let\WriteBookmarks\relax
\def\floatpagepagefraction{1}
\def\textpagefraction{.001}
\shorttitle{Response of a plate in turbulent channel flow: Analysis of
  fluid-solid coupling}
\shortauthors{S. Anantharamu and K. Mahesh}


\title [mode = title]{Response of a plate in turbulent channel flow: Analysis of
  fluid-solid coupling}                      




\author[1]{Sreevatsa Anantharamu}
\ead{anant035@umn.edu}


\address[1]{Aerospace Engineering and Mechanics, University of Minnesota - Twin Cities, Minneapolis, MN 55455, USA}

\author[1]{Krishnan Mahesh}
\cormark[1]
\ead{kmahesh@umn.edu}
%
%
%
%
%
%
%

\cortext[cor1]{Corresponding author}


\begin{abstract}

The paper performs simulation of a rectangular plate excited by turbulent channel flow at friction Reynolds numbers of 180 and 400. The fluid-structure interaction is assumed to be one-way coupled, i.e, the fluid affects the solid and not vice versa. We solve the incompressible Navier Stokes equations using finite volume direct numerical simulation  in the fluid domain. In the solid domain, we solve the dynamic linear elasticity equations using a time-domain finite element method. The obtained plate averaged displacement spectra collapse in the low frequency region in outer scaling. However, the high frequency spectral levels do not collapse in inner units. This spectral behavior is reasoned using theoretical arguments. We further study the sources of plate excitation using a novel formulation. This formulation expresses the average displacement spectra of the plate as an integrated contribution from the fluid sources within the channel.  Analysis of the net displacement source reveals that at the plate natural frequencies, the contribution of the fluid sources to the plate excitation peaks in the buffer layer. The corresponding wall-normal width is found to be $\approx 0.75\delta$. We analyze the decorrelated features of the sources using spectral Proper Orthogonal Decomposition (POD) of the net displacement source. We enforce the orthogonality of the modes in an inner product with a symmetric positive definite kernel. The dominant spectral POD mode contributes to the entire plate excitation. The contribution of the remaining modes from the different wall-normal regions undergo destructive interference resulting in zero net contribution. The envelope of the dominant mode further shows that the location and width of the contribution depend on inner and outer units, respectively. 

\end{abstract}

%


\maketitle

\section{Introduction}

The coupling between a turbulent flow and the resulting structural
excitation is a problem of interest in marine, civil and aerospace
engineering. In this paper, we investigate this coupling in a
canonical setting - linear one-way coupled (fluid affects solid, but
not vice versa) response of an elastic plate in turbulent channel flow
\citep{pope2001turbulent} due to wall-pressure fluctuations
alone. Specifically, we address the question - how much do the fluid
sources at different wall-normal locations contribute to plate
excitation for different frequencies, and what are the salient
features of these fluid sources?  We answer this question with a novel
formulation that combines Direct Numerical Simulation (DNS) data,
Green's function formulation and spectral Proper Orthogonal
Decomposition (POD). For brevity, we will sometimes refer to
wall-pressure fluctuations as just wall-pressure.

The one-way coupling between the fluid sources and plate excitation
can be broken into two parts: i) fluid source - wall-pressure
fluctuation coupling, and ii) wall-pressure fluctuation - plate
excitation coupling. Note that we neglect the wall-shear stress
contribution to the plate forcing. We further classify the techniques
to investigate the fluid source - wall-pressure fluctuation coupling
into - scaling variables-based, Green's function-based and conditional
averaging-based techniques. We discuss some features of the
wall-pressure fluctuation sources identified by each of these
techniques.

Identification of the scaling variables for the power-spectral density
(PSD) / wavenumber spectrum of wall-pressure fluctuation yields
qualitative information of the wall-normal region of the fluid
sources. The wall-pressure PSD in the low ($\omega\delta/u_{\tau}<5$),
mid ($5<\omega\delta/u_{\tau}<100$) and high frequency ranges
($\omega\delta/u_{\tau}>0.3Re_{\tau}$) scale with the potential flow
variables ($\rho_f,\delta^*,U_o$), outer flow variables
($\rho_f,\delta,\tau_w$), and inner flow variables
($\rho_f,\nu,\tau_w$), respectively \citep{farabee1991spectral}, where
$\omega$ is the angular frequency, $\rho_f$ is the fluid density,
$\delta$ is the boundary layer thickness, $\delta^*$ is the
displacement thickness of the boundary layer, $U_o$ is the centerline
velocity, $\tau_w$ is the wall-shear stress,
$u_{\tau}=\sqrt{\tau_w/\rho_f}$ is the friction velocity, and the
friction Reynolds number $Re_{\tau}$ is defined as
$u_{\tau}\delta/\nu_f$. Thus, the sources responsible for the low, mid
and high frequency wall-pressure fluctuations are predominantly in the
potential, outer and inner region of the turbulent boundary layer,
respectively.

The Green's function-based techniques
\citep{chang1999relationship,anantharamu2019analysis} yield
quantitative information of the sources of wall-pressure
fluctuation. The premultiplied streamwise wavenumber spectrum and the
PSD of the wall-pressure fluctuations in a turbulent channel show
peaks at $\lambda_x^+=300$ \citep{panton2017correlation} and
$\omega^+\approx 0.35$ \citep{hu2006wall} for $Re_{\tau}=180-5000$,
respectively, where $\lambda_x$ is the streamwise wavelength, and $+$
indicates normalization with viscous units ($\nu$ and $u_{\tau}$). The
dominant contributors to this inner peak are in the buffer region of
the channel \citep{anantharamu2019analysis}. The approach of
\cite{anantharamu2019analysis} that identified this dominant
contribution i) combines DNS data with the Green's function
formulation to express the wall-pressure PSD ($\phi_{pp}(\omega)$) as
integrated contribution ($\Gamma(r,s,\omega)$) from all wall-parallel
plane pairs,
$\phi_{pp}(\omega)=\iint_{-\delta}^{+\delta}\Gamma(r,s,\omega)\,\mathrm{dr}\,\mathrm{ds}$,
ii) accounts for the relative phase difference between the
contributions from different wall-parallel planes neglected in the
previous Green's function approach of \cite{chang1999relationship},
and iii) yields a distribution of sources in the wall-normal direction
instead of a wall-normal region as indicated by the scaling
variables. Further, the methodology identified decorrelated features
of wall-pressure fluctuation sources using spectral POD. The
identified dominant wall-pressure source at the linear and
premultiplied wall-pressure PSD peak frequency resembled tall and
inclined patterns, respectively.


The conditional averaging-based technique \citep{ghaemi2013turbulent}
yield patterns of the flow structure responsible that are correlated
to a particular wall-pressure fluctuation event. The time history of
the wall-pressure fluctuation signal at a point on the wall shows
occasional positive and negative high amplitude wall-pressure
peaks. The conditionally averaged flow fields show coupling between a
hairpin vortex and the high amplitude peaks
\citep{ghaemi2013turbulent}. The flow structure responsible for the
positive and negative high amplitude wall-pressure peak at a point are
the sweep and ejection event occuring above it, respectively. The
ejection event responsible for the negative peak occurs upstream of
the haripin head in between the quasi-streamwise vortices. The sweep
event that leads to the positive peak occurs downstream of the hairpin
head.

The dynamic linear elasticity equations describe the wall-pressure
fluctuation - plate excitation coupling. This one-way coupled FSI
approach is valid for small linear deformation ($du_{\tau}/\nu_f<1$)
of the plate, where $d$ is the wall-normal displacement. The approach
generally uses i) plate theories (e.g. Poisson Kirchoff) to describe
the deformation, and modal superposition to obtain the response, ii)
frequency domain since steady state response is usually the quantity
of interest, and iii) a model wavenumber-frequency spectrum
\citep{corcos1964structure,chase1980modeling,hwang1998discrete} for
the spatially homogenous wall-pressure fluctuations as input. Note
that the model wavenumber-frequency spectrum usually requires a model
PSD \citep{bull1967wall,smol1991models,goody2004empirical}. The mode
shapes and natural frequencies of the plate required to perform modal
superposition can be obtained analytically for simple boundary
conditions and geometry. For complicated boundary conditions and
geometry, Finite Element Method (FEM) is used to compute the modal
decomposition.

The wall-pressure fluctuation - plate excitation coupling has been
previously investigated in wavenumber space
\citep{hwang1990wavenumber,blake2017mechanics}. The modal force PSD of
the plate can be expressed as the wavenumber integral
\citep{blake2017mechanics}
\begin{equation} \label{eqn:modforcepkw}
\begin{split}
  \phi_{f_jf_j}(\omega)&=\iint_{-\infty}^{+\infty}\varphi_{pp}(k_1,k_3,\omega)|S_{j}(k_1,k_3)|^2\mathrm{dk_1}\,\mathrm{dk_3},\\
  S_{j}(k_1,k_3)&=\int_a^{a+L_x}\int_b^{b+L_z}S_j(x,z)e^{i(k_1x+k_3z)}\,\mathrm{dx}\,
  \mathrm{dz},
\end{split}
\end{equation}
where $a$ and $b$ are the origins of the plate in the streamwise and
spanwise directions, $L_x$ and $L_z$ are the lengths of the plate in
the streamwise and spanwise directions, $\phi_{f_jf_j}(\omega)$ is the
modal force PSD of the $j^{th}$ mode shape,
$\varphi_{pp}(k_1,k_3,\omega)$ is the wall-pressure
wavenumber-frequency spectrum and $|S_{j}(k_1,k_3)|^2$ is the modal
shape function. From the above equation, we observe that the modal
shape function couples the wall-pressure wavenumber-frequency spectrum
to the modal force. The relative contribution of different wavenumber
regions to the modal force spectra depends on the mode order ($j$),
boundary conditions, and the ratio of the streamwise modal wavenumber
($k_{m,j}$) to the convective wavenumber at the natural frequency of
the mode \citep{hwang1990wavenumber}. The high streamwise wavenumber
($k_1/k_{m,j} \gg 1$) portion of $|S_{j}(k_1,k_3)|^2$ decays as
$k_1^{-6}$, $k_1^{-4}$ and $k_1^{-2}$ for clamped, simply supported
and free boundary conditions on all edges
\citep{blake2017mechanics}. Thus, plates with free boundary conditions
accept more of the high streamwise wavenumber component of the
wall-pressure fluctuations. Further, special wall-pressure fluctuation
models that separately approximate the high and low wavenumber portion
of the wall-pressure fluctuation wavenumber-frequency spectrum can be
derived and used to obtain the response of plates
\citep{hambric2004vibrations}. \cite{hambric2004vibrations} showed
good agreement between FEM response of a plate excited by the modified
Corcos model of \cite{hwang1998discrete} and an equivalent edge
forcing model which only models the convective component in the
modified Corcos model for a plate with three edges clamped and one
edge free. This shows the importance of the convective region of
wall-pressure fluctuation spectrum for plates with free boundary
conditions. For a plate with all four edges clamped, FEM response from
a low wavenumber excitation model showed good agreement with the
modified Corcos model, thus highlighting the dominance of low
wavenumber contribution for clamped boundary condition.

Experiments by \cite{zhang2017deformation} have shown coupling between
flow structures and the response of a compliant wall in a turbulent
channel flow. The large positive and negative deformation of the
compliant wall is coupled to the ejection and sweep events,
respectively, occuring above it
\citep{zhang2017deformation}. Conditionally averaged flow fields show
that these events are related to the high amplitude pressure peaks and
hairpin vortices that surround the local deformation of the compliant
wall. For large deformation of the compliant wall, the plate
deflection affects the near-wall turbulence. The compliant wall
deflection into the buffer layer breaks the near-wall streaks and the
associated quasi-streamwise vortices, and induces more spanwise
coherence \citep{rosti2017numerical}.

In this paper, we develop a formulation to obtain the wall-normal
distribution of intensity and relative phase of the fluid sources
responsible for the plate excitation. Previous research works do not
yield such quantitative information of the fluid sources. The main
idea is to express the plate averaged displacement PSD as a double
wall-normal integral of the `net displacement source' cross-spectral
density (CSD) $\Gamma^a(r,s,\omega)$ across the height of the
channel. The analysis framework combines the volumetric DNS data,
Green's function solution of the pressure fluctuation and modal
superposition, and builds on the previous work of
\cite{anantharamu2019analysis}.  We then apply the framework to
explain the one-way coupled FSI simulation results of an elastic plate
in turbulent channel flow at $Re_{\tau}=180$ and $400$. The fluid and
solid simulations make use of finite volume DNS and time-domain FEM,
respectively. Further, the decorrelated fluid sources that contribute
the most to plate response are obtained using spectral Proper
Orthogonal Decomposition (POD) of the net displacement source CSD.


The organization of the paper is as follows: Section
\ref{sec:validmpcsolid} discusses the validation of the in-house FEM
solid solver - MPCUGLES-SOLID. In section \ref{sec:numsimdetails}, we
describe the computational domain, mesh resolution, and the FSI
simulation details. Section \ref{sec:anafram} discusses the novel
one-way coupling analysis framework. In section
\ref{subsec:onewayresp}, we discuss the obtained one-way coupled FSI
results. Section \ref{subsec:wallnormdist} discusses the spectral
features of the net displacement source CSD and in section
\ref{subsec:spectralpod}, we identify the decorrelated features of the
fluid source using spectral POD. Finally, we summarize the results in
section \ref{sec:summary}.

Note that $x$, $y$ and $z$ denote the streamwise, spanwise and
wall-normal coordinates, respectively. Superscripts/subscripts $f$ and
$s$ denote fluid and solid quantities, respectively.


\section{Validation of the in-house FEM solid solver -
  MPCUGLES-SOLID} \label{sec:validmpcsolid}

The in-house FEM solid solver - MPCUGLES-SOLID - is a time-domain
solver that uses the continuous Galerkin method to solve the dynamic
linear elasticity equations. We validate the solver's ability to
simulate random vibration problems by simulating the
\cite{han1999prediction} experiment using synthetically generated
loads. \cite{han1999prediction} measured the response of a rectangular
steel plate excited by a turbulent boundary layer at
$Re_{\tau}\approx 2000$. Table \ref{tab:validation_plate_prop} shows
the dimensions of the plate and the boundary layer properties in the
experiment. Note that the plate lies in the $x-z$ plane. We first
generate the wall-pressure fluctuations synthetically using a Fourier
series methodology based on the experimental conditions. Then, we
compare the obtained time-domain response of the plate from the solver
to the experiment. The generated fluctuations obey the Corcos
\citep{corcos1964structure} CSD model and the Smolyakov-Tkachenko
\citep{smol1991models} PSD model.

\begin{table}
  \begin{center}
\def~{\hphantom{0}}
  \begin{tabular}{lc}
    Plate length ($L_x^s$) & $0.47m$\\
    Plate width ($L_z^s$) & $0.37m$\\
    Plate thickness ($L_y^s$) & $1.59\times 10^{-3}m$\\
    Displacement thickness & $2.4\times 10^{-3}m$\\
    Flow velocity & $44.7ms^{-1}$\\
    Plate material & Steel
  \end{tabular}
  \caption{Plate properties, dimensions and experimental conditions of
    the \cite{han1999prediction} experiment used to validate the
    solver.}
  \label{tab:validation_plate_prop}
  \end{center}
\end{table}

We express the wall-pressure fluctuation $(p_w(x,z,t))$ as the Fourier series,
\begin{equation} \label{eqn:fourierseries}
  \begin{split}
  p_w(x,z,t)&=\sum_{l=-N^f_x/2}^{N^f_x/2-1}\sum_{m=-N^f_z/2}^{N^f_z/2-1}\sum_{n=-N^f_t/2}^{N^f_t/2-1}\hat{p}_{l,m,n}e^{i\left(k_lx+k_mz+\omega_nt\right)}, \\
  \hat{p}_{l,m,n}&=\left(\frac{2\pi}{L^f_x}\frac{2\pi}{L^f_z}\frac{2\pi}{T^f}\phi_{pp}(k_l,k_m,\omega_n)\right)^{1/2}e^{i\theta}, \\
  k_l=\frac{2\pi l}{L^f_x}&;\,k_m=\frac{2\pi m}{L^f_z};\,\omega_n=\frac{2\pi n}{T^f}.
\end{split}
\end{equation}
Here, $L^f_x$ and $L^f_z$ are the length and width of the domain used
to generate the fluctuations, respectively, $T^f$ is the timespan of
the generated fluctuations, $\theta$ is a uniformly distributed random
number between $0$ and $2\pi$, $N^f_x$, $N^f_z$, and $N^f_t$ are the
number of terms used to truncate the Fourier series in each dimension,
and $\phi_{pp}(k_l,k_m,\omega_n)$ is the wavenumber-frequency spectrum
of the wall-pressure fluctuations. The length and width of the domain
used to generate the load is ten times the size of the plate, i.e.,
$L^f_x=10L^s_x$ and $L^f_z=10L^s_z$.  In this way, we include the
contribution of the low wavenumber wall-pressure fluctuations
($|\overrightarrow{\mathbf{k}}|<|\overrightarrow{\mathbf{k}}_j|$) to
plate excitation. The timespan of the generated wall-pressure
fluctuations is $840$ times the period of the first mode of vibration
of the plate, i.e., $T^f=840\left(2\pi/\omega_1\right)$, where
$\omega_1$ is the first natural frequency of the plate. Hence, we
allow sufficient time for the transient response to decay. For the
wavenumber- frequency spectrum ($\phi_{pp}(k_l,k_m,\omega_n)$) in the
above equation, we set
\begin{equation}
  \begin{split}
    \phi_{pp}(k_x,k_z,\omega)&=\phi_{pp}(\omega)\frac{\alpha_x}{\pi} \frac{\alpha_z}{\pi} \left(\frac{\omega}{U_c}\right)^2 \left(\left(\frac{\alpha_x\omega}{U_c}\right)^2+\left(\frac{\omega}{U_c}+k_x\right)^2\right)^{-1}\left(\left(\frac{\alpha_x\omega}{U_c}\right)^2+k_z^2\right)^{-1}, \\
    \frac{\phi_{pp}(\omega)}{\tau_w^2\delta^*/U_{\infty}}&=\frac{1}{2}\frac{5.1}{1+0.44\left(\frac{|\omega|\delta^*}{U_{\infty}}\right)^{7/3}},
  \end{split}
\end{equation}
where we use the experimental conditions given in table for $\delta^*$
and $U_{\infty}$. For $U_c$, we use a constant value of
$0.89U_{\infty}$, and for $\tau_w$, we use the relation,
$\tau_w\approx0.0225\rho_fU_{\infty}^2Re_{\delta}^{-1/4}$ (equation
21.5 in \cite{schlichting1979boundary}), where
$Re_{\delta}=\frac{U_{\infty}8\delta^*}{\nu}$. We generate the Fourier
coefficients ($\hat{p}_{l,m,n}$) only in the right half-plane in
wavenumber space. To set the coefficients in the left-half plane, we
use the fact that the Fourier coefficients of a real function are
conjugate symmetric, i.e.,
$\hat{p}_{-l,-m,-n}=\hat{p}^*_{l,m,n}$. This ensures that the
generated wall-pressure fluctuations are
real. \cite{rogallo1981numerical} used a similar technique to generate
the initial velocity field in the isotropic turbulence decay
simulations.  \cite{maxit2016simulation} used a similar approach to
generate multiple realizations of the Fourier transform of the
pressure fluctuation for frequency-domain response. Here, we generate
only one time-domain realization of the space-time wall-pressure
fluctuations. Note that this approach applies to any spatially
homogenous wall-pressure cross spectral density model. Since this a
time-domain approach, even the response of nonlinear structures can be
obtained from the generated synthetic wall-pressure fluctuations.

We use a Cartesian mesh to discretize the solid domain. The number of
elements in the streamwise, spanwise and thickness directions is $32$,
$32$, and $1$, respectively. We use hexahedral elements of polynomials
of degree $2$. The number of frequencies ($N^f_t$) and wavenumbers
($N^f_x=N^f_z$) used to generate the load is $10000$ and $320$,
respectively. We make use of a parallel implementation to generate the
synthetic wall-pressure fluctuations that consists of $\approx 1$
billion terms. To efficiently compute the exact surface forces from
the generated wall-pressure fluctuations, we make use of $L^2$
orthogonal projection \citep{powell1981approximation} of the Fourier
series (equation \ref{eqn:fourierseries}) onto polynomials of degree
$2$ within each boundary surface element. For details of force
computation, we refer the reader to appendix \ref{sec:appA}.  The
timestep of the solid simulation is $5\times 10^{-5}\,s$, and the
total simulation time is $10\,s$.  We discard the first $5\,s$ since
they contain the transient response. We use the following $5\,s$ to
compute the statistics.

\begin{figure}
\centering
\begin{subfigure}{0.7\textwidth}
\begin{overpic}[width=\linewidth]{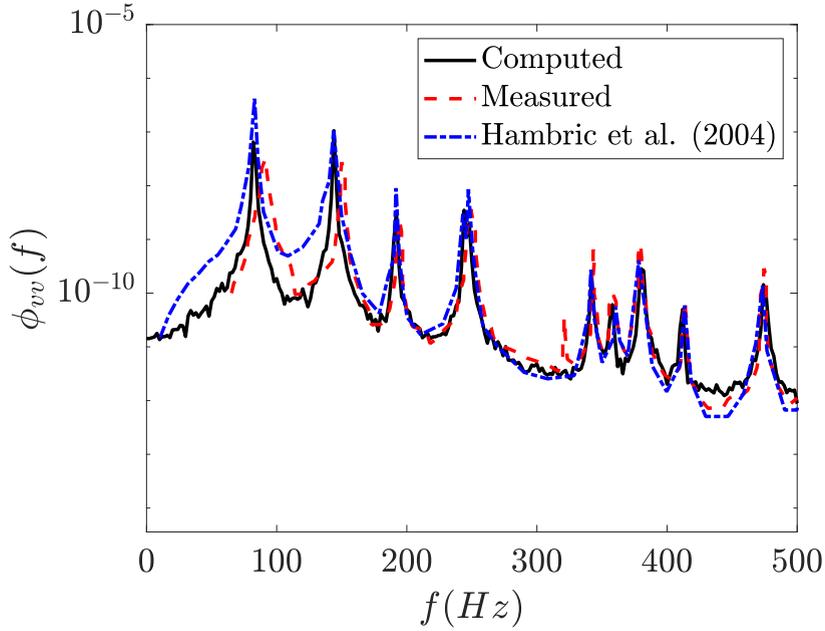}
\end{overpic}
\end{subfigure}
\caption{Comparison of the wall-normal velocity spectra at a point
  $(0.15m,0.12m)$ on the plate for the validation case.}
\label{fig:validation}
\end{figure}

Figure \ref{fig:validation} shows the comparison of the computed
velocity PSD at a point $(0.15m, 0.12m)$ on the top surface of the
plate to the measurement of \cite{han1999prediction}. The PSD agrees
well with the experiment. In the figure, we also compare our
time-domain results to the frequency domain results of
\cite{hambric2004vibrations}. \cite{hambric2004vibrations} used the
modified Corcos wall-pressure wavenumber frequency spectrum
\citep{hwang1990wavenumber} to compute the plate response. Even though
the spectral level of the standard Corcos \citep{corcos1964structure}
(that we use) is higher compared to the modified Corcos in the low
wavenumber range, our simulation results are closer to the experiment
than \cite{hambric2004vibrations}. Also, the low-frequency spectral
levels shown in figure \ref{fig:validation} are smaller than
\cite{hambric2004vibrations} when one might expect the opposite. We
believe this is due to the finite domain size used to generate the
load. The finite domain sets a lower bound on the wavenumbers
contributing to the excitation.

\section{FSI simulation details} \label{sec:numsimdetails}

\subsection{Computational domain}

\begin{figure}
\centering
\begin{subfigure}{0.8\textwidth}
\begin{overpic}[width=\linewidth]{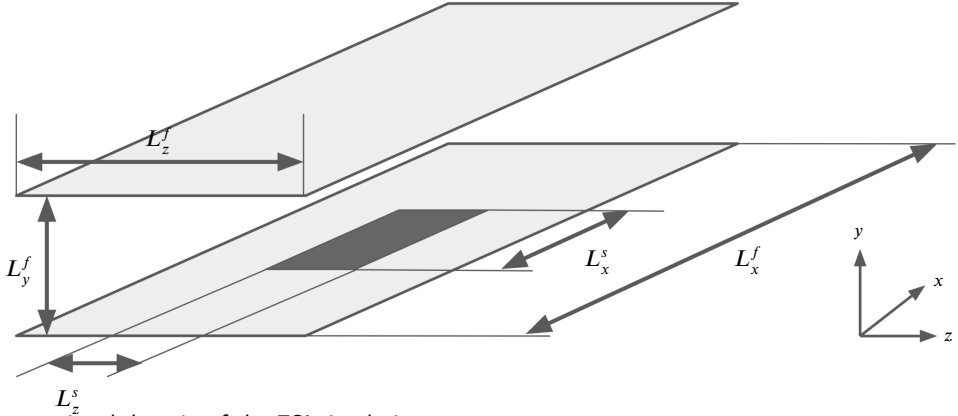}
\put(7,-1){$L_z^s$}
\put(2,12){$L_y^f$}
\put(16,25){$L_z^f$}
\put(60,13){$L_x^s$}
\put(75,13){$L_x^f$}
\put(87,16){\scriptsize$y$}
\put(96,5.5){\scriptsize$z$}
\put(95,11){\scriptsize$x$}
\end{overpic}
\end{subfigure}
\caption{Computational domain of the FSI simulation.}
\label{fig:compdomain}
\end{figure}

Figure \ref{fig:compdomain} shows a schematic of the fluid and solid
computational domain and table \ref{tab:domain} shows the domain
extents. The fluid computational domain is a Cartesian box of size
$L^f_x\times L^f_y\times L^f_z$. We choose $L^f_x=6\pi\delta$,
$L^f_y=2\delta$ and $L^f_z=2\pi\delta$, where $\delta$ is the half
channel height. Long streamwise and spanwise domains include the
contribution of large scale structures to pressure fluctuations. The
solid computational domain is a rectangular plate clamped on all sides
placed at the bottom wall of the channel. The plate is flush with the
bottom wall and centered. The length ($L^s_x$), width ($L^s_z$), and
thickness ($L^s_y$) of the plate is $6\pi\delta/5,\,2\pi\delta/5$ and
$0.004\delta$, respectively. The smaller dimension of the plate
ensures that the pressure fluctuations with wavelengths larger than
the plate dimensions are present in the computational box. Thus, we
include the low wavenumber ($k_1 \ll k_{m,j}$) wall-pressure
fluctuation contribution to plate excitation.

\begin{table}
  \begin{center}
\def~{\hphantom{0}}
  \begin{tabular}{cc}
    $L^f_x\times L^f_y\times L^f_z$ & $L^s_x\times L^s_y\times L^s_z$ \\[3pt]
    $6\pi\delta\times 2\delta\times 2\pi\delta$ & $\left(6\pi/5\right)\delta\times 0.004\delta\times \left(2\pi/5\right)\delta$
  \end{tabular}
  \caption{Fluid and solid domain extents.}
  \label{tab:domain}
  \end{center}
\end{table}

\begin{table}
  \begin{center}
\def~{\hphantom{0}}
  \begin{tabular}{lcccccc}
    $Re_{\tau}$  & $N^f_x\times N^f_y\times N^f_z$ & $N^s_x\times N^s_y\times N^s_z$ & $\Delta x^+$ & $\Delta z^+$ & $\Delta y_w^{f+}$ & $\Delta y_c^{f+}$ \\[3pt]
    180   & $720\times 176\times 330$ & $144\times 1\times 66$ & $4.7$ & $3.4$ & $0.27$ & $4.4$\\
    400   & $1388\times 288\times 660$ & $288 \times 1\times 132$ & $5.4$ & $3.8$ & $0.37$ & $5.9$\\
  \end{tabular}
  \caption{Fluid and solid mesh sizes and resolution of the FSI
    simulation. Note that the fluid and solid meshes match at the
    interface.}
  \label{tab:grid}
  \end{center}
\end{table}


\subsection{Fluid DNS}

We solve the incompressible Navier-Stokes equations in the fluid
domain using the collocated finite volume method of
\cite{mahesh2004numerical} in a frame of reference moving with the
bulk velocity of the fluid as done by
\cite{Bernardini2013turbulent}. This lead to better prediction of the
high frequency component of the pressure spectra. The walls in the
channel are assumed to be rigid in the fluid calculation. For time
integration, we use the Crank-Nicholson scheme. Overall, the method is
second order accurate in space and time, and non-dissipative. The
algorithm conserves kinetic energy discretely. This ensures stability
of the algorithm at high Reynolds numbers without adding numerical
dissipation. We perform the DNS using the in-house flow solver -
MPCUGLES.

The fluid mesh is cartesian. The mesh is uniform in the streamwise and
spanwise directions. In the wall-normal direction, we use a
non-uniform hyperbolic tangent spacing to cluster control volumes near
the wall. Table \ref{tab:grid} shows the fluid mesh sizes and
resolutions for both $Re_{\tau}$. The streamwise spacing
($\Delta x^+$), spanwise spacing ($\Delta z^+$), the wall-normal
spacing near the wall ($\Delta y_w^{f+}$) and channel centerline
($\Delta y_c^{f+}$) is fine enough to resolve the fine-scale features
of wall turbulence. The timestep of the fluid simulation is
$5\times 10^{-4}\delta/u_{\tau}$ for both $Re_{\tau}$. The velocity
($U_b/u_{\tau}$) of the moving reference frame is $15.8$ and $17.8$ in
the streamwise direction for $Re_{\tau}=180$ and $400$,
respectively. We employ a slip velocity boundary condition (equal to
-$U_b/u_{\tau}$) at the top and bottom wall. For pressure, we use a
zero Neumann boundary condition at the top and bottom wall. In the
streamwise and spanwise directions, we use periodic boundary
conditions for both velocity and pressure. For validation of the fluid
DNS, we refer the reader to \cite{anantharamu2019analysis}.


\subsection{Solid simulation}


We solve the three-dimensional dynamic linear elasticity equations in
the solid domain with the continuous Galerkin Finite Element Method
(abbreviated as just FEM). We perform the solid simulation using the
validated in-house solid solver - MPCUGLES-SOLID. We use second-order
polynomials in each element to represent the solution and trapezoidal
rule for the time integration of the equations. We precondition the
matrix problem using the scaled thickness preconditioner developed by
\cite{kloppel2011scaled} with the optimal scaling
$\frac{L^s_x/N^s_x+L^s_z/N^s_z}{2L^s_y}$. The preconditioner reduced
the simulation time by a factor of 3.


We non-dimensionalize the structural equations with the half-channel
height $(\delta)$, fluid density $(\rho_f)$, and friction velocity
$(u_{\tau})$. The non-dimensional properties of the plate are shown in
table \ref{tab:plateprop}.  We use a mass proportional Rayleigh
damping of $2.25$. The structural loss factor with the chosen mass
proportional damping is $0.05$ at the first natural frequency. The
solid simulation timestep is the same as the fluid DNS.

The solid mesh is Cartesian and composed of 27-node hexahedral
elements. Table \ref{tab:grid} gives the dimensions of the mesh. Since
the plate is of high aspect ratio, we only use one element in the
thickness direction. Further, the fluid and solid meshes match at the
interface. Thus, no special load transfer strategy is required. We set
the displacement of the nodes on all four sides of the plate to zero
and apply the rigid wall DNS wall-pressure fluctuations onto the top
surface of the plate.

The fluid DNS is first run until it reaches a statistically stationary
state. Then, the one-way coupled FSI simulation is run for a total time of
$16\delta/u_{\tau}$ units. We discard the first $8\delta/u_{\tau}$ time
units of the solid response as it contains the transient response of
the solid and use the remaining $8\delta/u_{\tau}$ time units to
compute the statistics of the plate reponse.


\begin{table}
  \begin{center}
\def~{\hphantom{0}}
  \begin{tabular}{lc}
    Young's modulus ($E/\left(\rho_fu_{\tau}^2\right)$) & $6.88\times 10^9$\\
    Poisson ratio ($\nu_s$) & $0.4$\\
    Solid density ($\rho_s/\rho_f$) & $1.17\times 10^3$\\
  \end{tabular}
  \caption{Non-dimensional properties of the plate.}
  \label{tab:plateprop}
  \end{center}
\end{table}

\section{Analysis framework} \label{sec:anafram}

\subsection{Theory}

The goal is to express the plate averaged displacement PSD as a double integral over all the wall-parallel plane pairs. To accomplish this, we first express the bottom wall displacement $d(x,-\delta,z,t)$ as a wall-normal integral,
\begin{equation} \label{eqn:d_fd}
  d(x,-\delta,z,t)=\int_{-\delta}^{+\delta}f_d(x,y,z,t)\mathrm{dy}.
\end{equation}
Here, $f_d(x,y,z,t)$ is called the `net displacement source' (exact form is
derived later). It gives the contribution of each wall-parallel plane to the surface displacement of the plate. We define the plate averaged displacement PSD $\phi_{dd}^a(\omega)$ as 
\begin{equation} \label{eqn:phia_dd_defn}
  \begin{split}
    \phi_{dd}^a(\omega)&=\frac{1}{A_p}\iint_{\Gamma_{fs}}\phi_{dd}(x,-\delta,z,\omega)\,\mathrm{dx}\,\mathrm{dz},\\
    \phi_{dd}(x,-\delta,z,\omega)&=\frac{1}{2\pi}\int_{-\infty}^{+\infty}d^*(x,-\delta,z,t)d(x,-\delta,z,t+\tau)e^{-i\omega \tau}\,\mathrm{d\tau},
  \end{split}
\end{equation}
where $\phi_{dd}(x,-\delta,z,\omega)$ is the displacement PSD at a point $(x,-\delta,z)$ on the surface of the plate, and $A_p$ is the area of the plate and $\Gamma_{fs}$ is the plate surface. We can then relate the plate averaged displacement PSD $\phi^a_{dd}(\omega)$ to the net displacement source $f_d(x,y,z,t)$ using equations \ref{eqn:d_fd} and \ref{eqn:phia_dd_defn} as 
\begin{equation}
  \begin{split}
    \phi_{dd}^a(\omega)&=\iint_{-\delta}^{+\delta}\Gamma^a(r,s,\omega)\,\mathrm{dr}\,\mathrm{ds},\\
    \Gamma^a(r,s,\omega)&=\frac{1}{A_p}\int_{\Gamma_{fs}}\left(\frac{1}{2\pi}\int_{-\infty}^{+\infty}\langle f^*_d(x,r,z,t) f_d(x,s,z,t+\tau)\rangle e^{-i\omega \tau}\,\mathrm{d\tau}\right)\,\mathrm{dx}\,\mathrm{dz},
  \end{split}
\end{equation}
where $\Gamma^a(r,s,\omega)$ is the plate averaged CSD of
$f_d(x,y,z,t)$. The function $\Gamma^a(r,s,\omega)$ yields the
contribution of each wall-parallel plane pair to the PSD
$\phi^a_{dd}(\omega)$ for different frequencies.

We obtain $f_d(x,y,z,t)$ as follows. Express $d(x,-\delta,z,t)$ in the modal basis as
\begin{equation} \label{eqn:d_modalsum}
  d(x,-\delta,z,t)=\sum_{j=1}^{\infty}d_j(t)\,\varphi_j(x,-\delta,z),
\end{equation}
where, $\varphi_j(x,-\delta,z)$ is the wall-normal component of the $j^{th}$ mode shape on the top surface of the plate, and $d_j(t)$ is the component of the solution along the $j^{th}$ mode shape. Assuming zero initial displacement and velocity of the plate, we write the solution for $d_j(t)$ using the Duhamel integral \citep{bathe2006finite} as
\begin{equation} \label{eqn:dj_soln}
d_j(t)=\frac{1}{\bar{\omega}_j}\int_{0}^tf_j(\tau)e^{-\xi_j\omega_j\left( t-\tau\right)}\mathrm{sin}\left(\bar{\omega}_j\left(t-\tau\right)\right)\,\mathrm{d\tau},
\end{equation}
where $\bar{\omega}_j=\omega_j\sqrt{1-\xi_j^2}$, and $f_j(t)$ is the modal force of the $j^{th}$ mode shape of the plate given by
\begin{equation} \label{eqn:fj_defn}
f_j(t)=-\iint_{\Gamma_{fs}}p(x,-\delta,z,t)\varphi_j(x,-\delta,z)\,\mathrm{dx}\,\mathrm{dz}.
\end{equation}
To account for only the steady state response, we use a large value for $t$ in equation \ref{eqn:phia_dd_defn}. For large enough t, the initial transient contribution to the response of the plate decays to small values, thus leaving only the steady state contribution. To express $p(x,-\delta,z,t)$ as a wall-normal integral, we use the pressure fluctuation Poisson equation,
\begin{equation}
  -\nabla^2p=f=\rho_f\left(2\frac{\partial U^f_i}{\partial x_j}\frac{\partial u^{f'}_j}{\partial x_i}+\frac{\partial^2}{\partial x_i \partial x_j}\left(u^{f'}_iu^{f'}_j-\overline{u^{f'}_iu^{f'}_j}\right)\right),
\end{equation}
where $U^f_i$ and $u^{f'}_i$ are the mean and fluctuating fluid
velocities, respectively. Neglecting the Stokes contribution, we use a
zero Neumann boundary condition at the top and bottom walls for the
pressure fluctuations. This is reasonable as the Stokes component of
wall-pressure fluctuations is small at high Reynolds number
\citep{hoyas2006scaling}. To obtain a unique solution, we set the
average of the pressure fluctuations at the top and bottom wall to
zero at all times. The solution to the pressure fluctuations
$p(x,-\delta,z,t)$ at the bottom wall is then,
\begin{equation} \label{eqn:wallpsource}
\begin{split}
  p(x,-\delta,z,t)=&\int_{-\delta}^{+\delta}f_G(x,y,z,t)\,\mathrm{dy},\\
  f_G(x,y,z,t)=&\iint_{-\infty}^{+\infty}G(-\delta,y,k)\hat{f}(k_1,y,k_3,t)e^{i(k_1x+k_3z)}\mathrm{dk_1}\,\mathrm{dk_3}, \\
  k=& \sqrt{k_1^2+k_3^2},\\
      G(r,s,k)&=\begin{cases}
      \frac{\rm{cosh}(k(s-\delta))\rm{cosh}(k(r+\delta))}{2k \rm{sinh}(k\delta)\rm{cosh}(k\delta)},  r\leq s,\\
      \frac{\rm{cosh}(k(s+\delta))\rm{cosh}(k(r-\delta))}{2k \rm{sinh}(k\delta)\rm{cosh}(k\delta)}, r > s,
    \end{cases}
\end{split}
\end{equation}
where $f_G(x,y,z,t)$ is termed the `net source' function
\citep{anantharamu2019analysis}, $G(-\delta,y,k)$ is the Green's
function, and $\hat{f}(k_1,y,k_3,t)$ is the multidimensional Fourier
transform of the source terms $f(x,y,z,t)$ defined as
\begin{equation}
  \hat{f}(k_1,y,k_3,t)=\frac{1}{(2\pi)^2}\iint_{-\infty}^{+\infty}f(x,y,z,t)e^{-i(k_1x+k_3z)}\mathrm{dx}\,\mathrm{dz}.
\end{equation}
We call $f_G(x,y,z,t)$ the `net source' function because it includes
contributions from all wavenumbers and the Green's function. Combining equations
\ref{eqn:d_modalsum}, \ref{eqn:dj_soln}, \ref{eqn:fj_defn} and
\ref{eqn:wallpsource}, we obtain the required expression for the net displacement source
$f_d(x,y,z,t)$ as \begin{equation}
  \begin{split}
    f_d(x,y,z,t)=\sum_{j=1}^{\infty}&\frac{1}{\bar{\omega}_j}\Bigg(\int_{0}^{t}\left(\iint_{\Gamma_{fs}}f_G(x,y,z,t)\varphi_j(x,-\delta,z)\,\mathrm{dx}\,\mathrm{dz}\right)\\
    &e^{-\xi_j\omega_j\left( t-\tau\right)}sin\left(\bar{\omega}_j\left(t-\tau\right)\right)\,\mathrm{d\tau}\Bigg)\varphi_j(x,-\delta,z)
  \end{split}
\end{equation}

To obtain the contribution from the cross-correlation of the fluid
sources with a particular plane $y=r$ to the plate averaged response
PSD, we integrate $\Gamma^a(r,s,\omega)$ along $s$ to obtain
$\Psi^a(r,\omega)$,
\begin{equation}
  \Psi^a(r,\omega)=\int_{-\delta}^{+\delta}\Gamma^a(r,s,\omega)\,\mathrm{ds}.
\end{equation}
It can be shown that $\Psi_a(r,\omega)$ is the plate averaged wall
displacement-net displacement source CSD,
\begin{equation}
  \Psi^a(r,\omega)=\frac{1}{A_p}\int_{\Gamma_{fs}}\left(\frac{1}{2\pi}\int_{-\infty}^{+\infty}\langle f^*_d(x,r,z,t) d(x,-\delta,z,t+\tau)\rangle e^{-i\omega \tau}\,\mathrm{d\tau}\right)\,\mathrm{dx}\,\mathrm{dz}.
\end{equation}
Further, the plate averaged wall displacement-net displacement source
CSD relates to the plate averaged displacement PSD
$\phi_{dd}^a(\omega)$ as
\begin{equation}
  \phi_{dd}^a(\omega)=\int_{-\delta}^{+\delta}\Psi^a(r,\omega)\,\mathrm{dr} = \int_{-\delta}^{+\delta}Re\left(\Psi^a(r,\omega)\right)\,\mathrm{dr},
\end{equation}
where $Re(\cdot)$ is the real part of $\cdot$.

We relate the plate averaged net displacement source CSD $\Gamma_a(r,s,\omega)$ to the four-dimensional CSD of the pressure fluctuation source terms $\varphi_{ff}(r,s,k_1,k_3,\omega)$ as follows. The four-dimensional CSD $\varphi_{ff}(r,s,k_1,k_3,\omega)$ is defined as
\begin{equation}
\begin{split}
&\varphi_{ff}(r,s,k_1,k_3,\omega)=\\
&\frac{1}{\left(2\pi\right)^3}\iiint_{-\infty}^{+\infty}\langle f^*(x,z,r,t)f(x+\xi_1,z+\xi_3,s,t+\tau)\rangle e^{-i\left(k_1\xi_1+k_3\xi_3+\omega\tau\right)}\mathrm{d\xi_1}\,\mathrm{d\xi_3}\,\mathrm{d\tau}.
\end{split}
\end{equation}
Neglecting the transient response of the plate, the modal displacement PSD $\phi_{d_jd_j}(\omega)$ relates to the modal force PSD $\phi_{f_jf_j}(\omega)$ as
\begin{equation} \label{eqn:phi_djdj_phi_fjfj}
  \begin{split}
    \phi_{d_jd_j}(\omega)&=|H_j(\omega)|^2\phi_{f_jf_j}(\omega),\\
    H_j(\omega)&=\frac{1}{\left(\omega_j^2-\omega^2\right)+i2\xi_j\omega_j\omega},
  \end{split}
\end{equation}
where $|H_j(\omega)|^2$ is the gain in the response of the $j^{th}$
mode. Further, the modal force $\phi_{f_jf_j}(\omega)$ relates to the
wall-pressure wavenumber frequency spectrum
$\phi_{pp}(k_1,k_3,\omega)$ as
\begin{equation} \label{eqn:phi_fjfj_phipp}
  \phi_{f_jf_j}(\omega)=\iint_{-\infty}^{+\infty}\phi_{pp}(k_1,k_3,\omega)|S_j(k_1,k_3)|^2\,\mathrm{dk_1}\,\mathrm{dk_3}.
\end{equation}
Relating $\phi_{pp}(k_1,k_3,\omega)$ to the four-dimensional CSD $\varphi_{ff}(r,s,k_1,k_3,\omega)$ using the Green's function, we obtain
\begin{equation} \label{eqn:phi_fjfj_phiff}
  \begin{split}
    \phi_{f_jf_j}(\omega)=\iint_{-\delta}^{+\delta}\iint_{-\infty}^{+\infty}&G^*(-\delta,r,k)G(-\delta,s,k)\varphi_{ff}(r,s,k_1,k_3,\omega)\\
    &|S_j(k_1,k_3)|^2\,\mathrm{dk_1}\,\mathrm{dk_3}\,\mathrm{dr}\,\mathrm{ds},
\end{split}
\end{equation}
where $S_j(k_1,k_3)=\iint_{\Gamma_{fs}}\varphi_j(x,0,z)e^{i\left(k_1x+k_3z\right)}\,\mathrm{dx}\,\mathrm{dz}$ is the Fourier transform of the mode shape, and $|S_j(k_1,k_3)|^2$ is the `modal shape function' \citep{hwang1990wavenumber}. Next, we relate the plate averaged displacement PSD $\phi^a_{dd}(\omega)$ to the modal displacement PSD $\phi_{d_jd_j}(\omega)$ as
\begin{equation} \label{eqn:phi_add_phi_djdj}
  \begin{split}
    \phi^a_{dd}(\omega)&=\frac{1}{A_p}\iint_{\Gamma_{fs}}\phi_{dd}(x,-\delta,z)\,\mathrm{dx}\,\mathrm{dz},\\
    &=\frac{1}{A_p}\iint_{\Gamma_{fs}}\sum_{i=1}^{\infty}\sum_{j=1}^{\infty}\phi_{d_id_j}(\omega)\varphi_i(x,-\delta,z)\varphi_j(x,-\delta,z)\,\mathrm{dx}\,\mathrm{dz},\\
    &\approx \frac{1}{\rho_s L_y^s A_p} \sum_{j=1}^{\infty}\phi_{d_jd_j}(\omega) \left(\because \iint_{\Gamma_{fs}}\varphi_i(x,-\delta,z)\varphi_j(x,-\delta,z)\,\mathrm{dx}\,\mathrm{dz}\approx\frac{1}{\rho L_y^s} \delta_{ij}\right)
  \end{split}
\end{equation}
Thus, combining equations \ref{eqn:phi_djdj_phi_fjfj}, \ref{eqn:phi_fjfj_phiff},
and \ref{eqn:phi_add_phi_djdj}, we obtain the required expression
\begin{equation} \label{eqn:gamma_a_phi_ff}
  \begin{split}
    \Gamma^a(r,s,\omega)\approx\frac{1}{\rho L_y^s A_p}\iint_{-\infty}^{+\infty}&G^*(-\delta,r,k)G(-\delta,s,k)\varphi_{ff}(r,s,k_1,k_3,\omega)\\
    &\left(\sum_{j=1}^{\infty}|S_j(k_1,k_3)|^2|H_j(\omega)|^2\right)\,\mathrm{dk_1}\,\mathrm{dk_3}.
  \end{split}
\end{equation}


We investigate the structure of the decorrelated contribution from
wall-parallel planes by performing spectral POD of the CSD
$\Gamma^a(r,s,\omega)$. We use the following inner product to define
the orthonormal relation between the eigenfunctions $\bar{\Phi}_i$ and
$\bar{\Phi}_j$ of $\Gamma^a(r,s,\omega)$,
\begin{equation}\label{eqn:spod_innerproduct} 
  \int_{-\delta}^{+\delta}\left(\left(-\left(1-\beta\right)\frac{\partial^2
  }{\partial y^2}+\beta\right)\bar{\Phi}_i\right)\bar{\Phi}_j^* \,\mathrm{dy}=\delta_{ij},
\end{equation}
where $\beta$ is a real number satisfying $0<\beta\leq 1$ and $\delta_{ij}$ is the Kroenecker delta. Further, the eigenfunctions $\bar{\Phi}_i(r,\omega)$ are assumed to satisfy the zero-Neumann boundary conditions at the wall $r=-\delta$ and $r=+\delta$. Following \cite{anantharamu2019analysis}, we call the above inner product as the Poisson inner product because the symmetric positive definite kernel $\left(-\left(1-\beta\right)\frac{\partial^2
  }{\partial y^2}+\beta\right)$ relates to the Poisson equation. The spectral POD of $\Gamma^a(r,s,\omega)$ is then
\begin{equation}\label{eqn:gamma_a_spod}
  \Gamma^a(r,s,\omega)=\sum_{j=1}^{\infty}\lambda_j(\omega)\Phi_j(r,\omega)\Phi^*_j(s,\omega),
\end{equation} 
where $\{\Phi_j,\lambda_j\}_{j=1}^{\infty}$ is the set of spectral POD modes and eigenvalues. The spectral POD mode $\Phi_j$ relates to the eigenfunction $\bar{\Phi}_j$ of $\Gamma^a(r,s,\omega)$ through the relation $\Phi_j=\left(-\left(1-\beta\right)\frac{\partial^2 }{\partial y^2}+\beta\right)\bar{\Phi}_j$. The associated eigenvalue problem for $\bar{\Phi}_j$ and $\lambda_j$ is
\begin{equation}
\int_{-\delta}^{+\delta}\Gamma^a(r,s,\omega)\,\bar{\Phi}_j(s,\omega)\,\mathrm{ds}=\lambda_j(\omega)\left(\left(-\left(1-\beta\right)\frac{\partial^2}{\partial
y^2}+\beta\right)\bar{\Phi}_j\right)(r,\omega).  
\end{equation} 
Further, the functions $\{\Phi_j\}_{j=1}^{\infty}$ and $\{\bar{\Phi}_j\}_{j=1}^{\infty}$ satisfy the orthonormality relation 
\begin{equation} 
  \int_{-\delta}^{+\delta}\Phi_i(y,\omega)\bar{\Phi}^*_j(y,\omega),\mathrm{dy}=\delta_{ij}.
\end{equation}

The sum of the obtained spectral POD eigenvalues gives ranked contribution from
each spectral POD mode to the following double integral,
\begin{equation} \label{eqn:sumofspod}
 \begin{split}
  \iint_{-\delta}^{+\delta}\frac{G\left(r,s,\frac{\beta}{1-\beta}\right)}{1-\beta}\Gamma^a(r,s,\omega)\,\mathrm{dr}\,\mathrm{ds}&=\sum_{j=1}^{\infty}\lambda_j(\omega)
 \end{split}
\end{equation}
where $G(r,s,\beta/(1-\beta))$ is the Green's function given by equation \ref{eqn:wallpsource}. For small values of $\beta$, the function $G(r,s,\beta/(1-\beta))$ becomes flatter and approaches a constant in $r$ and $s$, and the left hand side $\iint_{-\delta}^{+\delta}\frac{G\left(r,s,\frac{\beta}{1-\beta}\right)}{1-\beta}\Gamma^a(r,s,\omega)\,\mathrm{dr}\,\mathrm{ds}$ becomes a good proxy for the plate averaged displacement PSD $\phi_{dd}^a(\omega)=\iint_{-\delta}^{+\delta}\Gamma^a(r,s,\omega),\mathrm{dr}\,\mathrm{ds}$. Therefore, the obtained spectral POD modes isolate the dominant contributors to plate averaged displacement PSD. For more details about the effectiveness of the Poisson inner product, we refer the reader to \cite{anantharamu2019analysis}.


To obtain the contribution of each spectral POD mode to the plate averaged
displacement PSD, we doubly integrate equation \ref{eqn:gamma_a_spod} to obtain
\begin{equation} \label{eqn:phi_a_gamma}
  \begin{split}
    \phi^a_{dd}(\omega)=&\sum_{j=1}^{\infty}\gamma_j(\omega),\\
    \gamma_j(\omega)&=\lambda_j(\omega)|\int_{-\delta}^{+\delta}\Phi_j(y,\omega)\,\mathrm{dy}|^2;\,j=1,\dots,\infty,
  \end{split}
\end{equation}
where $\gamma_j(\omega)$ is the contribution of $j^{th}$ mode to PSD at frequency $\omega$. Further, we can show that 
\begin{equation} \label{eqn:int_net_phase}
  |\int_{-\delta}^{+\delta}\Phi_j(y,\omega)\,\mathrm{dy}|=\int_{-\delta}^{+\delta}|\Phi_j(y,\omega)|cos\left(\angle{\Phi_j(y,\omega)}-\angle{\Phi_i^n(\omega)}\right)\,\mathrm{dy},
\end{equation}
where $\angle\cdot$ is the argument of $\cdot$, and $\angle{\Phi_j^n(\omega)}$
is the argument of the integral $\int_{-1}^{+1}\Phi_j(y,\omega)\,\mathrm{dy}$. Using equation
\ref{eqn:int_net_phase} in equation \ref{eqn:phi_a_gamma}, we obtain
\begin{equation}
  \gamma_j(\omega)=\lambda_j(\omega)\left(\int_{-\delta}^{+\delta}|\Phi_j(y,\omega)|cos\left(\angle{\Phi_j(y,\omega)}-\angle{\Phi_i^n(\omega)}\right)\,\mathrm{dy}\right)^2;\,j=1,\dots,\infty.
\end{equation}
From the above equation, we observe that the eigenvalue, and both magnitude and
phase of the spectral POD mode all play a role in determining its contribution
to plate averaged displacement PSD. The contribution from different wall-normal
locations can constructively or destructively intefere based on the phase of the
spectral POD mode. Constructive inteference occurs between the contribution from
regions where the phase angle satisfies
$|\angle\Phi_j(y,\omega)-\angle\Phi_j^n(\omega)|<\pi/2$. Further, the
contribution from regions with the phase angle in the range
$|\angle\Phi_j(y,\omega)-\angle\Phi_j^n(\omega)|<\pi/2$ destructively interfere
with the regions where
$\pi/2<|\angle\Phi_j(y,\omega)-\angle\Phi_j^n(\omega)|<\pi$.

To obtain the contribution of each spectral POD mode to the integrated energy of
the net displacement source, we set $s=r$ in equation \ref{eqn:gamma_a_spod} and
integrate along $r$,
\begin{equation} \label{eqn:lambdabardefn}
  \begin{split}
    \int_{-\delta}^{+\delta}\Gamma^a(r,r,\omega)\,\mathrm{dr}&=\sum_{j=1}^{\infty}\bar{\lambda}_i(\omega);\\
    \bar{\lambda}_i(\omega)&=\lambda_i(\omega)\int_{-\delta}^{+\delta}|\Phi_i(r,\omega)|^2\,\mathrm{dr},
  \end{split}
\end{equation}
where $\bar{\lambda}_i$ is the contribution of the $i^{th}$ spectral POD mode to
the integrated net displacement source PSD.

\subsection{Implementation}

To compute the net displacement source CSD $\Gamma^a$, we need to
store the four-dimensional CSD $\varphi_{ff}$ from the fluid DNS
(equation \ref{eqn:gamma_a_phi_ff}). However, storing this function is
prohibhitively memory intensive. For the $Re_{\tau}=400$ case,
assuming $2000$ frequencies, approximately $1000$ TB is required to
store the four-dimensional function. To circumvent this issue, we use
a parallel, streaming methodology presented in
\cite{anantharamu2019analysis} with a small
modification. \cite{anantharamu2019analysis} presented the
implementation to compute the CSD
$\Gamma(r,s,\omega)=\iint_{-\infty}^{+\infty}G^*(-\delta,r,k)G(-\delta,s,k)\varphi_{ff}(r,s,k_1,k_3,\omega)\,\mathrm{dk_1}\,\mathrm{dk_3}$. We
modified their implementation to compute the CSD $\Gamma^a$ given by
equation \ref{eqn:gamma_a_phi_ff} instead. We use the first six mode
shapes of the plate to perform the summation in equation
\ref{eqn:gamma_a_phi_ff}. The first six mode shapes are sufficient to
analyze the fluid sources responsible for the first four peaks in the
plate averaged displacement spectra.

We use a total time of $8\delta/u_{\tau}$ ($16000$ timesteps) to
compute the net displacement source CSD $\Gamma^a$ for both
$Re_{\tau}$. The sampling interval is same as timestep of the FSI
simulation. We divide the temporal data into chunks of size
$1\delta/u_{\tau}$ ($2000$ timesteps) and use $50\%$ overlap between
the chunks to increase statistical convergence. We use Hanning window
to reduce spectral leakage.


\section{Results and discussion} \label{sec:results}

\subsection{FSI simulation results} \label{subsec:onewayresp}

\begin{figure}[htbp] 
  \centering 
  \begin{overpic}[width=\linewidth]{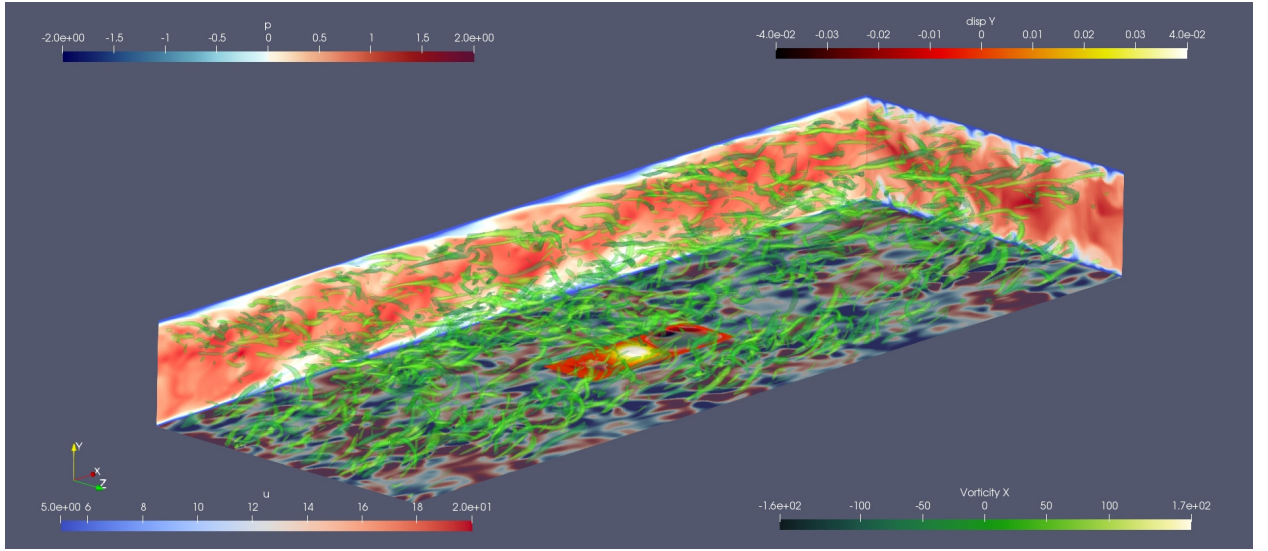}
  \end{overpic}
  \caption{Instantaneous visualization of the FSI simulation at $Re_{\tau}=180$.} 
  \label{fig:snapshot} 
\end{figure}

\begin{table}
  \begin{center}
\def~{\hphantom{0}}
  \begin{tabular}{lcc}
    $Re_{\tau}$  & $\left(\langle d^{+^2}\rangle\right)^{1/2}$ & $\left(\langle v^{+^2}\rangle\right)^{1/2}$ \\[3pt]
    180   & $1.81\times 10^{-2}$ & $5.32\times 10^{-3}$\\
    400   & $5.32\times 10^{-2}$ & $7.87\times 10^{-3}$\\
  \end{tabular}
  \caption{Plate averaged Root Mean Square (RMS) displacement and velocity of the plate.}
  \label{tab:rmsresults}
  \end{center}
\end{table}

Figure \ref{fig:snapshot} shows an instantaneous visualization of the
FSI simulation. The vertical and horizontal slices show the fluid
streamwise velocity and wall-pressure fluctuations, respectively. The
center patch denotes the deformed plate. The isosurfaces are of
Q-criterion at non-dimensional values of 500 and 1000. The colored
overlayed on the isosurface denotes the streamwise component of
vorticity. We use different colormaps for each quantity. The
instantaneous field clearly shows the fine scales features of wall
turbulence.

The plate averaged root mean square (RMS) wall-normal displacement and velocity
for both $Re_{\tau}$ is given in table \ref{tab:rmsresults}. Since, the RMS
displacement and velocity is much lesser than $1$ in viscous units, the one-way
coupling is justified.

\begin{figure} \centering \begin{subfigure}{0.49\textwidth}
  \begin{overpic}[width=\linewidth]{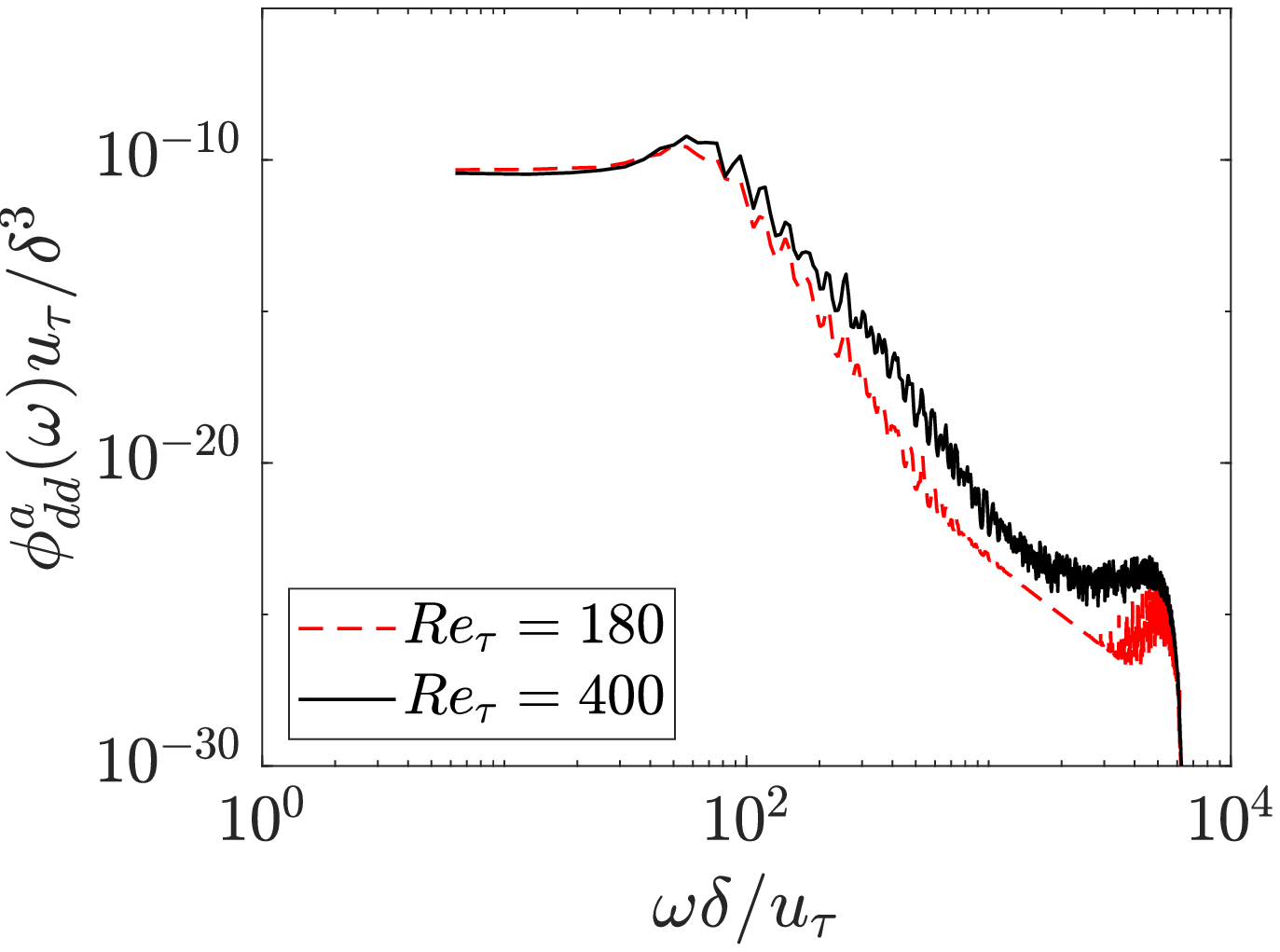} \put(1,70){$(a)$}
\end{overpic}
\end{subfigure} \begin{subfigure}{0.49\textwidth}
  \begin{overpic}[width=\linewidth]{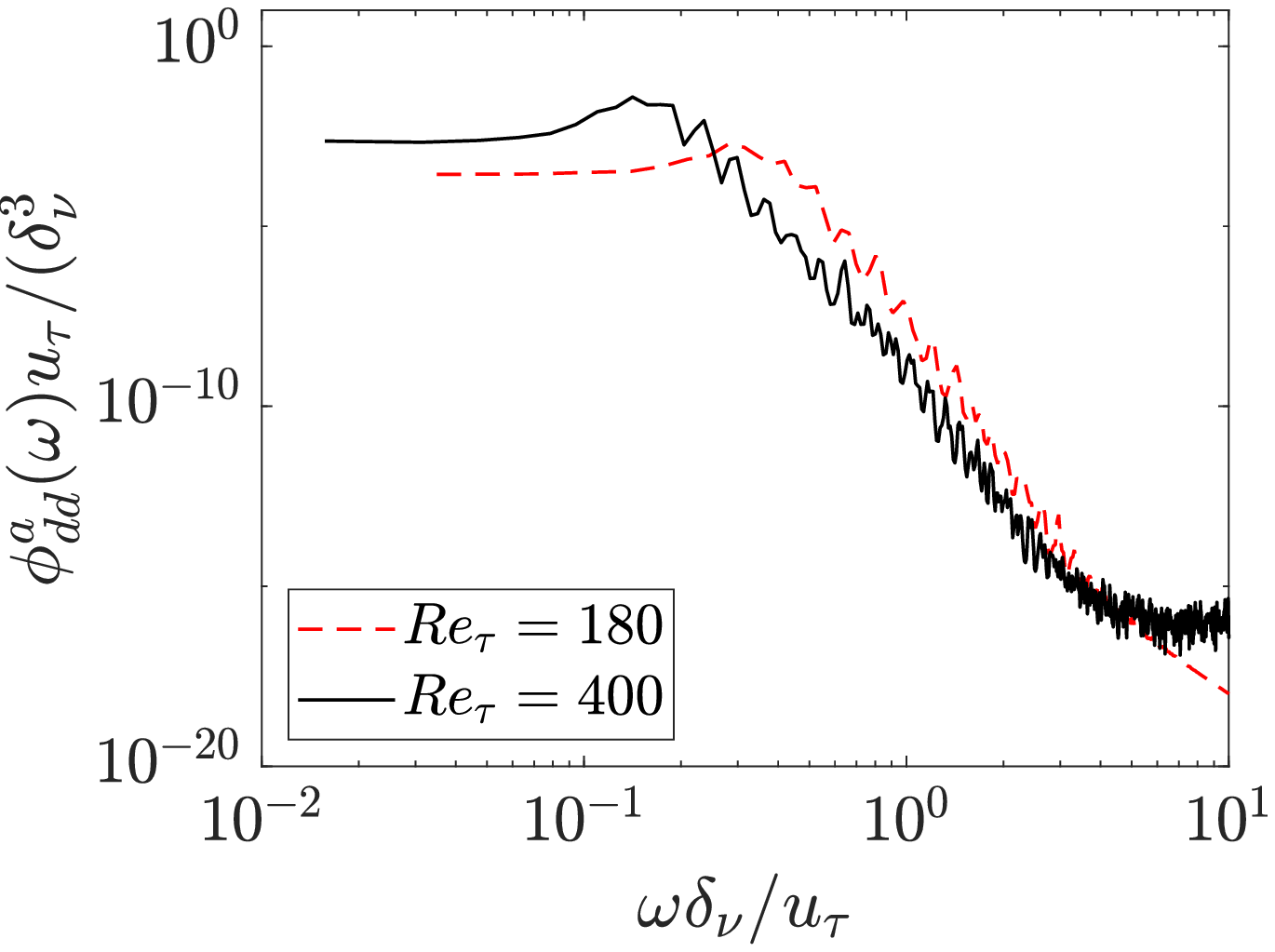} \put(1,70){$(b)$}
\end{overpic}
\end{subfigure}
  \caption{Plate averaged wall-normal displacement power spectra in a)
outer units (normalized by $\delta$ and $u_{\tau}$) and b) inner units
(normalized by $\delta_{\nu}$ and
$u_{\tau}$.} \label{fig:plate_average} \end{figure}


\begin{figure} \centering \begin{subfigure}{0.49\textwidth}
  \begin{overpic}[width=\linewidth]{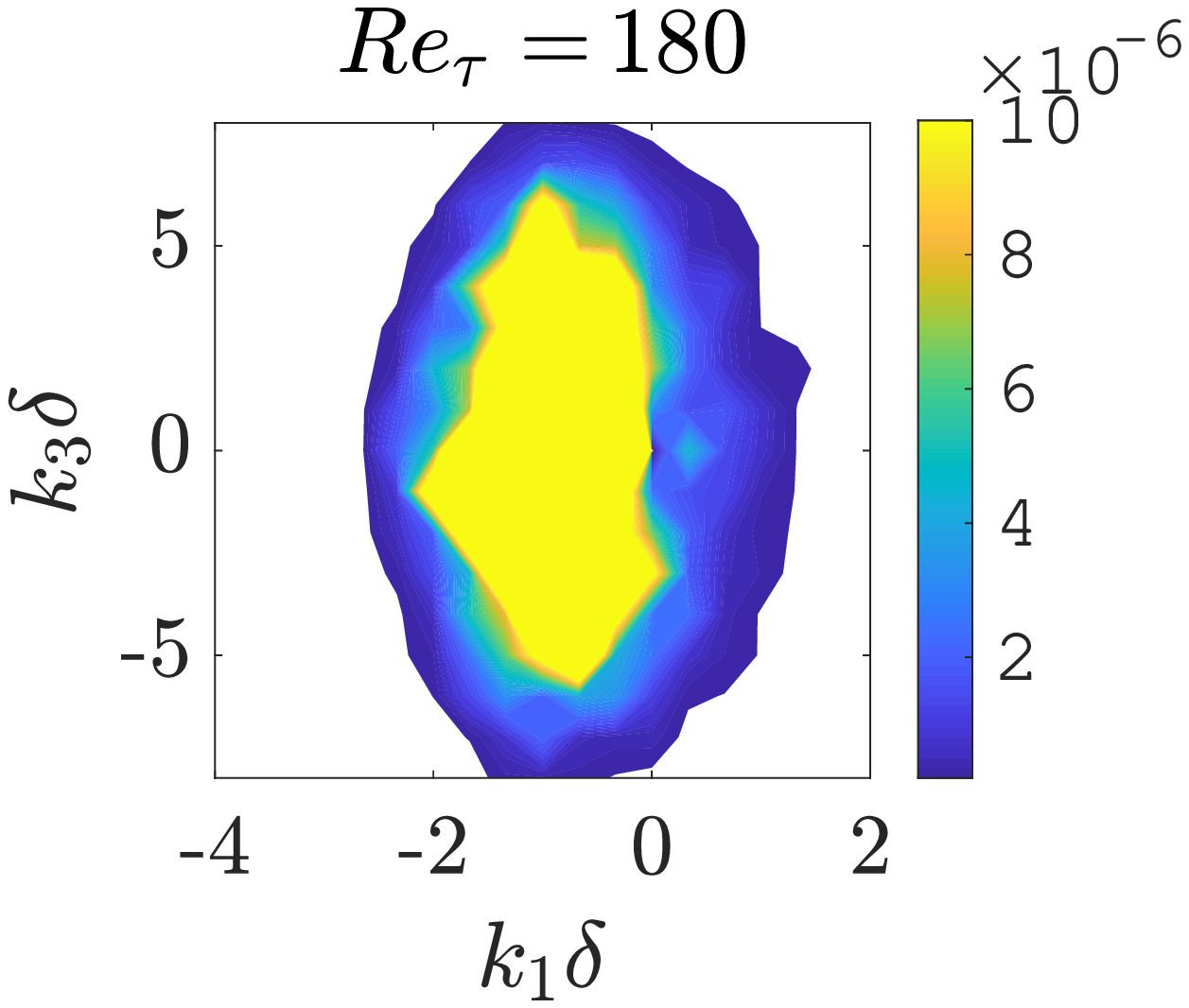} \put(1,70){$(a)$}
\end{overpic}
\end{subfigure} \begin{subfigure}{0.49\textwidth}
  \begin{overpic}[width=\linewidth]{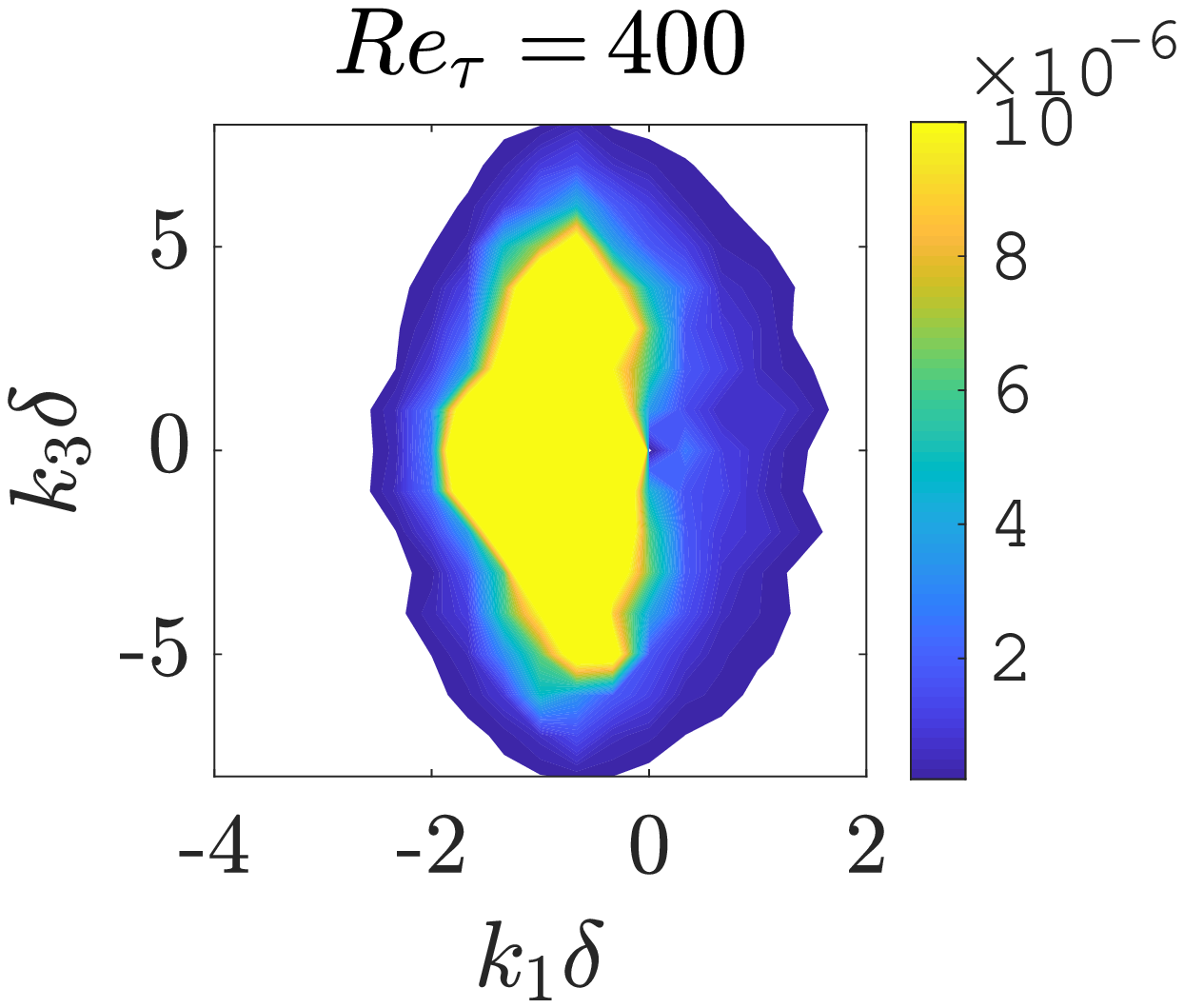} \put(1,70){$(b)$}
\end{overpic}
\end{subfigure}
  \caption{Product
  $|S_1(k_1,k_3)|^2\phi_{pp}(k_1,k_3,\omega=12.6u_{\tau}/\delta)$ for (a)
  $Re_{\tau}=180$ and (b) $Re_{\tau}=400$. Contours are 100 equally spaced
  values between $2\times 10^{-7}$ and $10^{-5}$} \label{fig:low_freq_pkwmacc}
  \end{figure}

  Figures \ref{fig:plate_average}a and b show the plate averaged
  wall-normal displacement spectra
  $\phi_{dd}^a(\omega)u_{\tau}/\delta^3$ (non-dimensionalized with
  $\delta$ and $u_{\tau}$) for both $Re_{\tau}$ in outer and inner
  units, respectively. The time span of the temporal data used to
  compute the spectra is $8\delta/u_{\tau}$. We divide the temporal
  data into chunks of size $1\delta/u_{\tau}$ for averaging. To
  increase convergence and reduce spectral leakage, we use $50\%$
  overlap and Hanning window \citep{bendat2011random},
  respectively. The peaks in the spectra correspond to the natural
  frequencies ($\omega_j\delta/u_{\tau}$) of the plate. Further, these
  natural frequencies coincide in outer units for both $Re_{\tau}$
  since the properties of the plate are the same in outer units for
  both the Reynolds numbers.

  The low frequency ($\omega << \omega_1$) spectral levels overlap for
  both $Re_{\tau}$. This is because i) the non-dimensional Young's
  modulus of the plate is the same for both $Re_{\tau}$ and ii) the
  low wavenumber and frequency component of the wall-pressure
  wavenumber-frequency spectrum is approximately the same in outer
  units for both $Re_{\tau}$. We can understand this as follows.

Combining equations \ref{eqn:phi_djdj_phi_fjfj}, \ref{eqn:phi_fjfj_phipp}, and \ref{eqn:phi_add_phi_djdj} , we have
\begin{equation}
\begin{split}
\phi^a_{dd}(\omega)&\approx\frac{1}{\rho_sL_y^sA_p}\sum_{j=1}^{\infty}|H_j(\omega)|^2\iint_{-\infty}^{+\infty}\phi_{pp}(k_1,k_3,\omega)|S_j(k_1,k_3)|^2\,\mathrm{dk_1}\,\mathrm{dk_3}.
\end{split}
\end{equation}
For frequencies $\omega \ll \omega_1$, we can approximate the average spectra using only the first mode as
\begin{equation}
\begin{split}
\phi^a_{dd}(\omega)&\approx\frac{1}{\rho_sL_y^sA_p}|H_1(\omega)|^2\iint_{-\infty}^{+\infty}\phi_{pp}(k_1,k_3,\omega)|S_1(k_1,k_3)|^2\,\mathrm{dk_1}\,\mathrm{dk_3},\\
&\approx\frac{1}{\rho_sL_y^sA_p\omega_1^4}\iint_{-\infty}^{+\infty}\phi_{pp}(k_1,k_3,\omega)|S_1(k_1,k_3)|^2\,\mathrm{dk_1}\,\mathrm{dk_3}.
\end{split}
\end{equation}
Since, the first natural frequency ($\omega_1$) is proportional to the longitudinal wave speed ($c_l$) of the plate, we have
\begin{equation}
c_l^4\phi^a_{dd}(\omega)\propto \iint_{-\infty}^{+\infty}\phi_{pp}(k_1,k_3,\omega)|S_1(k_1,k_3)|^2\,\mathrm{dk_1}\,\mathrm{dk_3}.
\end{equation}
Note that we have absorbed $\rho_s,\,L_y^s,\,A_p$ into the proportionality constant. Non-dimensionalizing the above equation, we have
\begin{equation}
\left(\frac{c^4_l}{u^3_{\tau}}\right)\frac{\phi^a_{dd}(\omega)}{\delta^3}\approx C\left(\frac{\omega\delta}{u_{\tau}},Re_{\tau}\right),
\end{equation}
where $C$ is some function of $\omega\delta/u_{\tau}$ and $Re_{\tau}$ only. We
absorb the proportionality constant into $C$. Figures
\ref{fig:low_freq_pkwmacc}a and b show the product
$|S_1(k_1,k_3)|^2\phi_{pp}(k_1,k_3,\omega)$ for
$Re_{\tau}=180$ and $400$ in outer units, respectively for a typical frequency
$\omega\delta/u_{\tau}=12.6\ll\omega_1$. Overall, the contours are similar for
both $Re_{\tau}$. This similarity of contours occurs in the frequency range
$\omega\ll\omega_1$. Thus, the dependency on $Re_{\tau}$ can be dropped, and we
have 
\begin{equation}
\left(\frac{c^4_l}{u^3_{\tau}}\right)\frac{\phi^a_{dd}(\omega)}{\delta^3}\approx C\left(\frac{\omega\delta}{u_{\tau}}\right).
\end{equation}
Further, substituting for $c_l$ in terms of the Young's modulus $E$,
we have
\begin{equation}
  \left(\frac{E}{\rho_su_{\tau}^2}\right)^2\frac{\phi^a_{dd}(\omega)u_{\tau}}{\delta^3}\approx C\left(\frac{\omega\delta}{u_{\tau}}\right).
\end{equation}
Since, $\frac{E}{\rho_su_{\tau}^2}$ is the same for both $Re_{\tau}$, we have
the required result,
\begin{equation}
  \frac{\phi^a_{dd}(\omega)u_{\tau}}{\delta^3}\approx C\left(\frac{\omega\delta}{u_{\tau}}\right).
\end{equation}


Figure \ref{fig:plate_average}b shows the plate averaged displacement
PSD with inner scaling ($\delta_{\nu}=\nu_f/u_{\tau}$ and $u_{\tau}$ as
length and velocity scale, respectively). The PSD at the two
$Re_{\tau}$ do not overlap in the high-frequency region. This is
because for identical natural frequencies in inner units, the
corresponding modal wavenumbers do not match in inner units, i.e., if
$j$ and $k$ are two mode indices such that
\begin{equation}
  \left(\omega_{j}\delta_{\nu}/u_{\tau}\right)_{Re_{\tau}=180}=\left(\omega_{k}\delta_{\nu}/u_{\tau}\right)_{Re_{\tau}=400},
\end{equation}
then
\begin{equation}
    \left(k_{m,j}\delta_{\nu}\right)_{Re_{\tau}=180}\neq
    \left(k_{m,k}\delta_{\nu}\right)_{Re_{\tau}=400}.
\end{equation}
Therefore, the plate filters different wavenumbers from the
wall-pressure wavenumber frequency spectra in viscous units leading to
dissimilar high-frequency spectral levels.

\begin{figure} \centering \begin{subfigure}{0.49\textwidth}
  \begin{overpic}[width=\linewidth]{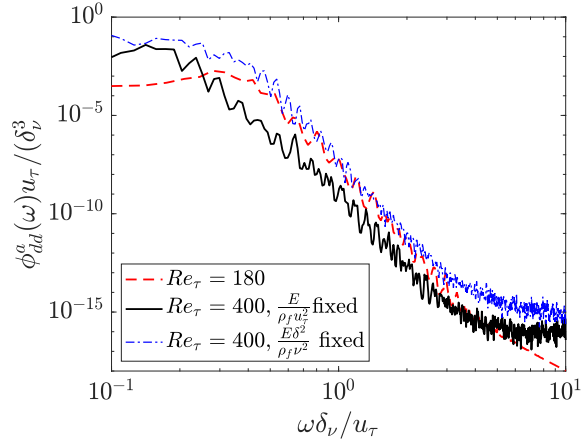}
\end{overpic}
\end{subfigure}
  \caption{High frequency plate averaged displacement spectra comparison by
  fixing $E\delta^2/(\rho_f\nu_f^2)$ and $E/{\rho_fu_{\tau}^2}$ between the two Reynolds
  numbers.} \label{fig:high_freq_comparison}
  \end{figure}

A better overlap of high-frequency spectral levels is observed (shown in figure
\ref{fig:high_freq_comparison}) if $E\delta^2/\left(\rho_f\nu_f^2\right)$
(velocity scale is $\nu_f/\delta$) is fixed for the two Reynolds numbers instead
of $E/{\rho_fu_{\tau}^2}$ (velocity scale is $u_{\tau}$). This is because for
fixed $E\delta^2/\left(\rho_f\nu_f^2\right)$ and coinciding natural frequencies in
inner units, the corresponding modal wavenumbers also coincide in inner units.
We explain this as follows. Let $j$ and $k$ be the mode indices with coinciding
natural frequencies in inner units for $Re_{\tau}=180$ and $400$, respectively,
i.e., \begin{equation}
  \left(\omega_{j}\delta_{\nu}/u_{\tau}\right)_{Re_{\tau}=180}\approx\left(\omega_{k}\delta_{\nu}/u_{\tau}\right)_{Re_{\tau}=400}.
\end{equation} 
We can show that for fixed $E\delta^2/\left(\rho_f\nu_f^2\right)$ and
$L_y^s/\delta$, we have
\begin{equation}
  \left(k_{m,j}\delta\right)^2\approx \left(k_{m,k}\delta\right)^2\left(\frac{180}{400}\right)^2.  
\end{equation}
Further, non-dimensionalizing in viscous units, we obtain the desired relation,
\begin{equation}
  \left(k_{m,j}\delta_{\nu}\right)_{Re_{\tau}=180}\approx\left(k_{m,k}\delta_{\nu}\right)_{Re_{\tau}=400}
\end{equation}


\subsection{Wall-normal distribution of fluid sources} \label{subsec:wallnormdist}

\begin{figure} \centering \begin{subfigure}{\textwidth}
  \begin{overpic}[width=\linewidth]{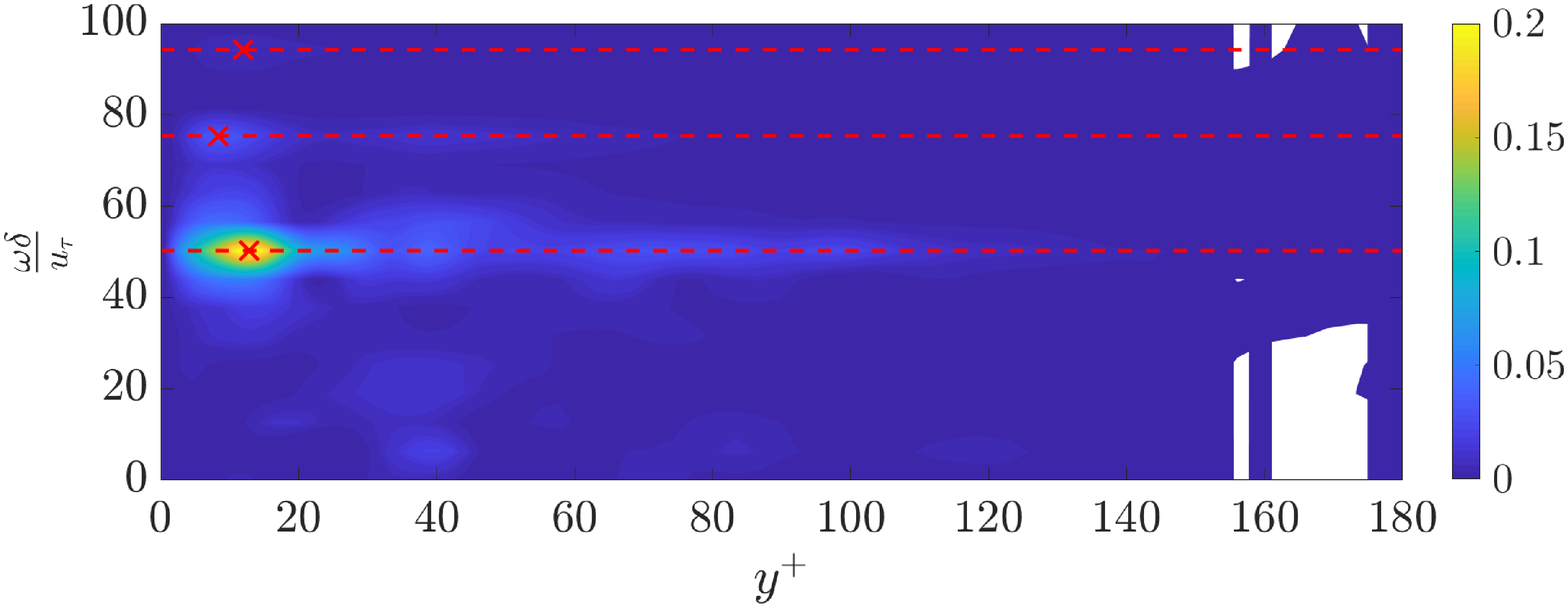} \put(1,70){$(a)$}
\end{overpic}
\end{subfigure} \begin{subfigure}{\textwidth}
  \begin{overpic}[width=\linewidth]{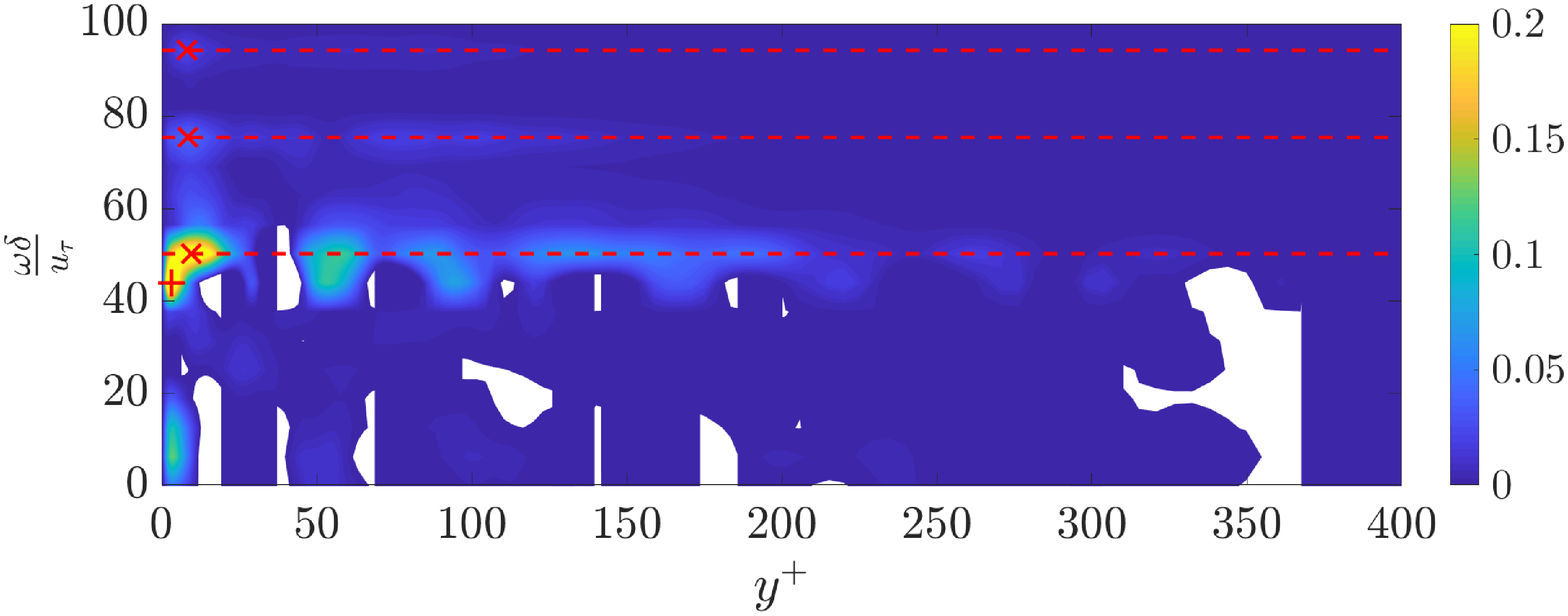} \put(1,70){$(b)$}
\end{overpic}
\end{subfigure}
\caption{Real part of the normalized wall displacement-net
  displacement source CSD
  $\frac{Re(\Psi_a(y^+,\omega\delta/u_{\tau}))}{\int_{-\infty}^{+\infty}\int_{0}^{2Re_{\tau}}\Psi_a(y^+,\omega\delta/u_{\tau})\,\mathrm{dy^+}\,\mathrm{d\omega\delta/u_{\tau}}}$
  for (a) $Re_{\tau}=180$ and (b) $Re_{\tau}=400$. Contours are 100
  equally spaced values between 0 and 0.2. Blank regions have negative
  value of $Re(\Psi^a)$.  Horizontal dashed red lines denote the peak
  frequencies of the plate averaged displacement PSD. Red crosses
  indicate the wall-normal coordinate with maximum value of $\Psi^a$
  at the peak frequencies. Red plus in (b) indicates the location of
  global maximum of $\Psi^a$ for $Re_{\tau}=400$.}
\label{fig:plate_average_psi} \end{figure}

Figures \ref{fig:plate_average_psi}a and \ref{fig:plate_average_psi}b
show the contours of the computed plate averaged wall displacement-net
displacement source CSD $\Psi_a(y^+,\omega\delta/u_{\tau})$
(normalized by its integral) for $Re_{\tau}=180$ and $400$,
respectively. $y^+$ is the distance from the wall. The three
horizontal dashed red lines in both figures denote the first three
peak frequencies ($\omega\delta/u_{\tau}=50.2,\,75.4$ and $94.25$) of
the plate averaged displacement PSD (figure \ref{fig:plate_average}a)
and the red markers `$\times$' denote the wall-normal coordinate with
maximum value of $\Psi^a$ at the peak frequencies. From a visual
inspection of the contours, we see that the location of maximum
intensity and width of the fluid sources approximately depend on inner
and outer units, respectively.

\begin{figure} 
  \centering 
  \begin{subfigure}{0.49\textwidth}
    \begin{overpic}[width=\linewidth]{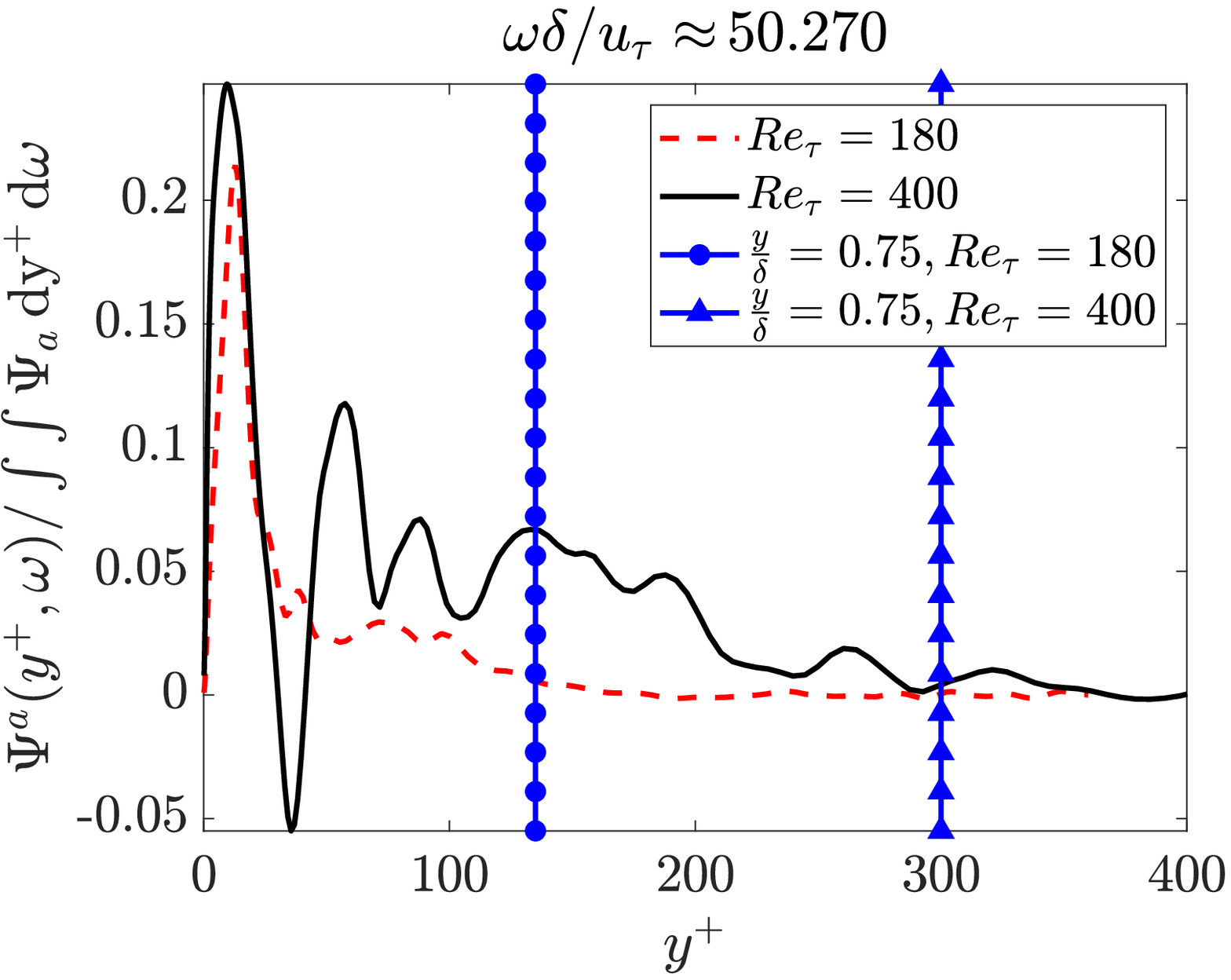} 
      \put(1,70){$(a)$}
    \end{overpic}
  \end{subfigure} 
  \begin{subfigure}{0.49\textwidth}
    \begin{overpic}[width=\linewidth]{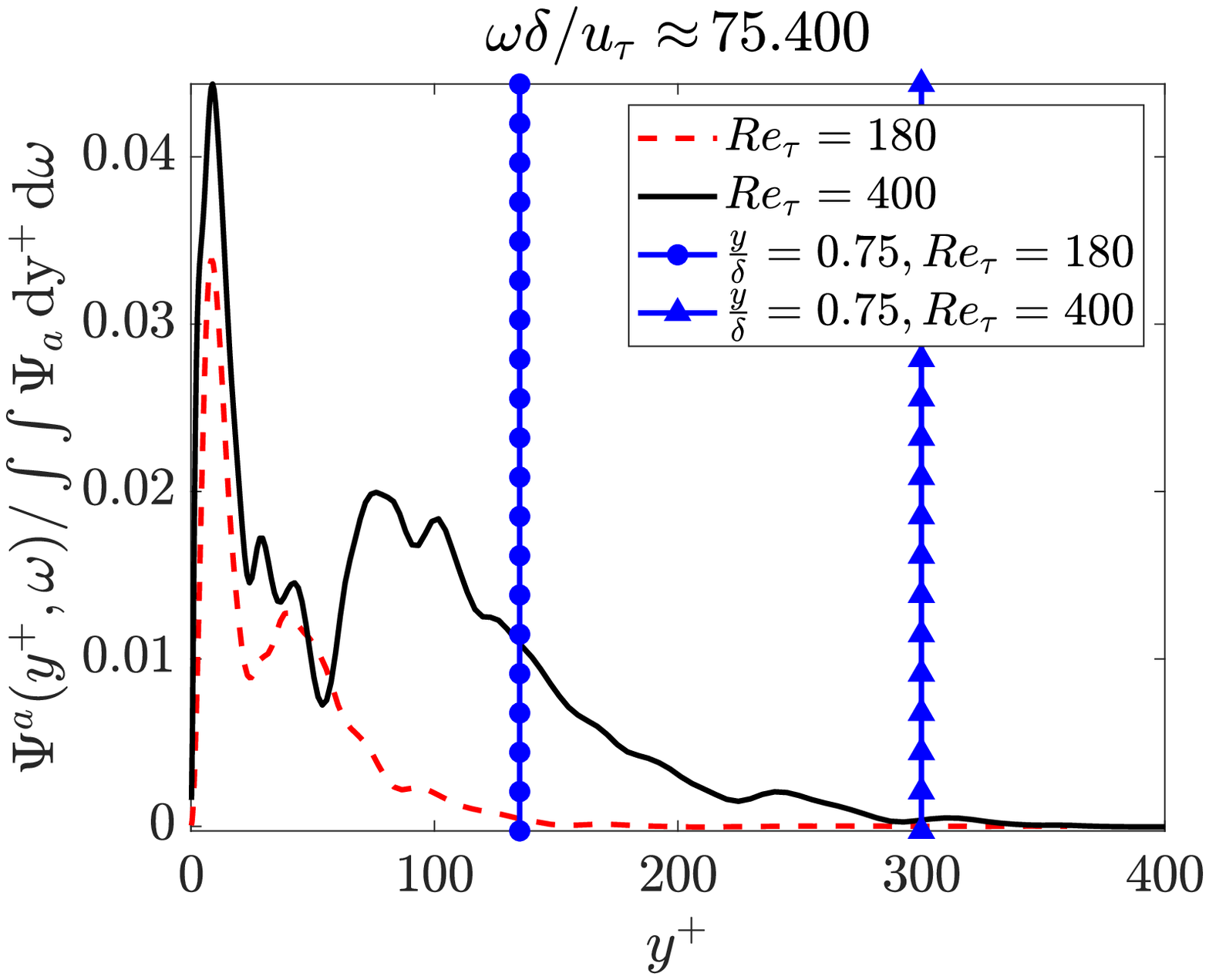} 
      \put(1,70){$(b)$}
    \end{overpic}
  \end{subfigure} 
  \begin{subfigure}{0.49\textwidth}
    \begin{overpic}[width=\linewidth]{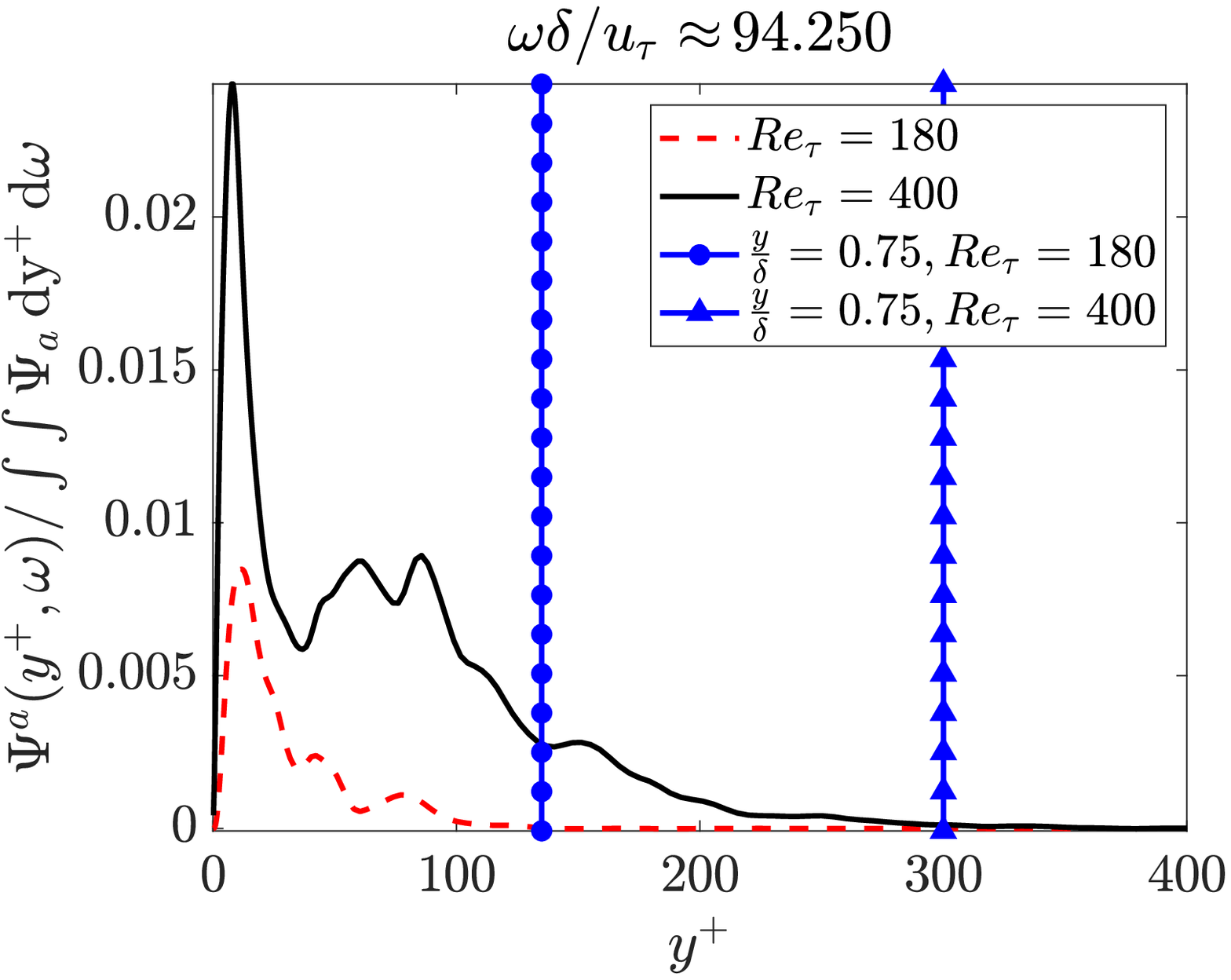} 
      \put(1,70){$(c)$}
    \end{overpic}
  \end{subfigure} 
  \caption{Comparison plate averaged wall displacement-net
    displacement source CSD (normalized by its integral) at the peak
    frequencies (a) $\omega\delta/u_{\tau}=50.2$, (b)
    $\omega\delta/u_{\tau}=75.4$ and (c) $\omega\delta/u_{\tau}=94.25$
    for $Re_{\tau}=180$ (dashed red line) and $Re_{\tau}=400$ (solid
    black line). Vertical dashed-dotted red and dotted black lines
    indicate $y/\delta=0.5$ for $Re_{\tau}=180$ and $400$,
    respectively. }
  \label{fig:psia_sel_freq} 
\end{figure}

To investigate this further, we plot the CSD $\Psi^a$ at the peak frequencies in
figure \ref{fig:psia_sel_freq} for both $Re_{\tau}$. All three frequencies have
a peak in buffer layer (around $y^+\approx 10$). This implies that the
correlation of the fluid sources with the buffer layer is a dominant contributor
to the response of the plate. The CSD has reasonable values for
$y/\delta\lessapprox 0.75$. Thus, the correlation of fluid sources with
wall-parallel planes within $y/\delta\approx 0.75$ have a sizeable contribution
to the response. In other words, the width of the fluid sources contributing to
the plate response depends on outer units and is approximately $y/\delta\approx
0.75$.

Further, $\Psi^a$ is negative for small wall-normal regions around
$y^+\approx 40$ for $\omega\delta/u_{\tau}\approx 50.2$ (shown by
white region). Thus, in a plate averaged sense, the plate displacement
$\hat{d}(x,-\delta,z,\omega)$ and $\hat{f}_d(x,y,z,\omega)$ have phase
difference $\theta$ satisfying $\pi/2<|\theta|<\pi$. However, we do
not observe such negative regions for $Re_{\tau}=180$. This negative
contribution comes from the coupling of the $Re_{\tau}=400$ fluid
sources with the lower mode shapes of the plate. More such negatively
correlated regions close to the wall (shown by white regions) are seen
in figure \ref{fig:plate_average_psi}b. Further, the global peak for
$Re_{\tau}=400$ in figure \ref{fig:plate_average_psi}b is at the
coordinate $(y_p,\omega_p\delta/u_{\tau})=(3,44)$ (indicated by `+'
symbol) which is much closer to the wall than the $Re_{\tau}=180$ peak
location at $(y_p,\omega_p\delta/u_{\tau})=(13,50)$. These differences
in the near wall region is because the natural frequencies and the
modal wavenumbers are different in viscous units for the two
$Re_{\tau}$. But, the four-dimesional CSD
$\varphi_{ff}(r,s,k_1,k_3,\omega)$ can be expected to be similar in
viscous units near to the wall for the two Reynolds
numbers. Therefore, the plate inherently filters different wavenumbers
and frequencies in viscous units from
$\varphi_{ff}(r,s,k_1,k_3,\omega)$ for the two $Re_{\tau}$ (equation
\ref{eqn:gamma_a_phi_ff}), thus leading to different near-wall
coupling with the fluid sources.

The Corcos form \citep{corcos1964structure} of the wall-pressure
wavenumber frequency spectrum has an interesting implication on the
associated one-way coupling. We show that for a Corcos type wavenumber
frequency spectrum, the plate averaged displacement spectra and the
wall-pressure PSD couple in a similar manner with the channel fluid
sources upto a multiplicative constant. The Corcos type
wavenumber-frequency spectrum $\varphi_{pp}(k_1,k_3,\omega)$ takes the
form,
\begin{equation} \label{eqn:phipp_kw_corcos}
  \varphi_{pp}(k_1,k_3,\omega)=\phi_{pp}(\omega)A\left(\frac{k_1U_c}{\omega}\right)B\left(\frac{k_3U_c}{\omega}\right),
\end{equation} 
where $A\left({k_1U_c}/{\omega}\right)$ and
$B\left({k_3U_c}/{\omega}\right)$ are functions that describe the
self-similar form of the streamwise and spanwise wavenumber
dependence, respectively. The wall-pressure PSD $\phi_{pp}(\omega)$
can be expressed as the wall-normal integral
\citep{anantharamu2019analysis} using the Green's function
formulation,
\begin{equation} \label{eqn:phipp_gamma_csd}
  \begin{split}
    \phi_{pp}(\omega)&=\iint_{-\delta}^{+\delta}\Gamma(r,s,\omega)\,\mathrm{dr}\,\mathrm{ds},\\
    \Gamma(r,s,\omega)&=\iint_{-\infty}^{+\infty}G^*(-\delta,r,k)G(-\delta,s,k)\varphi_{ff}(r,s,k_1,k_3,\omega)\,\mathrm{dk_1}\,\mathrm{dk_3}.
  \end{split}
\end{equation}
where $\Gamma(r,s,\omega)$ is the net source CSD. Net source is a
function $f_G(x,y,zt)$ whose integral in the wall-normal direction gives
the instantaneous wall-pressure fluctuation
$p(x,-\delta,z,t)=\int_{-1}^{+1}f_G(x,y,z,t)\,\mathrm{dy}$. Combining
equations \ref{eqn:phi_djdj_phi_fjfj}, \ref{eqn:phi_fjfj_phipp}, and
\ref{eqn:phi_add_phi_djdj} \ref{eqn:phipp_gamma_csd}, we obtain the
desired result,
\begin{equation}
  \begin{split}
    \phi_{dd}^a(\omega)=&=\iint_{-\delta}^{+\delta}\Gamma^a(r,s,\omega)\,\mathrm{dr}\,\mathrm{ds}=\iint_{-\delta}^{+\delta}\Gamma(r,s,\omega)\alpha(\omega)\,\mathrm{dr}\,\mathrm{ds},\\
    \alpha(\omega)=&\iint_{-\infty}^{+\infty}A\left(\frac{k_1U_c}{\omega}\right)B\left(\frac{k_3U_c}{\omega}\right)\\
    &\left(\sum_{j=1}^{\infty}|\hat{H}_j(\omega)|^2|S_j(k_1,k_3)|^2\right)\,\mathrm{dk_1}\,\mathrm{dk_3}.
  \end{split}
\end{equation}
Note that $\alpha(\omega)$ is a positive number. Thus, for a Corcos type
spectrum both plate averaged displacement PSD and wall-pressure PSD couple in a
similar manner with the fluid sources.

\subsection{Spectral POD of fluid sources} \label{subsec:spectralpod}

Before we present the spectral POD results of $\Gamma^a(r,s,\omega)$,
we discuss the relevance of the spectral POD modes and eigenvalues to
the plate surface displacement. Recall equation \ref{eqn:d_fd} that
relates the surface displacement at a point $(x,z)$ on the plate to
the net displacement source,
\begin{equation} \label{eqn:d_fd2}
  d(x,-\delta,z,t)=\int_{-\delta}^{+\delta}f_d(x,y,z,t)\,\mathrm{dy}.
\end{equation}
The Fourier transform of the net displacement source can be expanded
in the spectral POD basis $\{\Phi_j^*\}_{j=1}^{\infty}$ as
\begin{equation} \label{eqn:fd_spod}
  \begin{split}
    f_d(x,y,z,t)&=\int_{-\infty}^{+\infty}\hat{f}(x,y,z,\omega)e^{i\omega t}\,\mathrm{d\omega},\\
&=\int_{-\infty}^{+\infty}\sum_{j=1}^{\infty}\alpha_j(x,z,\omega)\Phi^*_j(y,\omega)\,e^{i\omega t}\,\mathrm{d\omega},
\end{split}
\end{equation}
where $\{\alpha_j(x,z,\omega)\}_{j=1}^{\infty}$ are the coefficients
of expansion of $\hat{f}_d(x,y,z,\omega)$. Using equation
\ref{eqn:fd_spod} in equation \ref{eqn:d_fd2}, and rearranging the
integral, we have
\begin{equation}
  d(x,-\delta,z,t)=\int_{-\infty}^{+\infty}\alpha_j(x,z,\omega)e^{i\omega t}\left(\int_{-\delta}^{+\delta}\Phi_j^*(y,\omega)\,\mathrm{dy}\right)\mathrm{d\omega}.
\end{equation}
Using the expression $\Phi_j^*(y,\omega)=|\Phi_j(y,\omega)|e^{-\angle \Phi_j(y,\omega)}$ in the above equation, we have
\begin{equation}
  d(x,-\delta,z,t)=\int_{-\infty}^{+\infty}\alpha_j(x,z,\omega)e^{i\omega t}\left(\int_{-\delta}^{+\delta}|\Phi_j(y,\omega)|e^{-\angle \Phi_j(y,\omega)}\,\mathrm{dy}\right)\mathrm{d\omega}.
\end{equation}
The above equation expresses the plate displacement as sum of contributions from
each spectral POD mode. Further, the coefficients
$\{\alpha_j(x,z,\omega)\}_{j=1}^{\infty}$ are decorrelated in the
plate averaged sense, i.e.,
\begin{equation}
\frac{1}{A_p}\int_{\Gamma_{fs}}\langle \alpha_j(x,z,\omega)\alpha_k(x,z,\omega_o) \rangle\,\mathrm{dx}\,\mathrm{dz}=\lambda_j(\omega)\delta_{jk}\delta(\omega-\omega_o),
\end{equation}
where $\delta_{ij}$ is the Kronecker delta and $\delta$ is the Dirac
Delta function. We include the effect of structures of all length
scales because we integrate over all wavenumbers in equation
\ref{eqn:gamma_a_phi_ff}. 

We set the parameter $\beta$ in the Poisson inner product (equation
\ref{eqn:spod_innerproduct}) to a small value of $0.5$ to compute the
spectral POD modes and eigenvalues. We did not observe a change in the
computed mode shapes or eigenvalues for values smaller than
$0.5$. Further, we will see that the value $\beta=0.5$ identifies a
single dominant mode of the net displacement source responsible for
the plate excitation.

\begin{figure}
\centering
\begin{subfigure}{0.49\textwidth}
\begin{overpic}[width=\linewidth]{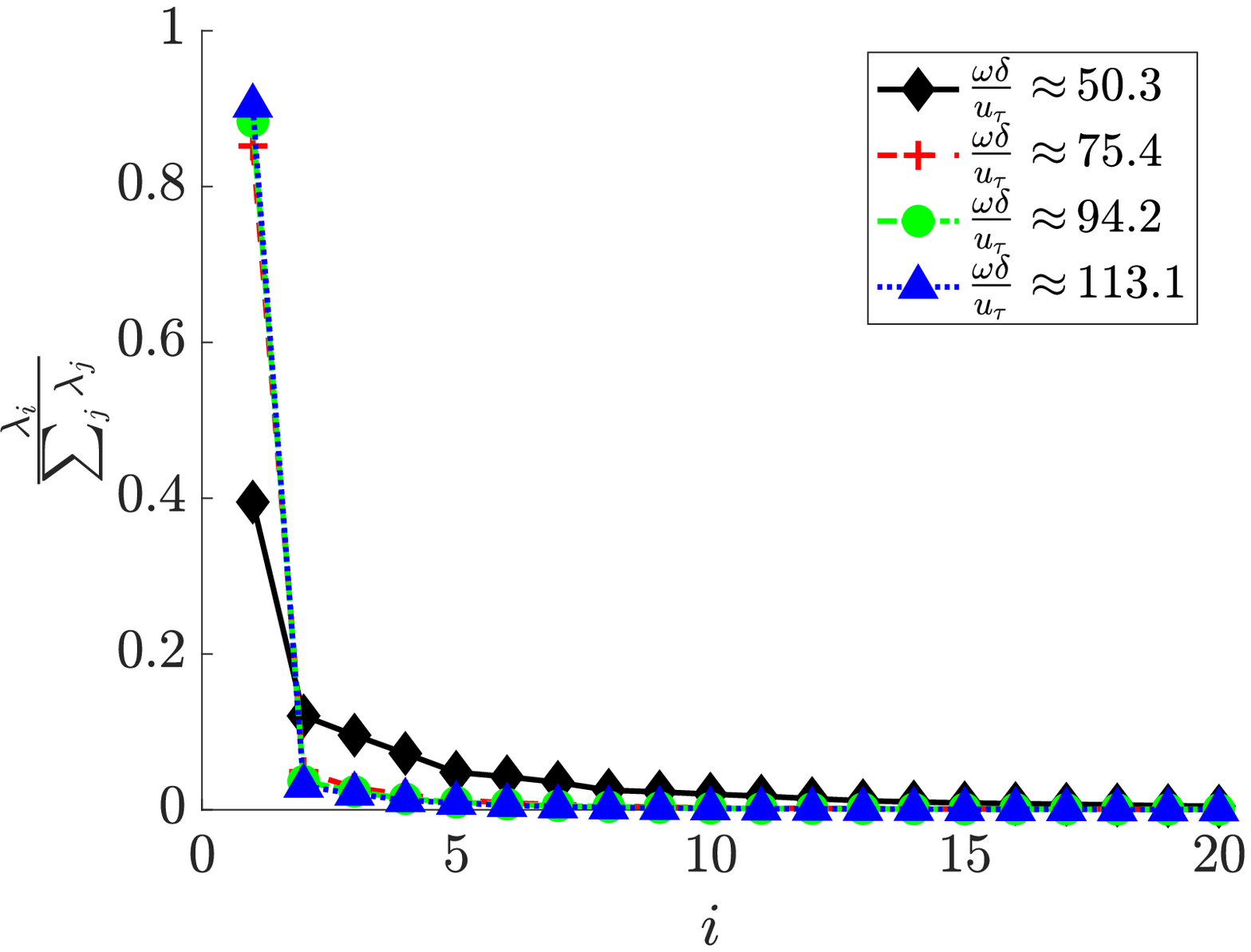}
\put(1,70){$(a)$}
\end{overpic}
\end{subfigure}
\begin{subfigure}{0.49\textwidth}
  \begin{overpic}[width=\linewidth]{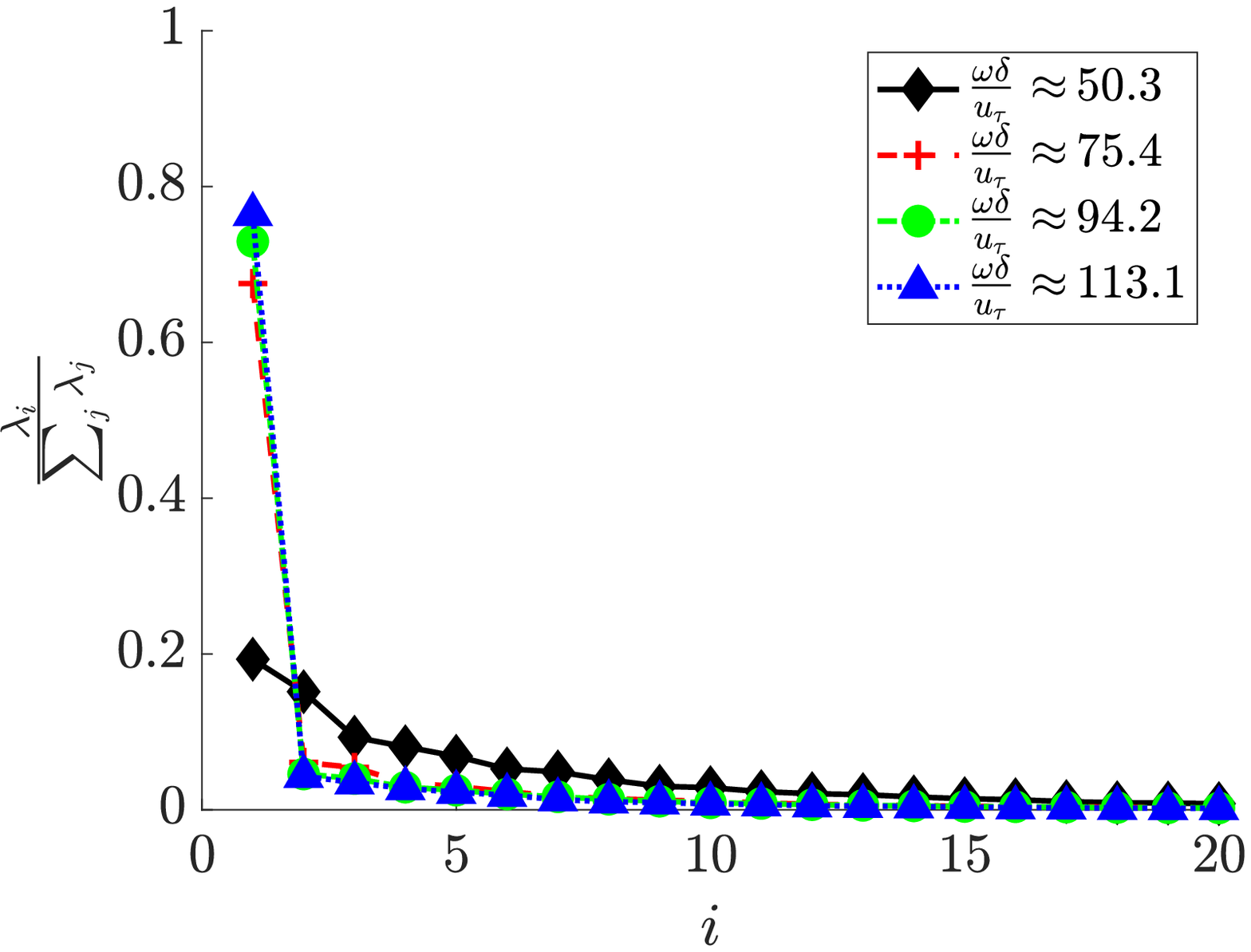}
\put(1,70){$(b)$}
\end{overpic}
\end{subfigure}
\caption{Spectral POD eigenvalues for a) $Re_{\tau}=180$ and b) $Re_{\tau}=400$ computed using the Poisson inner product ($\beta=0.5$).}
\label{fig:spod_eig}
\end{figure}

\begin{figure}
\centering
\begin{subfigure}{0.49\textwidth}
\begin{overpic}[width=\linewidth]{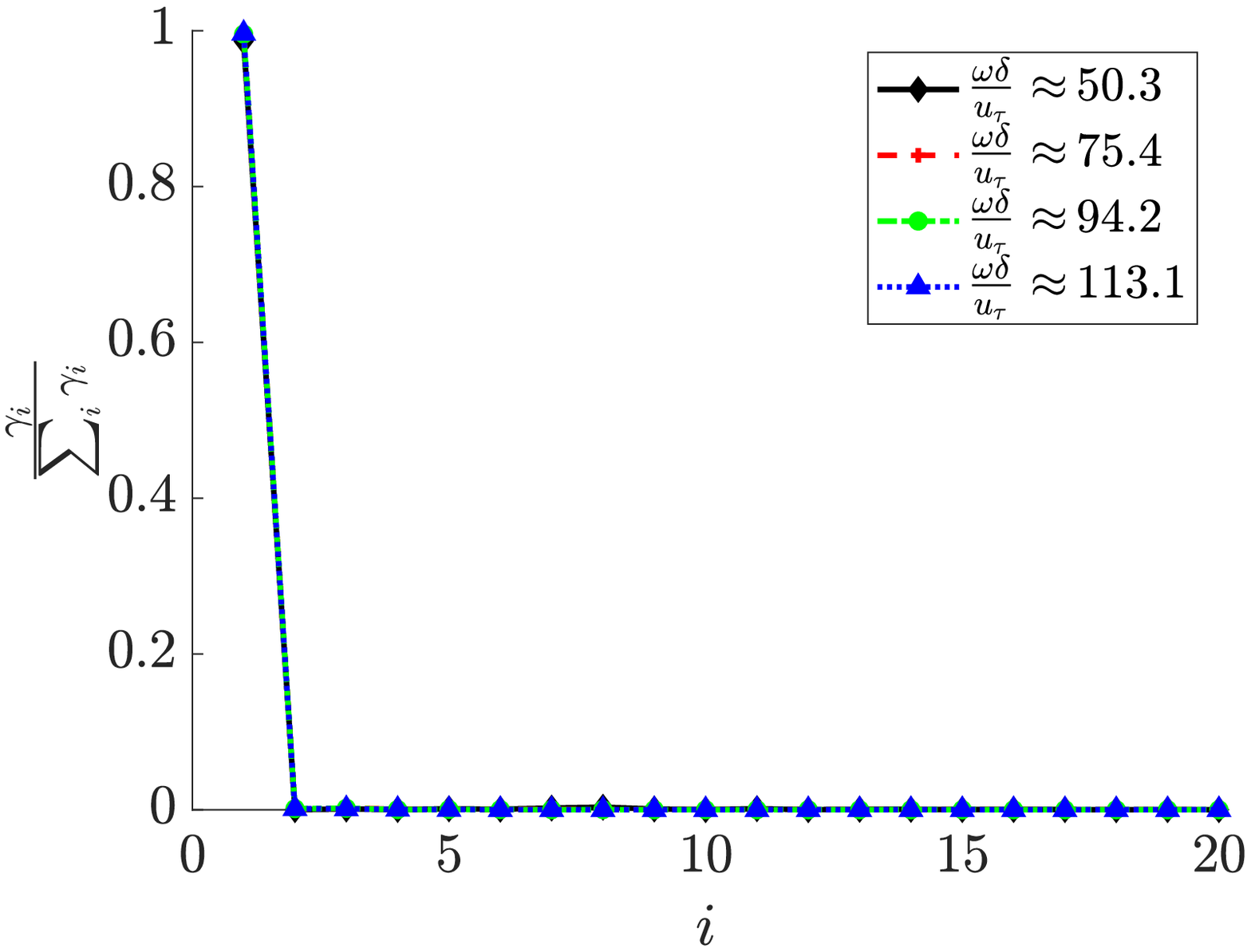}
\put(1,70){$(a)$}
\end{overpic}
\end{subfigure}
\begin{subfigure}{0.49\textwidth}
\begin{overpic}[width=\linewidth]{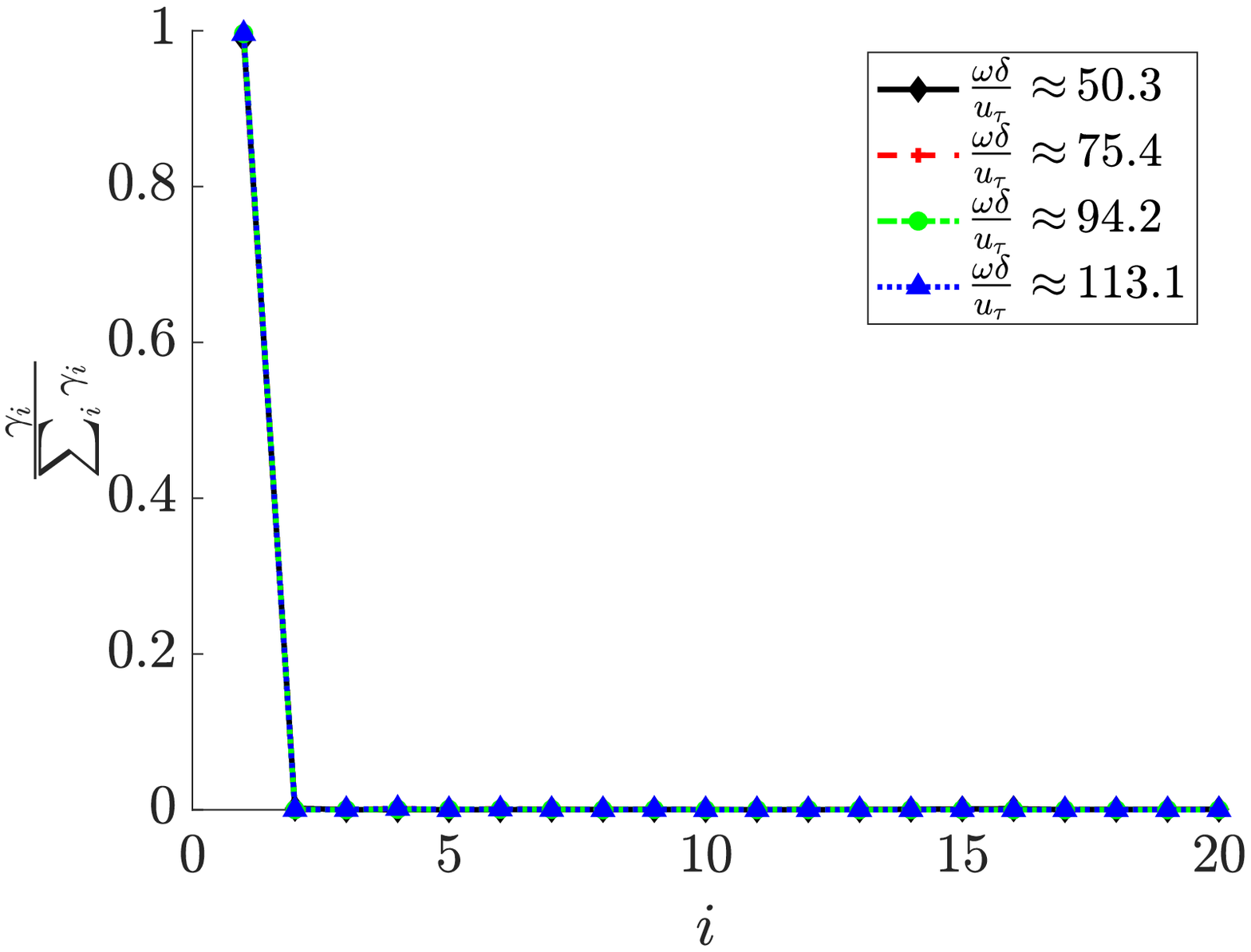}
\put(1,70){$(b)$}
\end{overpic}
\end{subfigure}
\caption{Contribution of each spectral POD mode to plate averaged displacement PSD for a) $Re_{\tau}=180$ and b) $Re_{\tau}=400$ computed using the Poisson inner product ($\beta=0.5$).}
\label{fig:spod_eigwt}
\end{figure}

\begin{figure}
\centering
\begin{subfigure}{0.49\textwidth}
\begin{overpic}[width=\linewidth]{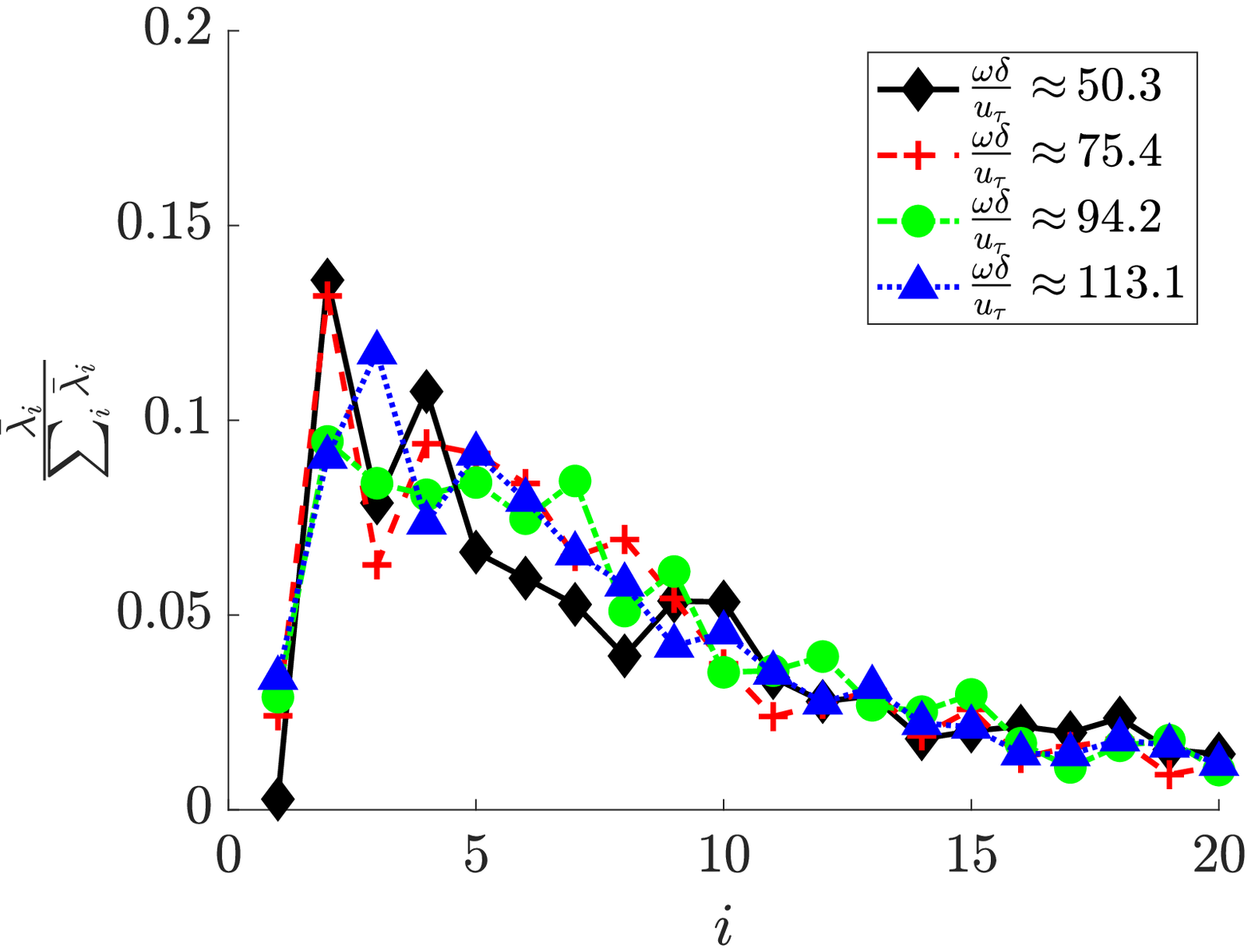}
\put(1,70){$(a)$}
\end{overpic}
\end{subfigure}
\begin{subfigure}{0.49\textwidth}
\begin{overpic}[width=\linewidth]{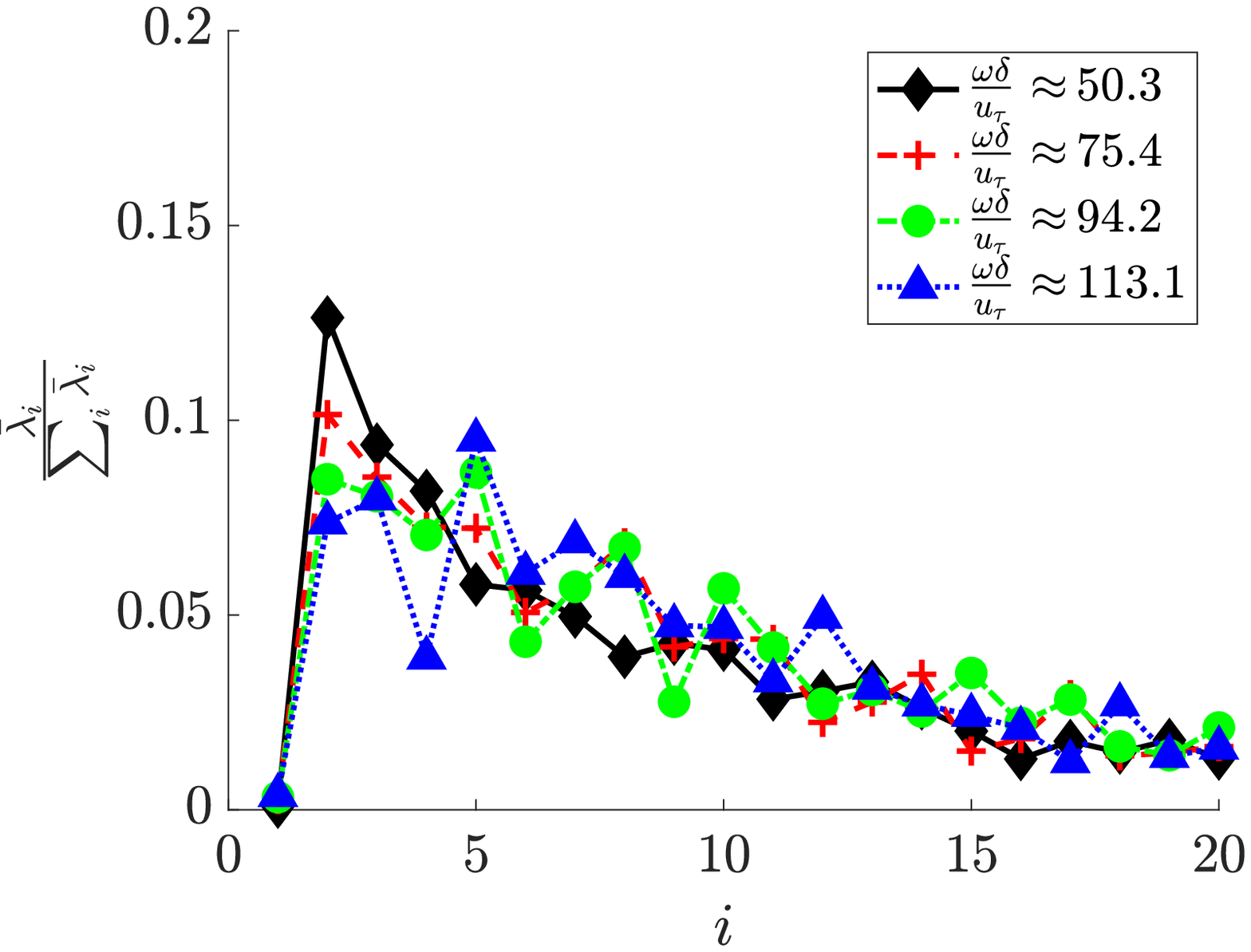}
\put(1,70){$(b)$}
\end{overpic}
\end{subfigure}
\caption{Contribution of each spectral POD mode computed using the
  Poisson inner product ($\beta=0.5$) to the net displacement source
  PSD for a) $Re_{\tau}=180$ and b) $Re_{\tau}=400$. For definition of
  $\bar{\lambda}_i$, see equation \ref{eqn:lambdabardefn}}.
\label{fig:spod_eigbar}
\end{figure}

\begin{figure}[htbp]
\centering
\begin{subfigure}{0.24\textwidth}
\begin{overpic}[width=\linewidth]{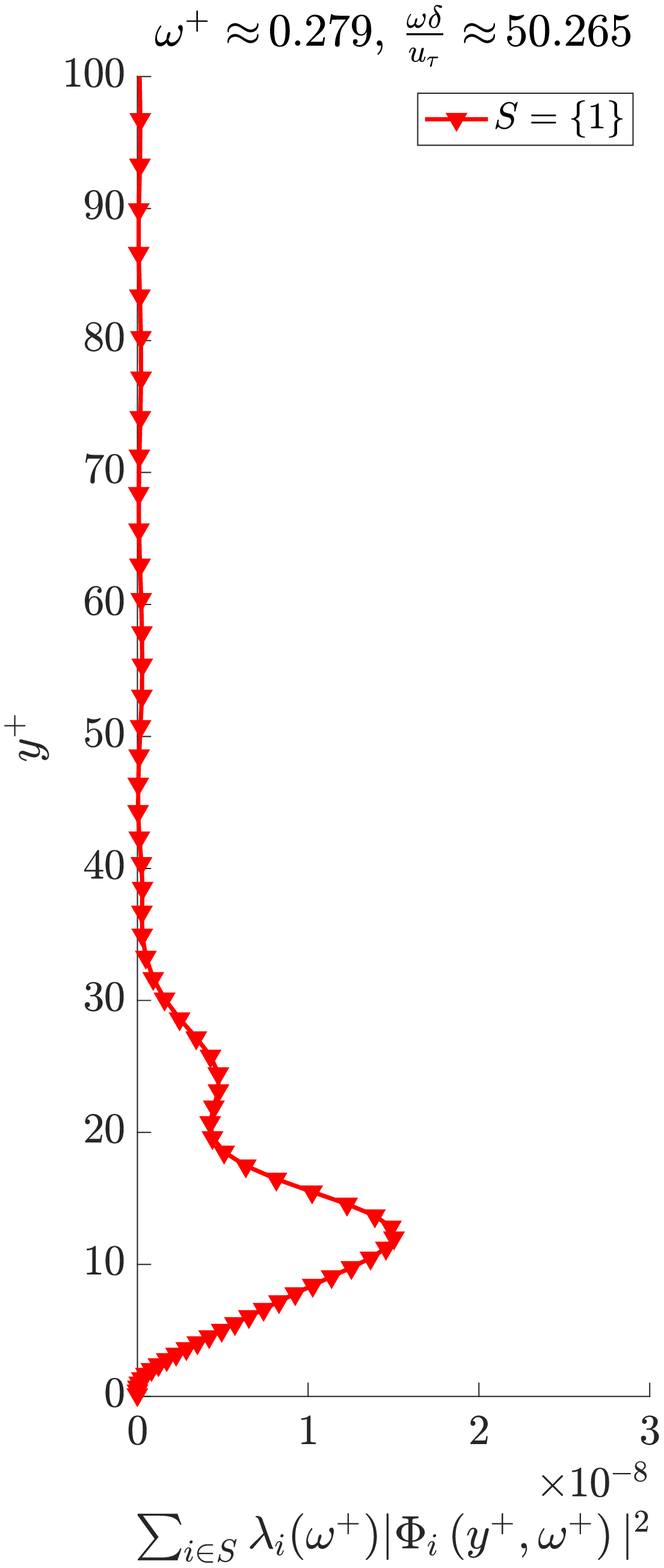}
\put(-2,1){$(a)$}
\end{overpic}
\end{subfigure}
\begin{subfigure}{0.24\textwidth}
\begin{overpic}[width=\linewidth]{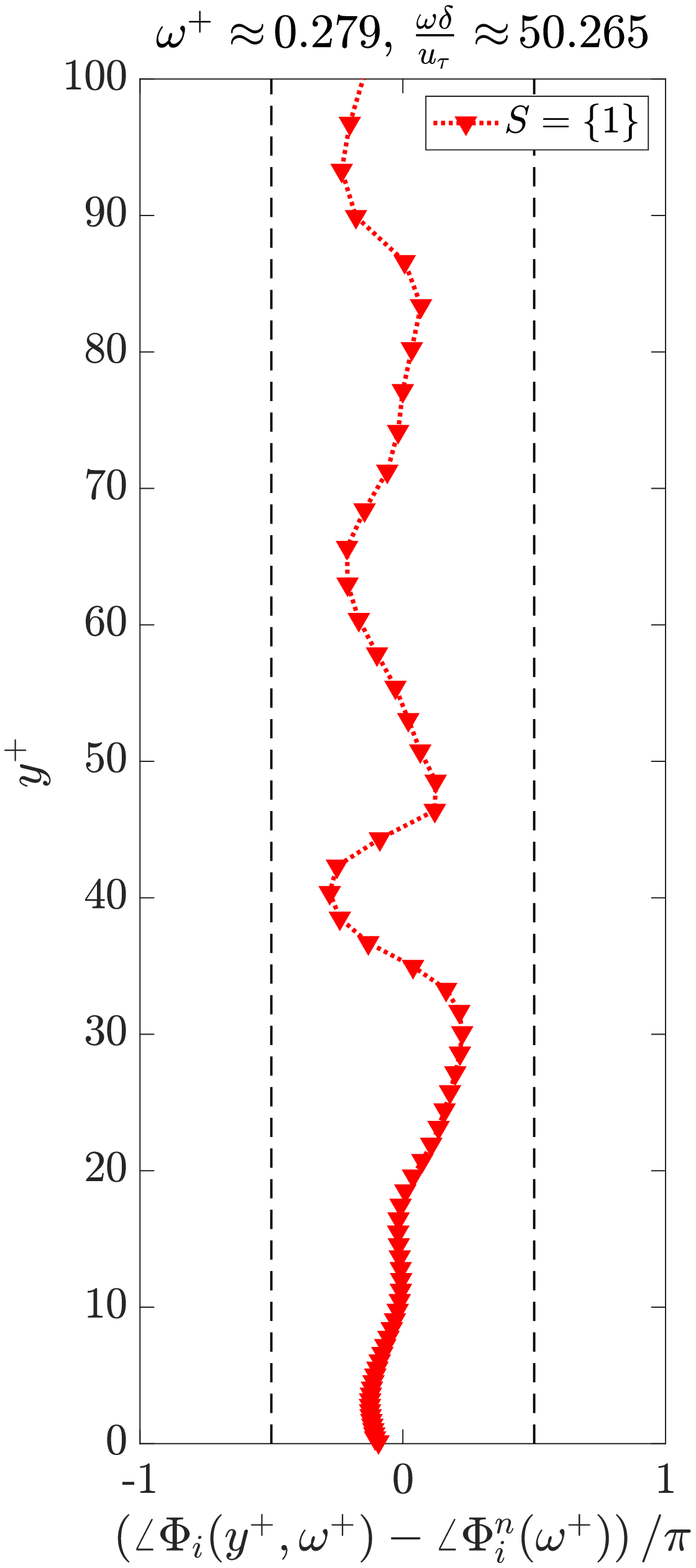}
\put(-2,1){$(b)$}
\end{overpic}
\end{subfigure}
\begin{subfigure}{0.24\textwidth}
\begin{overpic}[width=\linewidth]{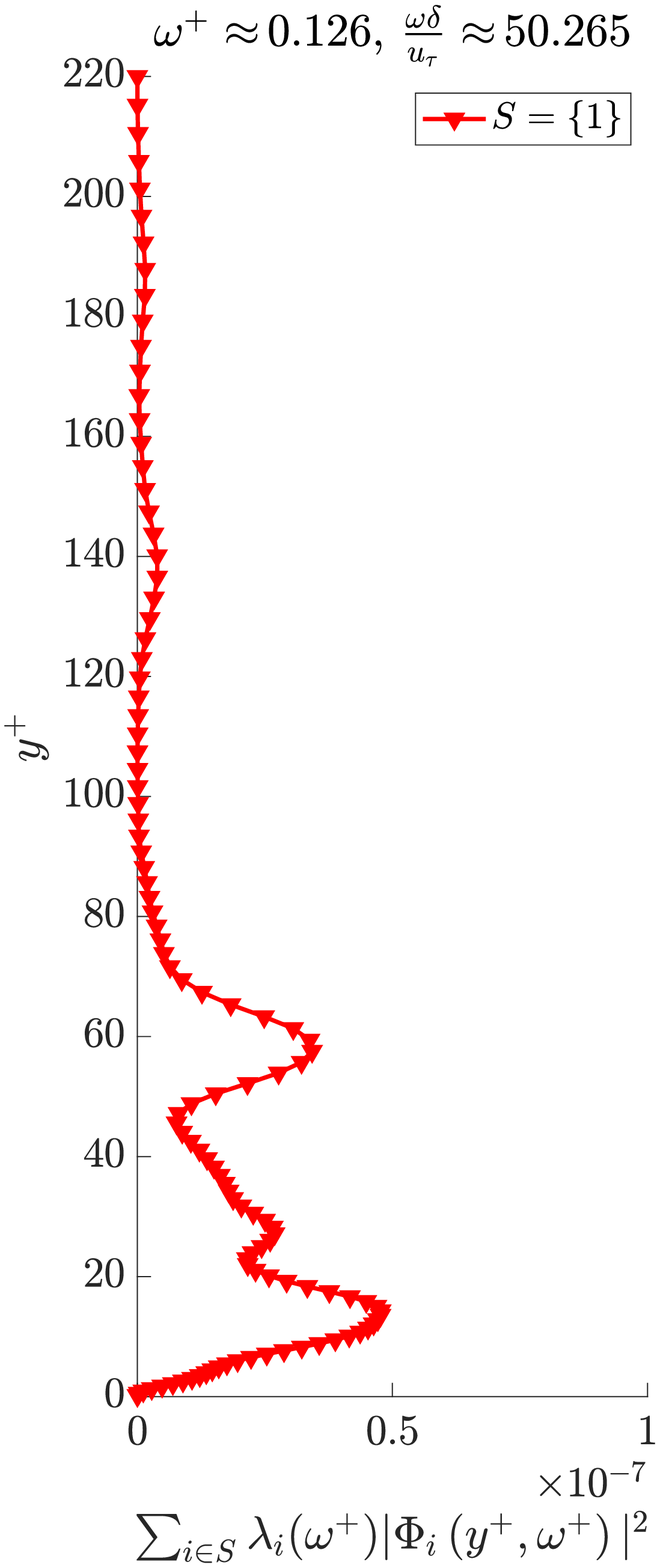}
\put(-2,1){$(c)$}
\end{overpic}
\end{subfigure}
\begin{subfigure}{0.24\textwidth}
\begin{overpic}[width=\linewidth]{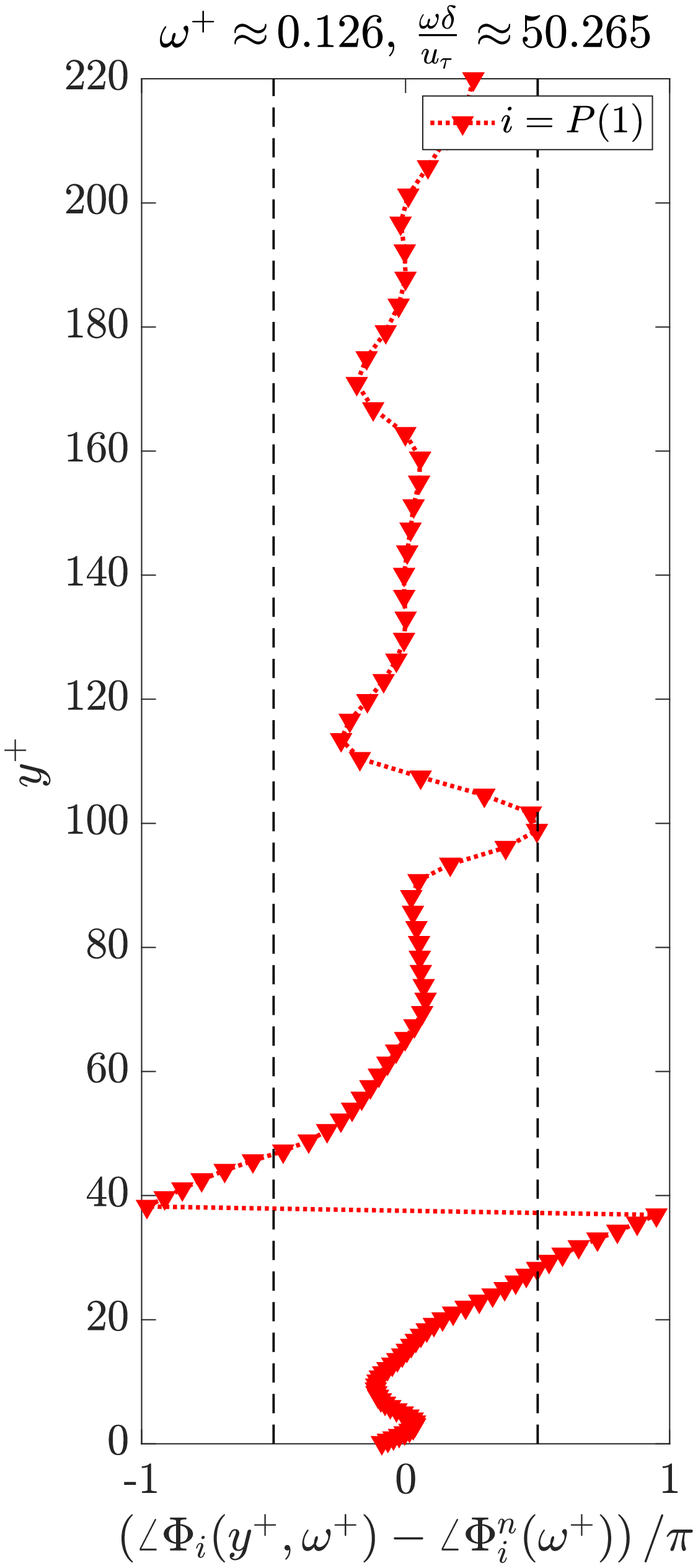}
\put(-2,1){$(d)$}
\end{overpic}
\end{subfigure}

\caption{Envelope and phase of the dominant spectral POD mode computed using the
  Poisson inner product ($\beta=0.5$) for $Re_{\tau}=180$
  ((a)-envelope, (b)-phase) and $Re_{\tau}=400$ ((c)-envelope,
  (d)-phase) at peak frequency $\omega\delta/u_{\tau}\approx 50.2$.}
\label{fig:dom_spod_mode_50}
\end{figure}

\begin{figure}[htbp]
\centering
\begin{subfigure}{0.24\textwidth}
\begin{overpic}[width=\linewidth]{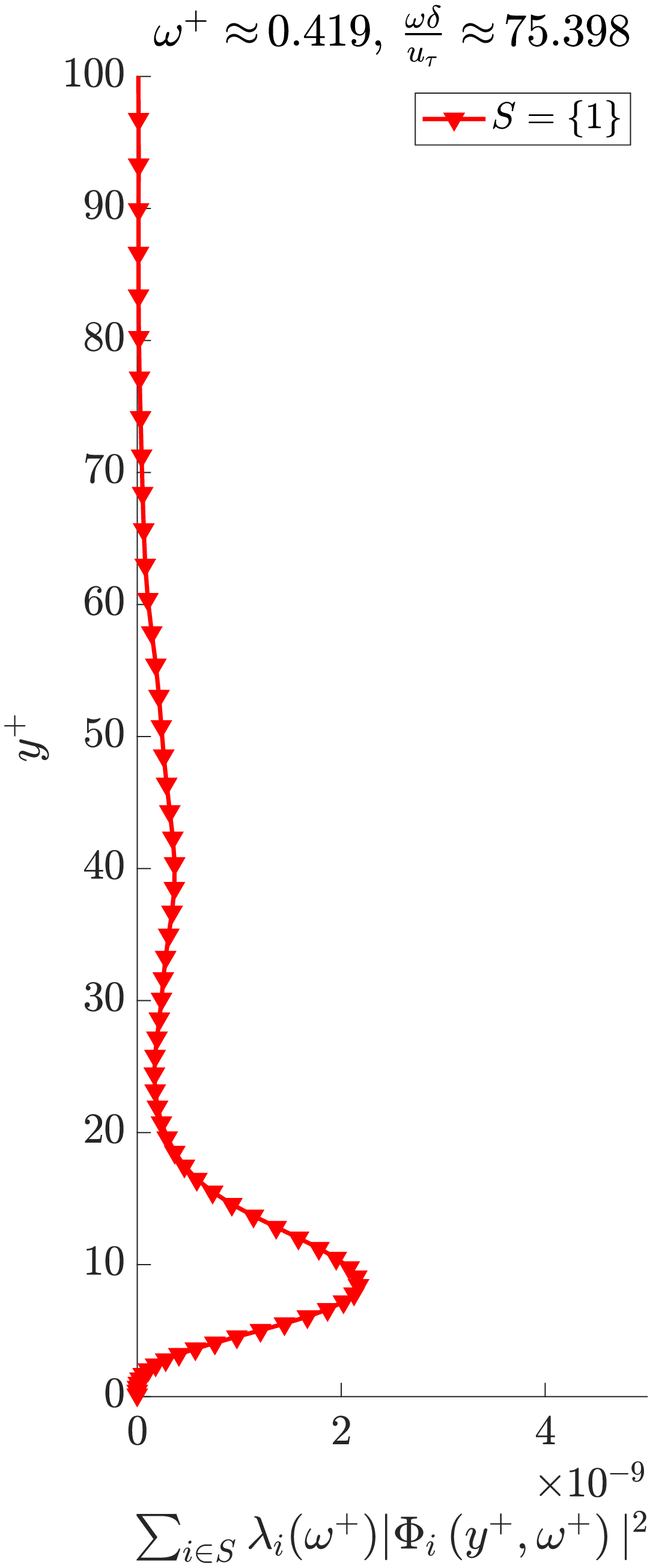}
\put(-2,1){$(a)$}
\end{overpic}
\end{subfigure}
\begin{subfigure}{0.24\textwidth}
\begin{overpic}[width=\linewidth]{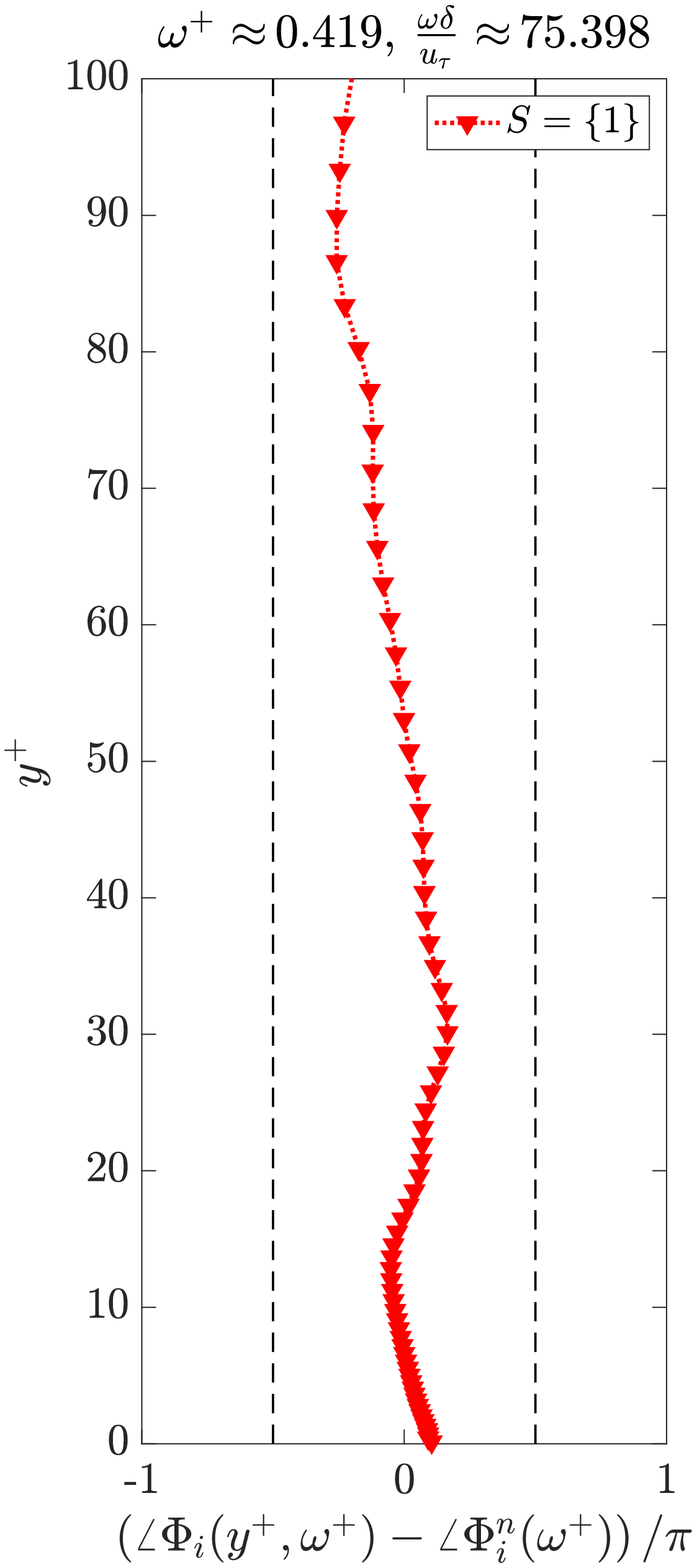}
\put(-2,1){$(b)$}
\end{overpic}
\end{subfigure}
\begin{subfigure}{0.24\textwidth}
\begin{overpic}[width=\linewidth]{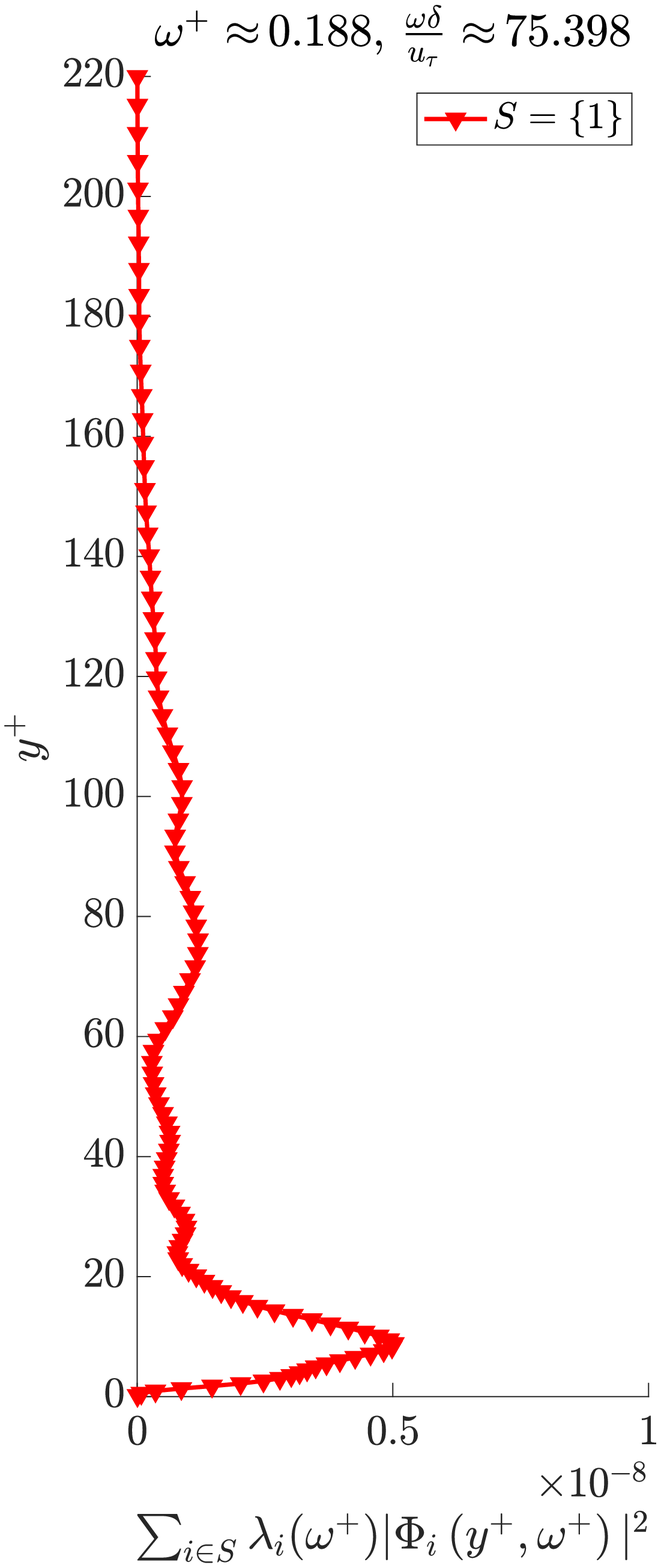}
\put(-2,1){$(c)$}
\end{overpic}
\end{subfigure}
\begin{subfigure}{0.24\textwidth}
\begin{overpic}[width=\linewidth]{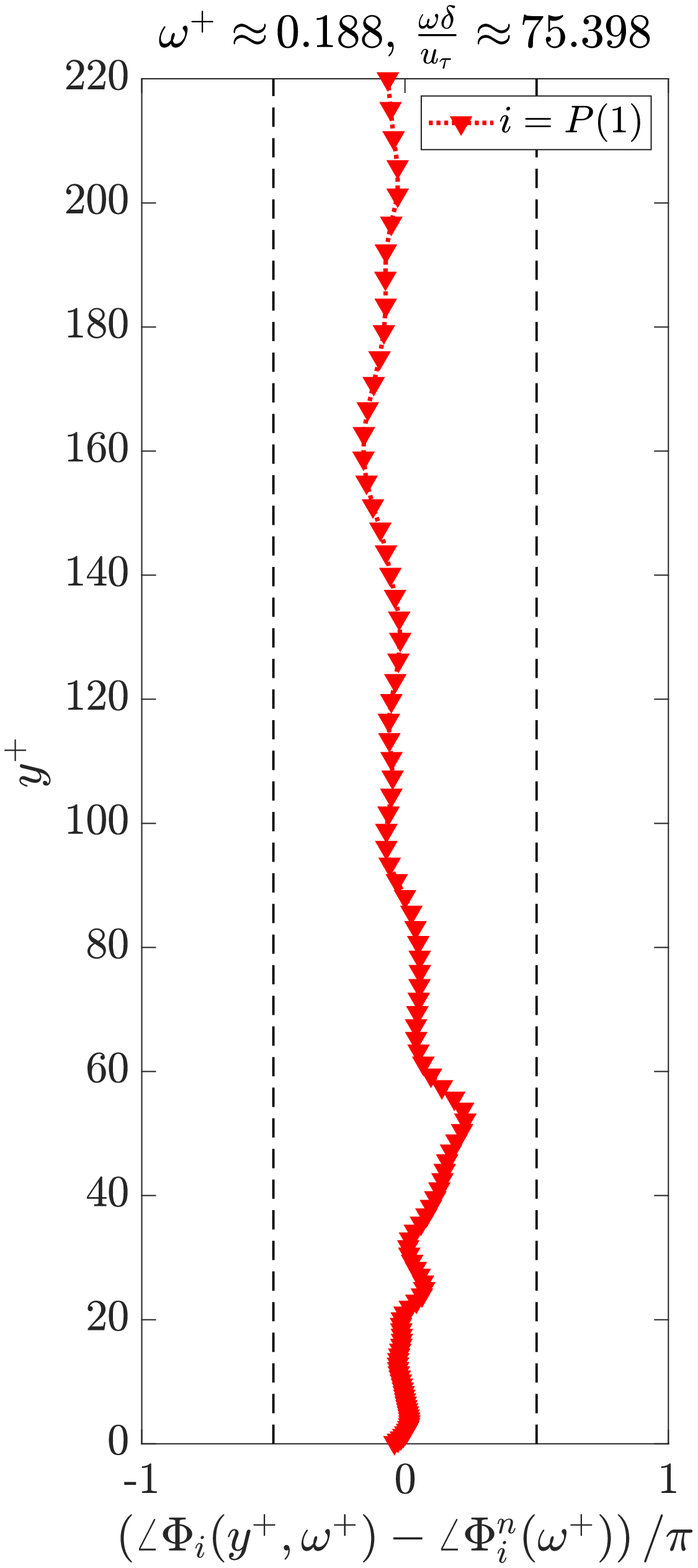}
\put(-2,1){$(d)$}
\end{overpic}
\end{subfigure}

\caption{Envelope and phase of the dominant spectral POD mode computed using the
  Poisson inner product ($\beta=0.5$) for $Re_{\tau}=180$
  ((a)-envelope, (b)-phase) and $Re_{\tau}=400$ ((c)-envelope,
  (d)-phase) at peak frequency $\omega\delta/u_{\tau}\approx 75$.}
\label{fig:dom_spod_mode_75}
\end{figure}

\begin{figure}[htbp]
\centering
\begin{subfigure}{0.24\textwidth}
\begin{overpic}[width=\linewidth]{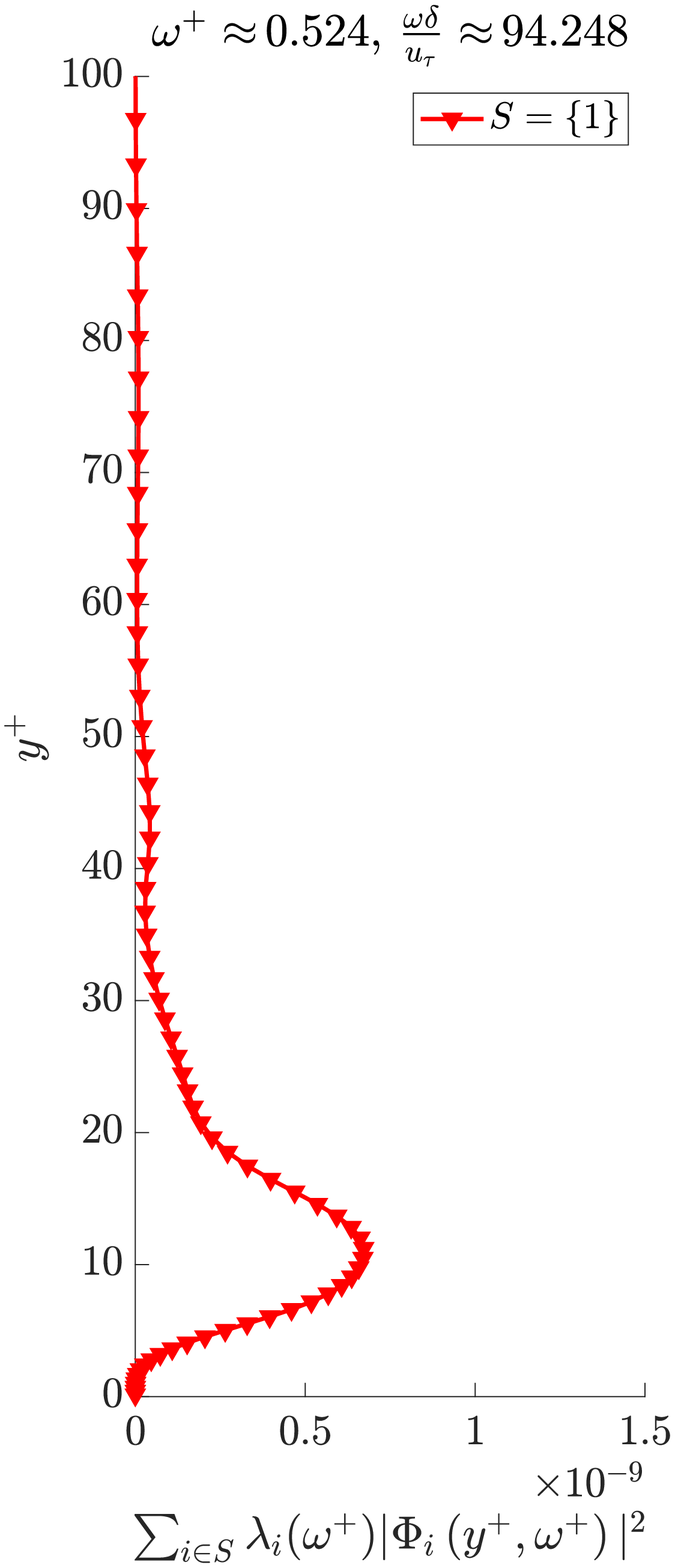}
\put(-2,1){$(a)$}
\end{overpic}
\end{subfigure}
\begin{subfigure}{0.24\textwidth}
\begin{overpic}[width=\linewidth]{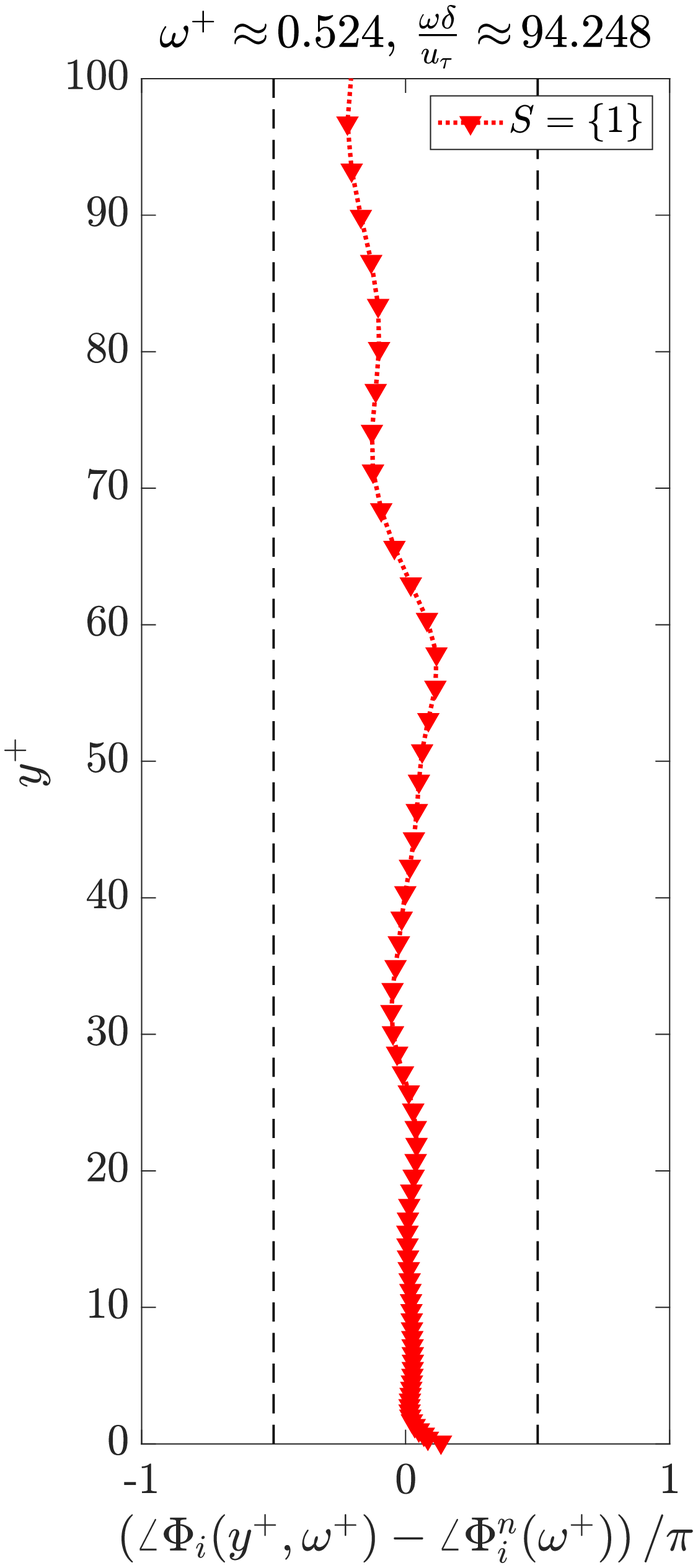}
\put(-2,1){$(b)$}
\end{overpic}
\end{subfigure}
\begin{subfigure}{0.24\textwidth}
\begin{overpic}[width=\linewidth]{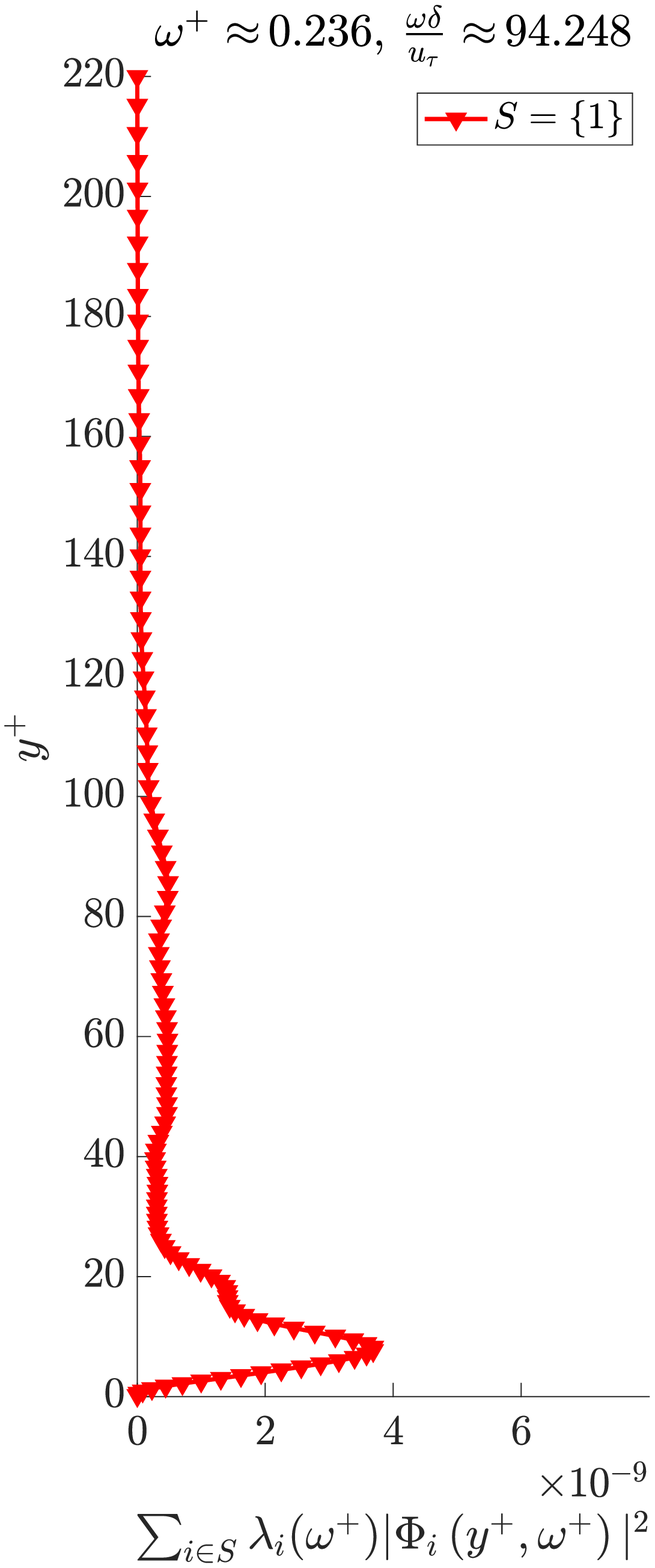}
\put(-2,1){$(c)$}
\end{overpic}
\end{subfigure}
\begin{subfigure}{0.24\textwidth}
\begin{overpic}[width=\linewidth]{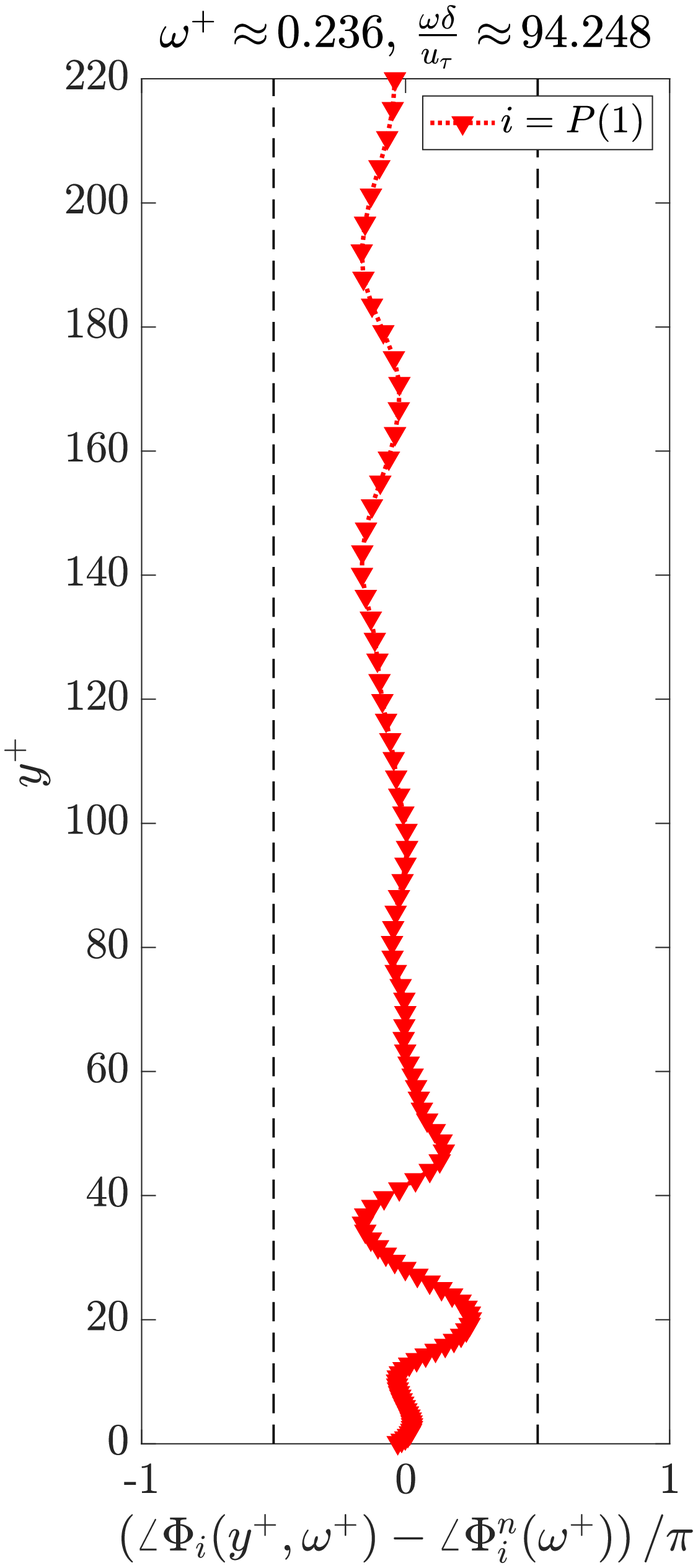}
\put(-2,1){$(d)$}
\end{overpic}
\end{subfigure}

\caption{Envelope and phase of the dominant spectral POD mode computed using the
  Poisson inner product ($\beta=0.5$) for $Re_{\tau}=180$
  ((a)-envelope, (b)-phase) and $Re_{\tau}=400$ ((c)-envelope,
  (d)-phase) at peak frequency $\omega\delta/u_{\tau}\approx 94$.}
\label{fig:dom_spod_mode_94}
\end{figure}

\begin{figure}[htbp]
\centering
\begin{subfigure}{0.24\textwidth}
\begin{overpic}[width=\linewidth]{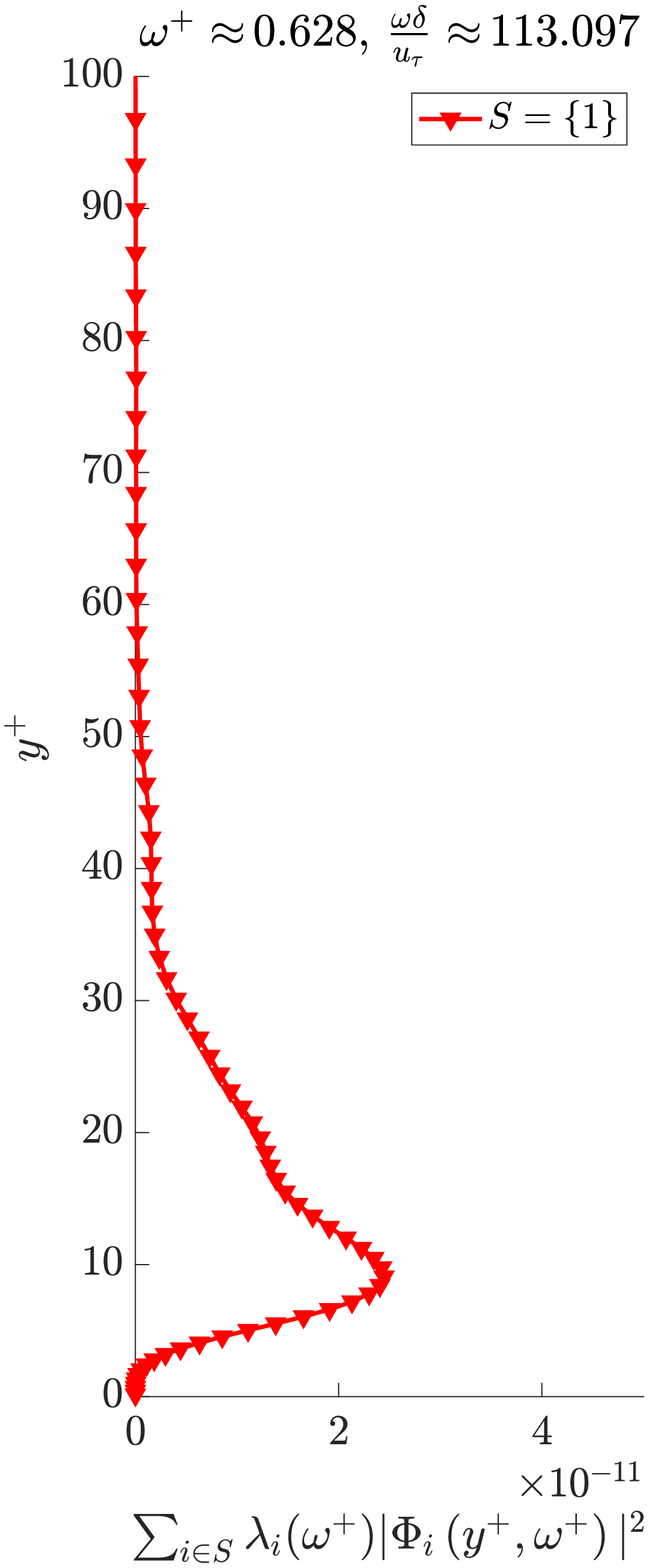}
\put(-2,1){$(a)$}
\end{overpic}
\end{subfigure}
\begin{subfigure}{0.24\textwidth}
\begin{overpic}[width=\linewidth]{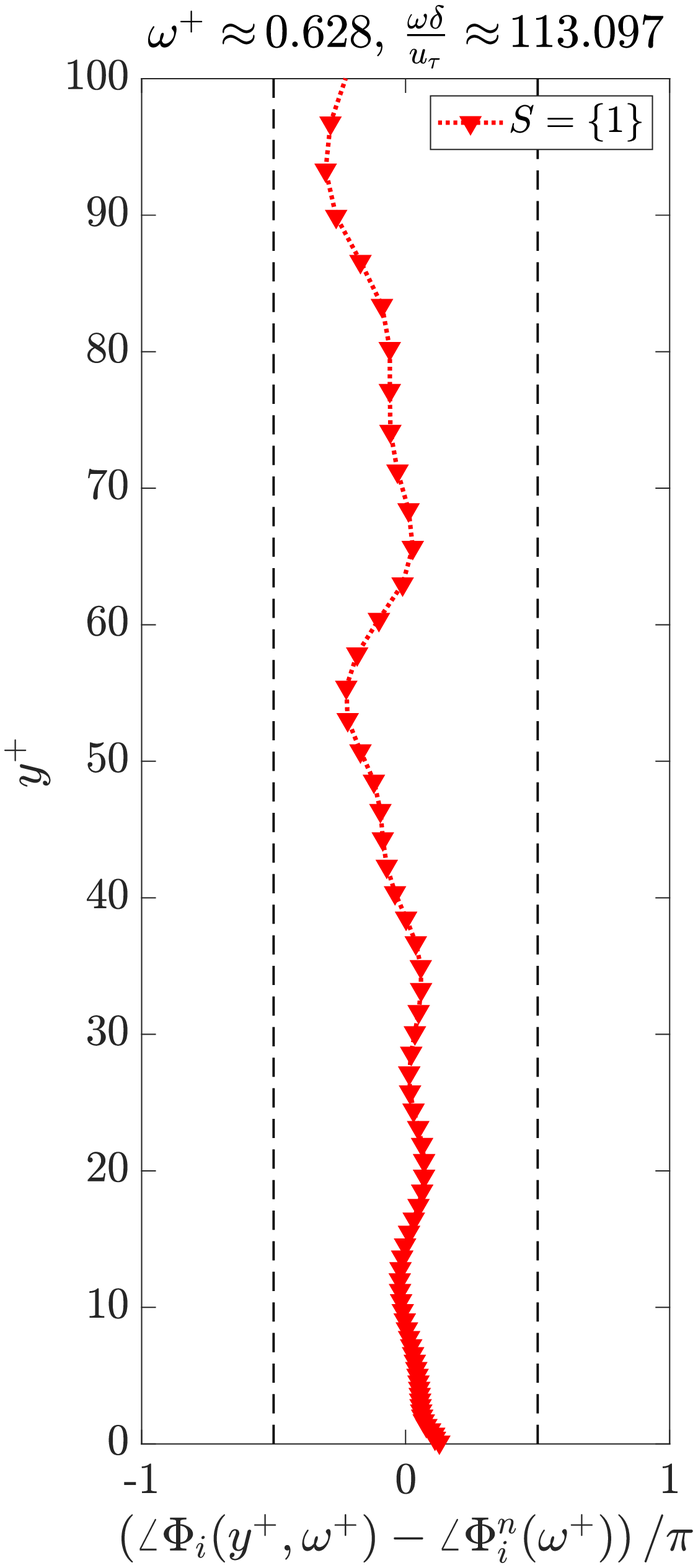}
\put(-2,1){$(b)$}
\end{overpic}
\end{subfigure}
\begin{subfigure}{0.24\textwidth}
\begin{overpic}[width=\linewidth]{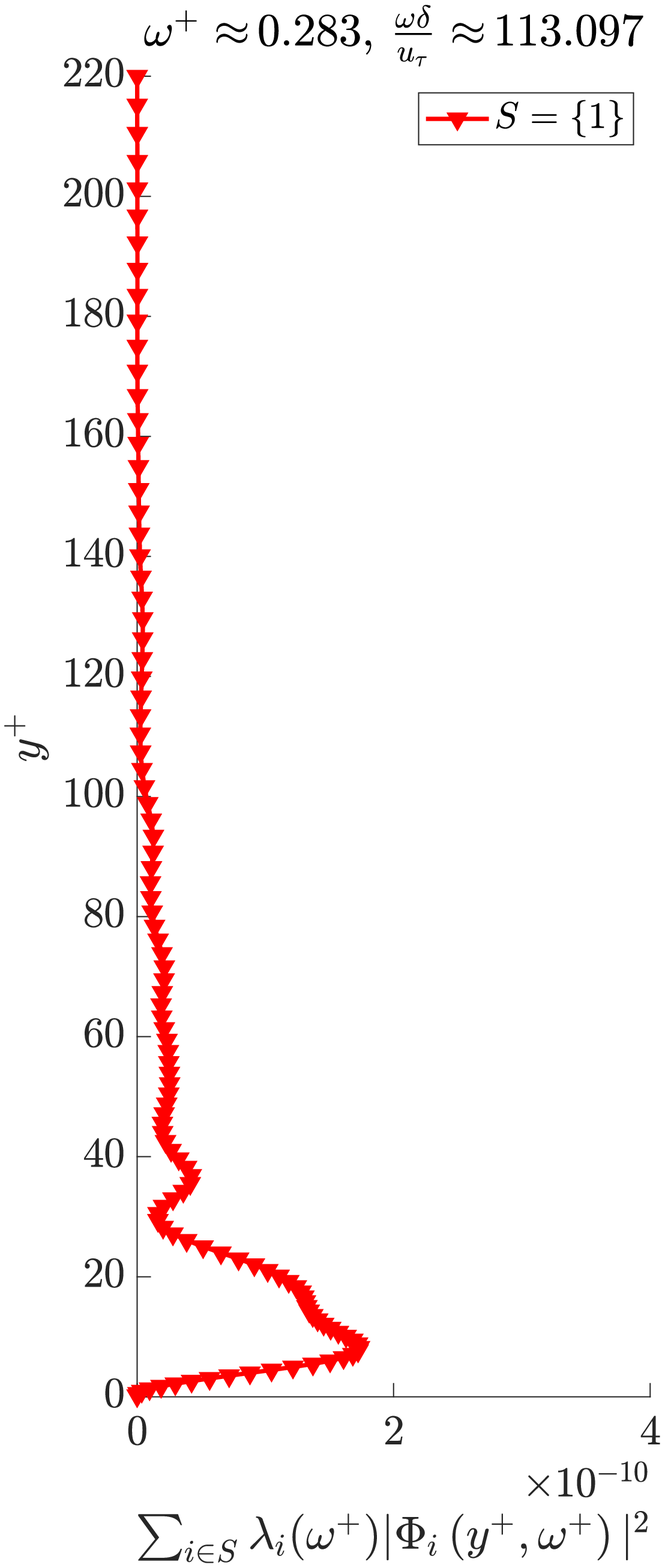}
\put(-2,1){$(c)$}
\end{overpic}
\end{subfigure}
\begin{subfigure}{0.24\textwidth}
\begin{overpic}[width=\linewidth]{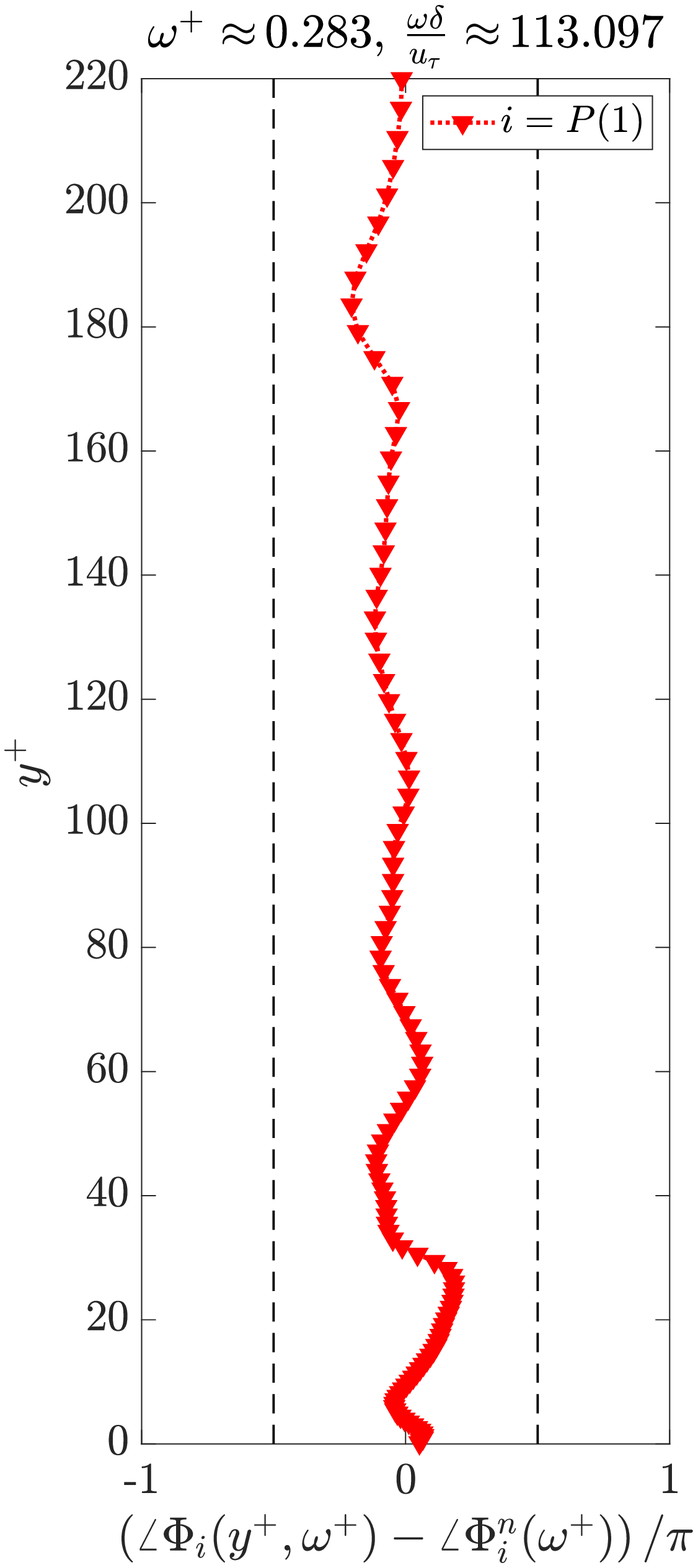}
\put(-2,1){$(d)$}
\end{overpic}
\end{subfigure}

\caption{Envelope and phase of the dominant spectral POD mode computed using the
  Poisson inner product ($\beta=0.5$) for $Re_{\tau}=180$
  ((a)-envelope, (b)-phase) and $Re_{\tau}=400$ ((c)-envelope,
  (d)-phase) at peak frequency $\omega\delta/u_{\tau}\approx 113$.}
\label{fig:dom_spod_mode_113}
\end{figure}

\begin{figure}[htbp]
\centering
\begin{subfigure}{0.24\textwidth}
\begin{overpic}[width=\linewidth]{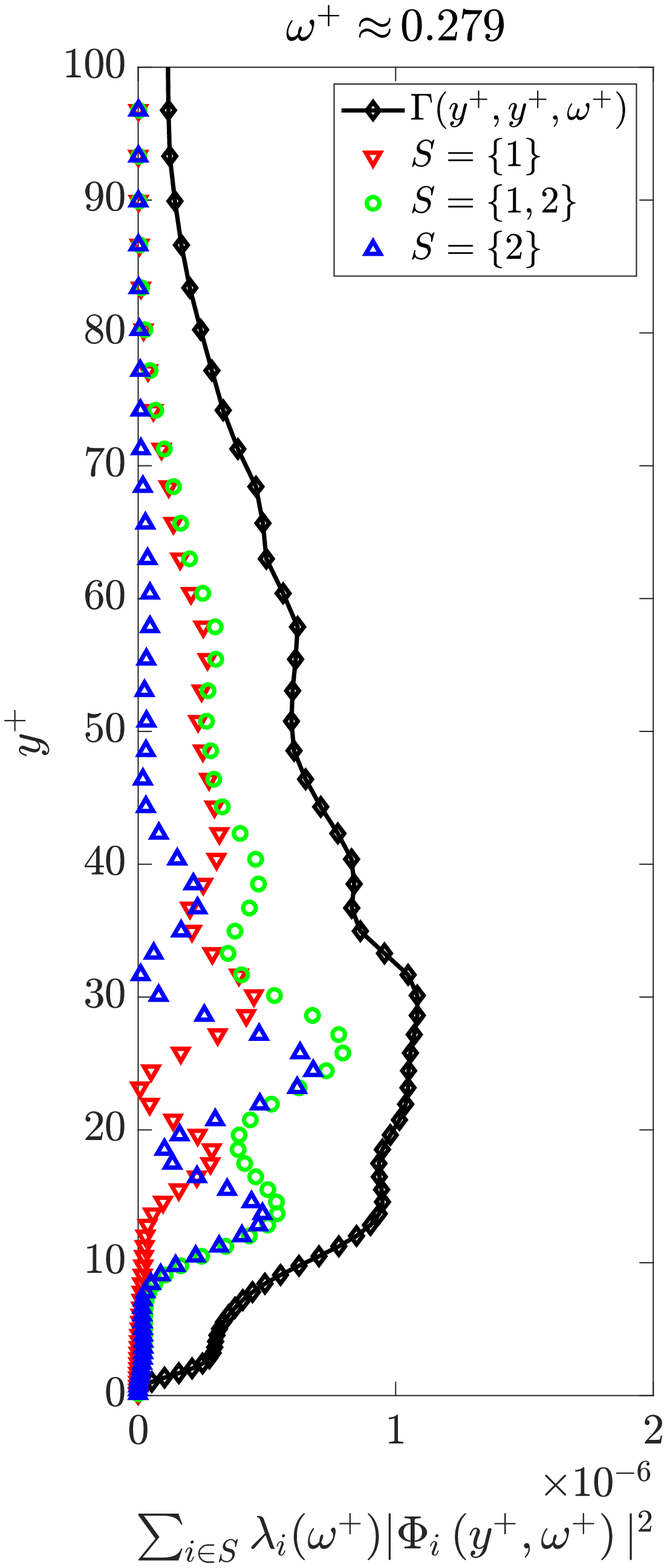}
\put(-2,1){$(a)$}
\end{overpic}
\end{subfigure}
\begin{subfigure}{0.24\textwidth}
\begin{overpic}[width=\linewidth]{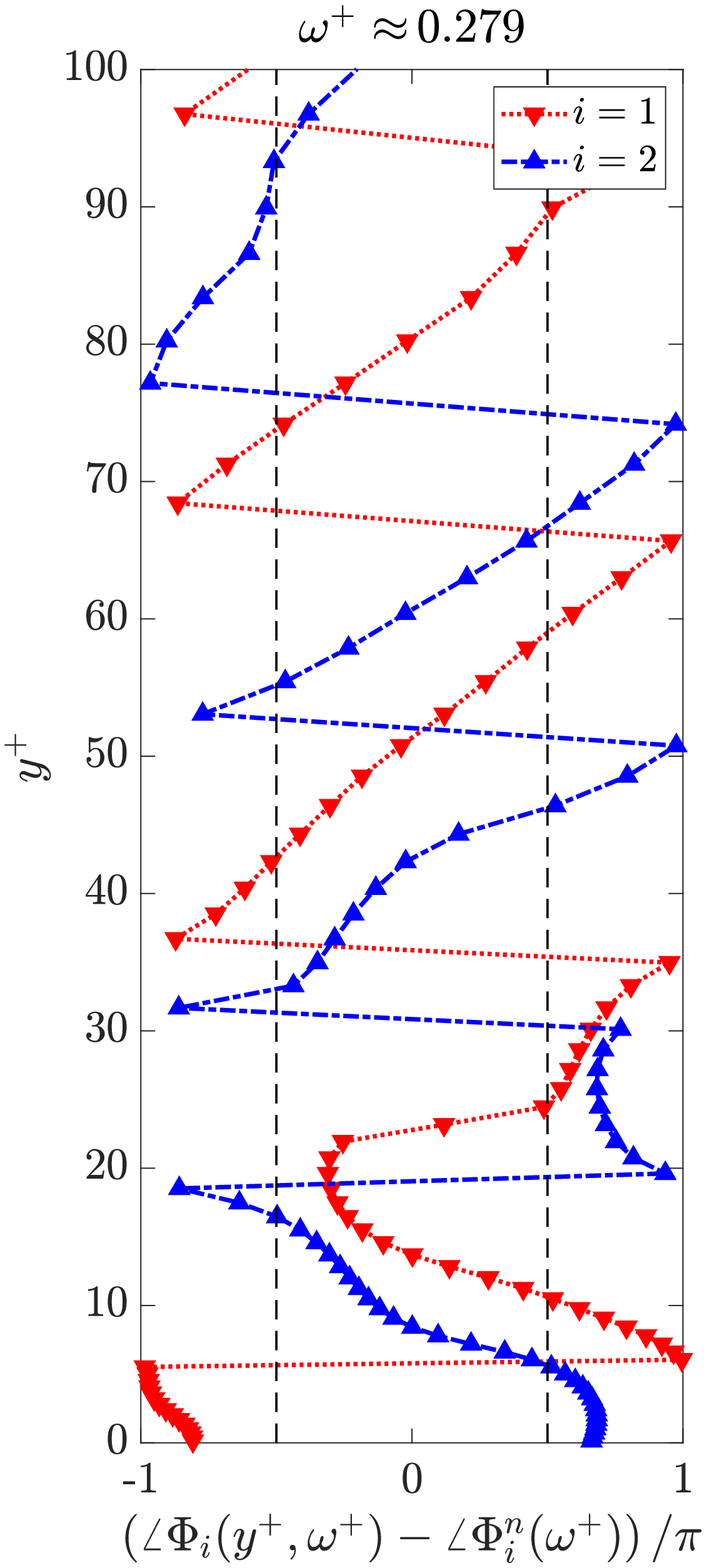}
\put(-2,1){$(b)$}
\end{overpic}
\end{subfigure}
\begin{subfigure}{0.24\textwidth}
\begin{overpic}[width=\linewidth]{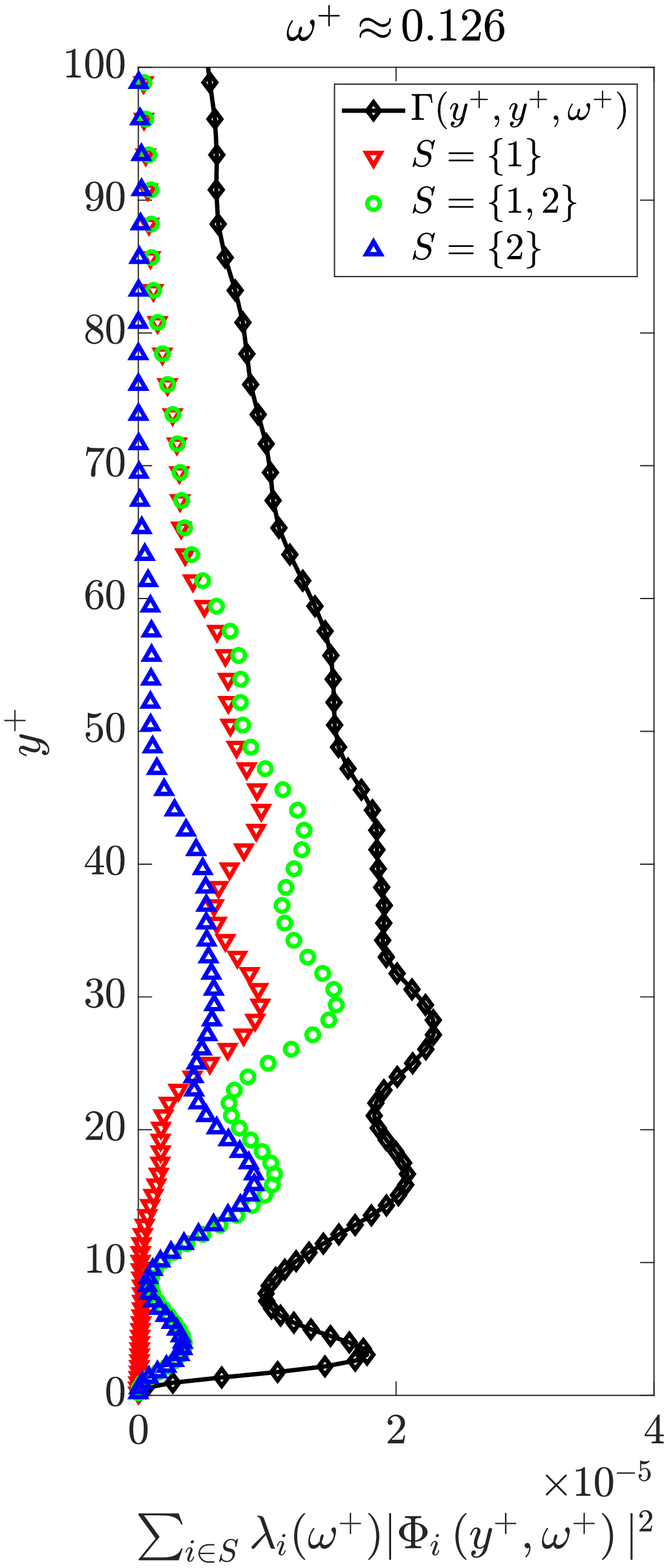}
\put(-2,1){$(c)$}
\end{overpic}
\end{subfigure}
\begin{subfigure}{0.24\textwidth}
\begin{overpic}[width=\linewidth]{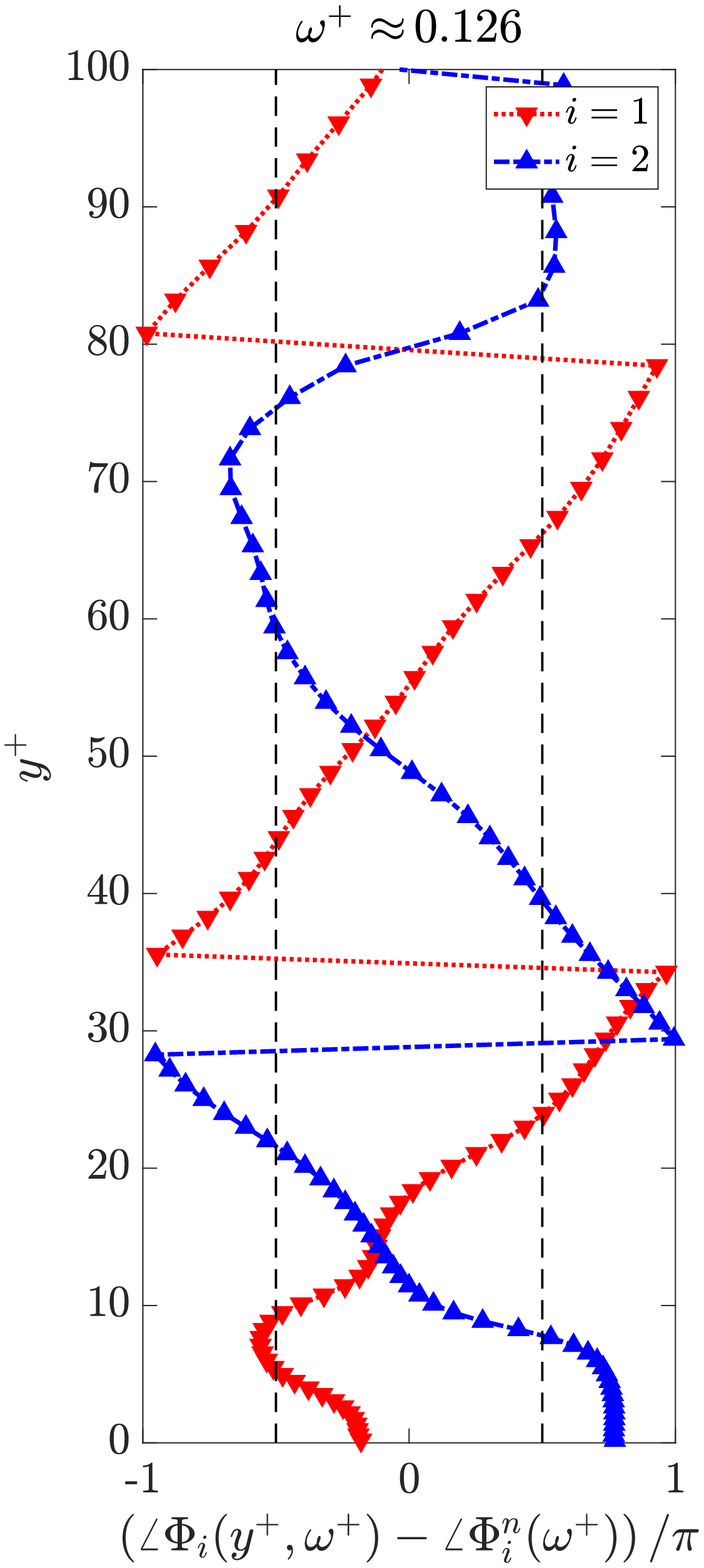}
\put(-2,1){$(d)$}
\end{overpic}
\end{subfigure}

\caption{Envelope and phase of the dominant spectral POD modes computed
  using the $L^2$ inner product for $Re_{\tau}=180$ ((a)-envelope,
  (b)-phase) and $Re_{\tau}=400$ ((c)-envelope, (d)-phase) at peak
  frequency $\omega\delta/u_{\tau}\approx
  50.2$.} \label{fig:dom_spod_mode_l2_50} \end{figure}

Figures \ref{fig:spod_eig}a and b show the computed spectral POD eigenvalues
(normalized by the sum of eigenvalues) for both $Re_{\tau}$ at the first four
peak frequencies in figure \ref{fig:plate_average}a. The eigenvalues decay
faster with increasing frequency indicating that fewer modes are required to
represent the double integral
$\iint_{-1}^{+1}G(r,s,\beta/(1-\beta))/(1-\beta)\,\Gamma^a(r,s,\omega)\,\mathrm{dr}\,\mathrm{ds}$
(equation \ref{eqn:sumofspod}).

Figures \ref{fig:spod_eigwt}a and b show the contribution of each spectral POD
mode to the plate averaged displacement PSD (equation \ref{eqn:phi_a_gamma}) for
the frequencies plotted in figure \ref{fig:spod_eig}. The first spectral POD
mode contributes nearly all the plate averaged displacement PSD at all
frequencies for both $Re_{\tau}$. Thus, the first spectral POD mode is the
dominant decorrelated contributor to plate response at all the peak frequencies.
To investigate the structure of this dominant fluid source, we plot the envelope
and phase of the first spectral POD mode in figures
\ref{fig:dom_spod_mode_50}-\ref{fig:dom_spod_mode_113} for the frequencies
plotted in figure \ref{fig:spod_eig}. For all the Reynolds numbers and
frequencies, the envelope is maximum in the buffer layer around
$y^+\approx 10$ and the modes have a similar wall-normal width for both
$Re_{\tau}$. This again reaffirms the observation in the previous section that
the location and width of the dominant fluid source is in the buffer layer and
depends on outer units, respectively. The phase variation of these dominant
modes is mostly in the range $-\pi/2$ to $\pi/2$. Thus, the contribution from
different wall-parallel planes interfere constructively. This constructive
interference is absent in the suboptimal spectral POD modes. Even though the second
spectral POD mode contains more energy than the first mode (figure
\ref{fig:spod_eigbar}), the contributions interfere destructively resulting in
very small net contribution. Therefore, the interference of the contributions
from different wall-parallel planes play a major role in determining the
dominance of a spectral POD mode.

Figure \ref{fig:spod_eigbar} shows that the dominant spectral POD mode
is not the dominant contributor to the integrated net displacement
source PSD
$(\int_{-\delta}^{+\delta}\Gamma^a(y,y,\omega)\,\mathrm{dy})$, i.e.,
they are not energetically dominant. The first two energetically
dominant modes identified by the inner product with $\beta=1$
(standard $L^2$ inner product) is shown in figure
\ref{fig:dom_spod_mode_l2_50} for $\omega\delta/u_{\tau}\approx
50$. The shape of the modes resemble a stationary wavepacket enclosing
a travelling wave (almost linear phase variation) for both
$Re_{\tau}$. However, these wavepackets do not contribute much to the
response of the plate as the contribution from different wall-normal
locations interfere destructively to produce no net contribution. This
behavior of the energetically dominant mode is also true at higher
frequencies (not shown).  Further, the spectral POD modes identified
by the inner-products with $\beta=1$ (standard $L^2$ inner-product)
and $\beta=0.5$ are inherently different because they optimize the
modes based on their contribution to
$\int_{-\delta}^{+\delta}\Gamma^a(y,y,\omega)\,\mathrm{dy}$ and
$\iint_{-\delta}^{+\delta}2G(y,r,1))\Gamma^a(r,y,\omega)\,\mathrm{dr}\mathrm{dy}$,
respectively. The former picks the energetically dominant mode whereas
the latter identifies the mode that contributes the most to the double
integral (which is a proxy for $\phi_{dd}^a(\omega)$).

Overall, spectral POD identifies a single dominant contributor to the
plate excitation at each of the first four peak frequencies of the
plate averaged displacement PSD. All the identified dominant plate
excitation modes have an envelope that has a peak in the buffer layer
around $y^+\approx 10-13$ and has a width that scales in outer units for
the two $Re_{\tau}$.

\section{Summary} \label{sec:summary}

In summary, we present a novel framework to investigate the
fluid-solid coupling in a canonical setting - linear one-way coupled
excitation of an elastic plate in turbulent channel flow. We apply the
framework to explain the response of a clamped plate obtained using
the in-house FSI solver - MPCUGLES-SOLID at $Re_{\tau}=180$ and $400$.

The structural solver is first validated using the experiment of
\cite{han1999prediction}. They measured the response of a rectangular
steel plate excited by a turbulent boundary layer at
$Re_{\tau}\approx 2000$. We generate synthetic space-time
wall-pressure fluctuations at the experimental conditions using a
Fourier series methodology. The generated fluctuations satisfy the
Corcos CSD and Smolyakov-Tkachenko PSD models. To compute the surface
forces accurately and efficiently, we perform $L^2$ orthogonal
projection of the generated wall-pressure fluctuations onto the
Legendre polynomial basis within each surface element of the
plate. The obtained time-domain FEM response of the plate shows good
agreement with the measured velocity PSD at a point on the plate.

The obtained plate response at $Re_{\tau}=180$ and $400$ with fixed
non-dimensional Young's modulus $E/(\rho_fu_{\tau}^2)$ have
overlapping plate averaged low frequency spectrum in outer units.
But, the high frequency component of the spectrum does not show
overlap in inner units. Fixing $E\delta^2/(\rho_f\nu^2)$ instead of
$E/(\rho_fu_{\tau}^2)$ for the two $Re_{\tau}$ yields a better
collapse of the high-frequency region in inner units. We show that
this high-frequency behavior is due to the inner scaling of the modal
wavenumber of the plate.

In the proposed fluid-solid coupling framework, we express the
displacement at a point on the surface of the plate $d(x,-1,z,t)$ as a
wall-normal integral of the net displacement source $f_d(x,y,z,t)$,
i.e.,
$d(x,-\delta,z,t)=\int_{-\delta}^{+\delta}f_d(x,y,z,t)\,\mathrm{dy}$. To
quantify the statistical features of fluid sources of plate
excitation, we compute the plate averaged CSD of the net displacement
source using the DNS database and modal superposition. We use the
first six mode shapes of the plate to investigate the fluid sources
responsible for the first four peaks of the plate averaged
displacement spectra. The computed plate averaged WD-NDS CSD
$\Psi^a(y,\omega)$ shows a dominant buffer layer contribution at the
peak frequencies of the plate averaged displacement PSD. The CSD has
large values for $y/\delta<0.75$ indicating that the width of the
fluid source depends on outer units. Further, we show that the Corcos
form of the wavenumber frequency spectrum implies identical coupling
of the plate averaged displacement PSD and the wall-pressure PSD with
the fluid sources in the channel upto a multiplicative constant.

We perform spectral POD of the net displacement source CSD to identify
the decorrelated dominant fluid sources responsible for the plate
excitation. To accomplish this, we require the modes to be orthogonal
in a Poisson inner product (with $\beta=0.5$ instead of the commonly
used $L^2$ inner product). The envelope of the dominant spectral POD
mode (obtained with $\beta=0.5$) peaks in the buffer region around
$y^+\approx 10-13$ for both $Re_{\tau}$ and the width of the envelope
scales in outer units. This reaffirms the previous observation that
the location and wall-normal width of the dominant source is a
function of inner and outer units, respectively. The dominance of such
a fluid source is mainly due to the constructive interference of the
contributions from different wall-parallel planes. However, this
dominant contribution to plate excitation is not energetically
dominant. The energetically dominant fluid sources obtained from
spectral POD with the $L^2$ inner product ($\beta=1$) resemble
stationary wall-normal wave packets. But these wavepackets do not
contribute much to the plate response as the contribution from
different wall-normal locations to the plate response undergo
destructive interference.

Overall, the buffer region sources are dominant contributors to plate
excitation. In FSI simulations that use wall-modeled Large Eddy
Simulation (LES) in the fluid domain, the first point in the fluid
domain will be in the logarithmic layer. Thus, wall-modeling will fail
to account for this dominant buffer region contribution. With
increasing Reynolds numbers, the contribution of the outer layer
structures to the wall-pressure fluctuation increases
(\cite{panton2017correlation}). The proposed framework can be used to
quantitatively investigate this outer layer contribution to the plate
excitation. Further, due to the dominance of the buffer layer
structures, modulating the buffer region can be a possible way to
control structural excitation.

\section*{Acknowledgements}
This work is supported by the United States Office of Naval Research
(ONR) under grant N00014-17-1-2939 with Dr. Ki-Han Kim as the
technical monitor. The computations were made possible through
computing resources provided by the US Army Engineer Research and
Development Center (ERDC) in Vicksburg, Mississipi on the Cray
machines, Copper and Onyx of the High Performance Computing
Modernization Program. We also thank for the computing resources
provided by the US Air Force Research Laboratory DoD Supercomputing
Resource Center (DSRC) on the SGI ICE machine, Thunder of the High
Performance Computing Modernization Program.

%
%
%
%
%

\appendix
\section{Surface force evaluation for validation}\label{sec:appA}
In FEM, the integral to compute the surface force at a typical boundary node $i$ in a boundary element $e$ is 
\begin{equation} \label{eqn:surf_force}
\int_{\Gamma_e}p(x,-1,z,t)N_i(x,z)\,\mathrm{dx}\,\mathrm{dz},
\end{equation}
where $\Gamma_e$ is the boundary surface of element $e$ and $N_i$ is
the shape function of the node $i$. The cost of computing this
integral exactly using the standard Gauss-Legendre quadrature is high
because i) the generated $p(x,-1,z,t)$ consists of $\approx 1$ billion
terms and ii) the order of quadrature rule to exactly integrate the
high wavenumber component of the generated wall-pressure fluctuations
is high. To reduce this high cost, we write the projected pressure
fluctuation $\overline{p(x,-1,z,t)}$ within each surface element in
the Legendre polynomial basis, i.e.,
$\overline{p(x,-1,z,t)}\rvert_{\Gamma_e}=\sum_{j,k=0}^2\alpha^e_{j,k}(t)P_j(x)P_k(z)$,
where $\{P_j\}_{j=0}^2$ is the set of Legendre polynomials of degree
$\leq 2$. To find the coefficients
$\{\{\alpha_{j,k}(t)\}_{j,k=0}^2\}$, we require the error in
projection of $p(x,-1,z,t)$ (equation \ref{eqn:fourierseries}) to be
orthogonal to polynomials of degree $2$, i.e.,
\begin{equation} \label{eqn:proj}
\int_{\Gamma_e}\left((p(x,-1,z,t)-\overline{p(x,-1,z,t)}\Big\rvert_{\Gamma_e}\right)P_j(x)P_k(z)\,\mathrm{dx}\,\mathrm{dz}=0;\,j,k=0,1,2.
\end{equation}
We can show that the expression for $\alpha^e_{j,k}(t)$ is 
\begin{equation}
\alpha^e_{j,k}(t)=\sum_{n=-N^f_t/2}^{N^f_t/2-1}\left(\sum_{l=-N^f_x/2}^{N^f_x/2-1}\sum_{m=-N^f_z/2}^{N^f_z/2-1}\hat{p}_{l,m,n}\beta^{e,l,m}_{j.k}\right)e^{i\omega_nt},
\end{equation}
where $\beta^{e,l,m}_{j,k}$ is the coefficient of the projection of $e^{i\left(k_lx+k_mz\right)}$, i.e.,
\begin{equation}
\overline{e^{i\left(k_lx+k_mz\right)}}\Big\rvert_{\Gamma_e}=\sum_{j,k=0}^2\beta_{j,k}^{e,l,m}P_j(x)P_k(z),
\end{equation}
We define $\overline{e^{i\left(k_lx+k_mz\right)}}\Big\rvert_{\Gamma_e}$ by replacing $p(x,-1,z,t)$ in equation \ref{eqn:proj} by $e^{i\left(k_lx+k_mz\right)}$. For implementation details, see appendix. We use the obtained $\overline{p(x,-1,z,t)}\Big\rvert_{\Gamma_e}$ to compute the surface force instead of $p(x,-1,z,t)$ since
\begin{equation}
\int_{\Gamma_e}p(x,-1,z,t)N_i(x,z)\,\mathrm{dx}\,\mathrm{dz}=\int_{\Gamma_e}\overline{p(x,-1,z,t)}\Big\rvert_{\Gamma_e}N_i(x,z)\,\mathrm{dx}\,\mathrm{dz}.
\end{equation}
The above equality holds since we can express $N_i(x,z)$ as a combination of the polynomial basis functions $\{P_i(x)P_k(z)\}_{i,k=0}^2$ used to perform the projection (equation \ref{eqn:proj}). Such an expression for $N_i(x,z)$ is possible because i) the surface element is a Cartesian domain, and ii) the degree of polynomial used to represent the FEM solution is less than or equal to the degree of the polynomial used to perform the above projection.

\bibliographystyle{cas-model2-names}

\bibliography{papers}

\begin{thebibliography}{29}
\expandafter\ifx\csname natexlab\endcsname\relax\def\natexlab#1{#1}\fi
\providecommand{\url}[1]{\texttt{#1}}
\providecommand{\href}[2]{#2}
\providecommand{\path}[1]{#1}
\providecommand{\DOIprefix}{doi:}
\providecommand{\ArXivprefix}{arXiv:}
\providecommand{\URLprefix}{URL: }
\providecommand{\Pubmedprefix}{pmid:}
\providecommand{\doi}[1]{\href{http://dx.doi.org/#1}{\path{#1}}}
\providecommand{\Pubmed}[1]{\href{pmid:#1}{\path{#1}}}
\providecommand{\bibinfo}[2]{#2}
\ifx\xfnm\relax \def\xfnm[#1]{\unskip,\space#1}\fi
\bibitem[{Anantharamu and Mahesh(2019)}]{anantharamu2019analysis}
\bibinfo{author}{Anantharamu, S.}, \bibinfo{author}{Mahesh, K.},
  \bibinfo{year}{2019}.
\newblock \bibinfo{title}{Analysis of wall-pressure fluctuation sources from
  {DNS} of turbulent channel flow.} \bibinfo{note}{Under review in Journal of
  Fluid Mechanics (arXiv preprint arXiv:1911.08534v2)}.
\bibitem[{Bathe(2006)}]{bathe2006finite}
\bibinfo{author}{Bathe, K.J.}, \bibinfo{year}{2006}.
\newblock \bibinfo{title}{Finite element procedures}.
\newblock \bibinfo{publisher}{Klaus-Jurgen Bathe}.
\bibitem[{Bendat and Piersol(2011)}]{bendat2011random}
\bibinfo{author}{Bendat, J.S.}, \bibinfo{author}{Piersol, A.G.},
  \bibinfo{year}{2011}.
\newblock \bibinfo{title}{Random data: analysis and measurement procedures}.
  volume \bibinfo{volume}{729}.
\newblock \bibinfo{publisher}{John Wiley \& Sons}.
\bibitem[{Bernardini et~al.(2013)Bernardini, Pirozzoli, Quadrio and
  Orlandi}]{Bernardini2013turbulent}
\bibinfo{author}{Bernardini, M.}, \bibinfo{author}{Pirozzoli, S.},
  \bibinfo{author}{Quadrio, M.}, \bibinfo{author}{Orlandi, P.},
  \bibinfo{year}{2013}.
\newblock \bibinfo{title}{Turbulent channel flow simulations in convecting
  reference frames}.
\newblock \bibinfo{journal}{Journal of Computational Physics}
  \bibinfo{volume}{232}, \bibinfo{pages}{1--6}.
\bibitem[{Blake(2017)}]{blake2017mechanics}
\bibinfo{author}{Blake, W.K.}, \bibinfo{year}{2017}.
\newblock \bibinfo{title}{Mechanics of Flow-Induced Sound and Vibration, Volume
  1 and 2.}
\newblock \bibinfo{publisher}{Academic Press}.
\bibitem[{Bull(1967)}]{bull1967wall}
\bibinfo{author}{Bull, M.K.}, \bibinfo{year}{1967}.
\newblock \bibinfo{title}{Wall-pressure fluctuations associated with subsonic
  turbulent boundary layer flow}.
\newblock \bibinfo{journal}{Journal of Fluid Mechanics} \bibinfo{volume}{28},
  \bibinfo{pages}{719--754}.
\bibitem[{Chang~III et~al.(1999)Chang~III, Piomelli and
  Blake}]{chang1999relationship}
\bibinfo{author}{Chang~III, P.A.}, \bibinfo{author}{Piomelli, U.},
  \bibinfo{author}{Blake, W.K.}, \bibinfo{year}{1999}.
\newblock \bibinfo{title}{Relationship between wall pressure and velocity-field
  sources}.
\newblock \bibinfo{journal}{Physics of Fluids} \bibinfo{volume}{11},
  \bibinfo{pages}{3434--3448}.
\bibitem[{Chase(1980)}]{chase1980modeling}
\bibinfo{author}{Chase, D.M.}, \bibinfo{year}{1980}.
\newblock \bibinfo{title}{Modeling the wavevector-frequency spectrum of
  turbulent boundary layer wall pressure}.
\newblock \bibinfo{journal}{Journal of Sound and Vibration}
  \bibinfo{volume}{70}, \bibinfo{pages}{29--67}.
\bibitem[{Corcos(1964)}]{corcos1964structure}
\bibinfo{author}{Corcos, G.M.}, \bibinfo{year}{1964}.
\newblock \bibinfo{title}{The structure of the turbulent pressure field in
  boundary-layer flows}.
\newblock \bibinfo{journal}{Journal of Fluid Mechanics} \bibinfo{volume}{18},
  \bibinfo{pages}{353--378}.
\bibitem[{Farabee and Casarella(1991)}]{farabee1991spectral}
\bibinfo{author}{Farabee, T.M.}, \bibinfo{author}{Casarella, M.J.},
  \bibinfo{year}{1991}.
\newblock \bibinfo{title}{Spectral features of wall pressure fluctuations
  beneath turbulent boundary layers}.
\newblock \bibinfo{journal}{Physics of Fluids A: Fluid Dynamics}
  \bibinfo{volume}{3}, \bibinfo{pages}{2410--2420}.
\bibitem[{Ghaemi and Scarano(2013)}]{ghaemi2013turbulent}
\bibinfo{author}{Ghaemi, S.}, \bibinfo{author}{Scarano, F.},
  \bibinfo{year}{2013}.
\newblock \bibinfo{title}{Turbulent structure of high-amplitude pressure peaks
  within the turbulent boundary layer}.
\newblock \bibinfo{journal}{Journal of Fluid Mechanics} \bibinfo{volume}{735},
  \bibinfo{pages}{381--426}.
\bibitem[{Goody(2004)}]{goody2004empirical}
\bibinfo{author}{Goody, M.}, \bibinfo{year}{2004}.
\newblock \bibinfo{title}{Empirical spectral model of surface pressure
  fluctuations}.
\newblock \bibinfo{journal}{AIAA Journal} \bibinfo{volume}{42},
  \bibinfo{pages}{1788--1794}.
\bibitem[{Hambric et~al.(2004)Hambric, Hwang and
  Bonness}]{hambric2004vibrations}
\bibinfo{author}{Hambric, S.A.}, \bibinfo{author}{Hwang, Y.F.},
  \bibinfo{author}{Bonness, W.K.}, \bibinfo{year}{2004}.
\newblock \bibinfo{title}{Vibrations of plates with clamped and free edges
  excited by low-speed turbulent boundary layer flow}.
\newblock \bibinfo{journal}{Journal of Fluids and Structures}
  \bibinfo{volume}{19}, \bibinfo{pages}{93--110}.
\bibitem[{Han et~al.(1999)Han, Bernhard and Mongeau}]{han1999prediction}
\bibinfo{author}{Han, F.}, \bibinfo{author}{Bernhard, R.J.},
  \bibinfo{author}{Mongeau, L.G.}, \bibinfo{year}{1999}.
\newblock \bibinfo{title}{Prediction of flow-induced structural vibration and
  sound radiation using energy flow analysis}.
\newblock \bibinfo{journal}{Journal of Sound and Vibration}
  \bibinfo{volume}{227}, \bibinfo{pages}{685--709}.
\bibitem[{Hoyas and Jim{\'e}nez(2006)}]{hoyas2006scaling}
\bibinfo{author}{Hoyas, S.}, \bibinfo{author}{Jim{\'e}nez, J.},
  \bibinfo{year}{2006}.
\newblock \bibinfo{title}{Scaling of the velocity fluctuations in turbulent
  channels up to re $\tau$= 2003}.
\newblock \bibinfo{journal}{Physics of fluids} \bibinfo{volume}{18},
  \bibinfo{pages}{011702}.
\bibitem[{Hu et~al.(2006)Hu, Morfey and Sandham}]{hu2006wall}
\bibinfo{author}{Hu, Z.}, \bibinfo{author}{Morfey, C.L.},
  \bibinfo{author}{Sandham, N.D.}, \bibinfo{year}{2006}.
\newblock \bibinfo{title}{Wall pressure and shear stress spectra from direct
  simulations of channel flow}.
\newblock \bibinfo{journal}{AIAA Journal} \bibinfo{volume}{44},
  \bibinfo{pages}{1541--1549}.
\bibitem[{Hwang(1998)}]{hwang1998discrete}
\bibinfo{author}{Hwang, Y.F.}, \bibinfo{year}{1998}.
\newblock \bibinfo{title}{A discrete model of turbulence loading function for
  computation of flow-induced vibration and noise}, in:
  \bibinfo{booktitle}{Proceedings of the ASME International Mechanical
  Engineering Congress and Exposition, Anaheim, CA}.
\bibitem[{Hwang and Maidanik(1990)}]{hwang1990wavenumber}
\bibinfo{author}{Hwang, Y.F.}, \bibinfo{author}{Maidanik, G.},
  \bibinfo{year}{1990}.
\newblock \bibinfo{title}{A wavenumber analysis of the coupling of a structural
  mode and flow turbulence}.
\newblock \bibinfo{journal}{Journal of Sound and Vibration}
  \bibinfo{volume}{142}, \bibinfo{pages}{135--152}.
\bibitem[{Kl{\"o}ppel et~al.(2011)Kl{\"o}ppel, Gee and
  Wall}]{kloppel2011scaled}
\bibinfo{author}{Kl{\"o}ppel, T.}, \bibinfo{author}{Gee, M.W.},
  \bibinfo{author}{Wall, W.A.}, \bibinfo{year}{2011}.
\newblock \bibinfo{title}{A scaled thickness conditioning for solid-and
  solid-shell discretizations of thin-walled structures}.
\newblock \bibinfo{journal}{Computer Methods in Applied Mechanics and
  Engineering} \bibinfo{volume}{200}, \bibinfo{pages}{1301--1310}.
\bibitem[{Mahesh et~al.(2004)Mahesh, Constantinescu and
  Moin}]{mahesh2004numerical}
\bibinfo{author}{Mahesh, K.}, \bibinfo{author}{Constantinescu, G.},
  \bibinfo{author}{Moin, P.}, \bibinfo{year}{2004}.
\newblock \bibinfo{title}{A numerical method for large-eddy simulation in
  complex geometries}.
\newblock \bibinfo{journal}{Journal of Computational Physics}
  \bibinfo{volume}{197}, \bibinfo{pages}{215--240}.
\bibitem[{Maxit(2016)}]{maxit2016simulation}
\bibinfo{author}{Maxit, L.}, \bibinfo{year}{2016}.
\newblock \bibinfo{title}{Simulation of the pressure field beneath a turbulent
  boundary layer using realizations of uncorrelated wall plane waves}.
\newblock \bibinfo{journal}{The Journal of the Acoustical Society of America}
  \bibinfo{volume}{140}, \bibinfo{pages}{1268--1285}.
\bibitem[{Panton et~al.(2017)Panton, Lee and Moser}]{panton2017correlation}
\bibinfo{author}{Panton, R.L.}, \bibinfo{author}{Lee, M.},
  \bibinfo{author}{Moser, R.D.}, \bibinfo{year}{2017}.
\newblock \bibinfo{title}{Correlation of pressure fluctuations in turbulent
  wall layers}.
\newblock \bibinfo{journal}{Physical Review Fluids} \bibinfo{volume}{2},
  \bibinfo{pages}{094604}.
\bibitem[{Pope(2001)}]{pope2001turbulent}
\bibinfo{author}{Pope, S.B.}, \bibinfo{year}{2001}.
\newblock \bibinfo{title}{Turbulent flows}.
\newblock \bibinfo{publisher}{Cambridge University Press}.
\bibitem[{Powell(1981)}]{powell1981approximation}
\bibinfo{author}{Powell, M.J.D.}, \bibinfo{year}{1981}.
\newblock \bibinfo{title}{Approximation theory and methods}.
\newblock \bibinfo{publisher}{Cambridge university press}.
\bibitem[{Rogallo(1981)}]{rogallo1981numerical}
\bibinfo{author}{Rogallo, R.S.}, \bibinfo{year}{1981}.
\newblock \bibinfo{title}{Numerical experiments in homogeneous turbulence} .
\bibitem[{Rosti and Brandt(2017)}]{rosti2017numerical}
\bibinfo{author}{Rosti, M.E.}, \bibinfo{author}{Brandt, L.},
  \bibinfo{year}{2017}.
\newblock \bibinfo{title}{Numerical simulation of turbulent channel flow over a
  viscous hyper-elastic wall}.
\newblock \bibinfo{journal}{Journal of Fluid Mechanics} \bibinfo{volume}{830},
  \bibinfo{pages}{708--735}.
\bibitem[{Schlichting(1979)}]{schlichting1979boundary}
\bibinfo{author}{Schlichting, H.}, \bibinfo{year}{1979}.
\newblock \bibinfo{title}{Boundary-layer theory}, in:
  \bibinfo{booktitle}{Boundary-layer theory}. \bibinfo{publisher}{McGraw-Hill}.
\bibitem[{Smol'Iakov and Tkachenko(1991)}]{smol1991models}
\bibinfo{author}{Smol'Iakov, A.V.}, \bibinfo{author}{Tkachenko, V.M.},
  \bibinfo{year}{1991}.
\newblock \bibinfo{title}{Models of a field of pseudoacoustic turbulent wall
  pressures and experimental data}.
\newblock \bibinfo{journal}{Akusticheskii Zhurnal} \bibinfo{volume}{37},
  \bibinfo{pages}{1199--1207}.
\bibitem[{Zhang et~al.(2017)Zhang, Wang, Blake and Katz}]{zhang2017deformation}
\bibinfo{author}{Zhang, C.}, \bibinfo{author}{Wang, J.},
  \bibinfo{author}{Blake, W.}, \bibinfo{author}{Katz, J.},
  \bibinfo{year}{2017}.
\newblock \bibinfo{title}{Deformation of a compliant wall in a turbulent
  channel flow}.
\newblock \bibinfo{journal}{Journal of Fluid Mechanics} \bibinfo{volume}{823},
  \bibinfo{pages}{345--390}.

\end{thebibliography}

\end{document}